\documentclass[aps,prd,preprintnumbers,superscriptaddress,nofootinbib,amssymb,notitlepage,eqsecnum]{revtex4-2}
\usepackage{amsmath,amsthm,amssymb,amsfonts}
\usepackage{bm}
\usepackage{dcolumn}
\usepackage{graphicx}
\usepackage{hyperref}
\usepackage{color}
\allowdisplaybreaks[1]

\newcommand{\be}{\begin{equation}}
\newcommand{\ee}{\end{equation}}
\newcommand{\ba}{\begin{eqnarray}}
\newcommand{\ea}{\end{eqnarray}}

\newcommand{\Mpl}{M_{\rm Pl}}
\newcommand{\delr}{\Delta r}
\newcommand{\mF}{\mathcal{F}}
\newcommand{\mG}{\mathcal{G}}
\newcommand{\mH}{\mathcal{H}}
\newcommand{\mK}{\mathcal{K}}

\newcommand{\mA}{\mathcal{A}}

\newcommand{\matrixsym}[1]{\bm{\mathit{#1}}}

\begin{document}

\preprint{YITP-26-118, WUCG-26-09}

\title{Perturbations and stability of black holes with static scalar hair \\
in general GLPV theories}

\author{Petarpa~Boonserm}
\email{Petarpa.Boonserm@gmail.com}
\affiliation{Department of Mathematics and Computer Science, Faculty of Science,
Chulalongkorn University, Phayathai Road, Pathumwan,
Bangkok 10330, Thailand}

\author{Antonio~De~Felice}
\email{antonio.defelice@yukawa.kyoto-u.ac.jp}
\affiliation{Center for Gravitational Physics and Quantum Information,
Yukawa Institute for Theoretical Physics, Kyoto University, Kyoto 606-8502, Japan}

\author{Ratchaphat~Nakarachinda}
\email{ratchaphat.n@rumail.ru.ac.th}
\affiliation{Quantum and Gravity Theory Research Group, Department of Physics,
Faculty of Science, Ramkhamhaeng University, 282 Ramkhamhaeng Road,
Hua Mak, Bang Kapi, Bangkok 10240, Thailand}

\author{Shinji~Tsujikawa}
\email{tsujikawa@waseda.jp}
\affiliation{Department of Physics, Waseda University,
3-4-1 Okubo, Shinjuku, Tokyo 169-8555, Japan}

\author{Pitayuth~Wongjun}
\email{pitayuthw@nu.ac.th}
\affiliation{The Institute for Fundamental Study (IF), Naresuan University,
99 Moo 9, Tha Pho, Mueang Phitsanulok, Phitsanulok 65000, Thailand}

\begin{abstract}
We derive the background equations and complete odd- and even-parity
quadratic actions for static, spherically symmetric black holes with a radial
scalar profile in general quartic--quintic
Gleyzes--Langlois--Piazza--Vernizzi (GLPV) theories, including 
Horndeski as a subclass.
On regular, nondegenerate branches, the formulation applies across Killing
horizons and yields local no-ghost conditions and radial and angular eikonal
characteristics. Tensor modes in both parity sectors share the same squared
radial speed. When the radial coordinate is timelike, generic antisymmetric
mixing induces nonstandard large-multipole scaling for the two even-parity
angular branches. Requiring both branches to have finite, nonzero 
eikonal phase speeds selects
the Horndeski-related compatibility condition between the quartic and quintic
beyond-Horndeski functions. This condition aligns the corresponding
covariant degeneracy directions and permits a local disformal map to
Horndeski, provided it is regular and invertible. For a finite, nonzero scalar
kinetic term at the horizon, every nontrivial branch of the known exact
quadratic GLPV black hole is locally unstable or has a degenerate tensor cone
near a simple outer horizon. 
Apart from identified exceptions, this obstruction extends to general
shift- and reflection-symmetric quadratic GLPV theories and the regular
Horndeski-related branch of the broader quartic--quintic class.
Finally, for scalar--Gauss--Bonnet black holes with a
vanishing horizon scalar kinetic term, we construct analytic power-law
quartic beyond-Horndeski deformations whose associated quintic function is
fixed by the same condition. At sufficiently small coupling, all local
no-ghost and high-frequency gradient-stability conditions hold throughout the
exterior. For a linear Gauss--Bonnet coupling, we estimate an interior scale
below which the background and stability expansions lose perturbative
control; this does not imply a physical instability.
\end{abstract}

\date{\today}
\maketitle

\section{Introduction}
\label{introsec}

General relativity (GR) has passed precision tests in the Solar System and
binary-pulsar systems~\cite{Will:2014kxa}, while gravitational-wave observations
and black-hole (BH) imaging now probe gravity in genuinely strong-field regimes
\cite{LIGOScientific:2017vwq,LIGOScientific:2017zic,
EventHorizonTelescope:2019dse}.  On cosmological scales, the discovery of
late-time acceleration
\cite{SupernovaSearchTeam:1998fmf,SupernovaCosmologyProject:1998vns} and
independent evidence from cosmic microwave background (CMB) and baryon acoustic
oscillation (BAO) measurements \cite{WMAP:2003elm,Eisenstein:2005su} point to a
dominant dark-energy (DE) component whose microscopic origin remains unknown.
The large hierarchy between its measured density and the natural vacuum-energy
scale \cite{Weinberg:1988cp,Copeland:2006wr} motivates dynamical alternatives
to a cosmological constant and extensions of GR.

Recent DESI BAO measurements, combined with CMB and Type Ia supernova data,
have indicated a preference for dynamical DE over a cosmological constant
\cite{DESI:2024mwx,DESI:2025DR2}.  A scalar field is one of the simplest
candidates for such dynamical DE.  Scalar fields can also constitute dark
matter (DM) and, particularly in the ultralight-mass regime, affect BH dynamics
and observables through accumulation and superradiance
\cite{Hu:2000ke,Hui:2016ltb,Hui:2019aqm,Brito:2015oca}.  Axion-like fields
provide concrete examples: they can form superradiant clouds around rotating
BHs \cite{Arvanitaki:2010sy,Yoshino:2012kn,Yoshino:2013ofa,
Arvanitaki:2014wva,Arvanitaki:2016qwi,Brito:2017wnc}, while an axion--photon
coupling can support charged BHs with axion hair whose quasinormal-mode spectra
and greybody factors carry characteristic signatures
\cite{DeFelice:2024axion,Nakarachinda:2025axion}.  Compact objects therefore
provide sensitive probes
of new fields and strong-field departures from GR
\cite{Berti:2015itd,Barack:2018yly,Berti:2018cxi,Berti:2018vdi}.  In particular,
a nontrivial scalar profile surrounding a BH directly probes gravity beyond GR.

Horndeski theories constitute the most general class of four-dimensional
scalar-tensor theories yielding second-order Euler--Lagrange equations
\cite{Horndeski:1974wa}.  In their modern formulation, they are characterized
by four arbitrary functions $G_i(\phi,X)$, with $i=2,3,4,5$, of the scalar
field $\phi$ and its kinetic term
$X=-\nabla^\mu\phi\nabla_\mu\phi/2$
\cite{Def11,Kobayashi:2011nu,Charmousis:2011bf}.
The second-order structure avoids the additional Ostrogradsky mode normally
associated with nondegenerate higher-derivative theories
\cite{Woodard:2015zca}, leaving one scalar degree of freedom in addition to
the two tensor polarizations.  Nonlinear derivative interactions such as
Galileons \cite{Nicolis:2008in,Deffayet:2009wt} can also suppress the
scalar-mediated fifth force through the Vainshtein mechanism
\cite{Vainshtein:1972sx,Burrage:2010rs,Kimura:2011dc,Koyama:2013paa}, making it
possible to reconcile cosmological modifications with local tests of gravity.

In GR, stationary asymptotically flat electrovacuum BHs are characterized by
their mass, angular momentum, and electric charge
\cite{Israel:1967wq,Carter:1971zc,Ruffini:1971bza,Hawking:1971vc}, while
related no-hair theorems constrain scalar hair
\cite{Bekenstein:1972ny,Hawking:1972qk,Bekenstein:1995un,
Graham:2014mda,Faraoni:2017ock}.  Similar restrictions follow from the
Hui--Nicolis no-hair argument in shift-symmetric Horndeski theories, provided
that the norm of the shift current is regular at the horizon and its radial
component $J^r$ is proportional to $\phi'\equiv {\rm d}\phi/{\rm d}r$, with
$r$ the areal radius and the proportionality coefficient finite and nonzero
as $\phi'\to0$.  Under these assumptions, a static,
spherically symmetric BH cannot support a nontrivial static scalar profile
approaching a constant at infinity \cite{Hui:2012qt}.

These assumptions can be evaded in several ways.  Nonminimal derivative
couplings to curvature admit BHs with static radial scalar profiles that are
not asymptotically Minkowski
\cite{Kolyvaris:2011fk,Rinaldi:2012vy,Anabalon:2013oea,
Minamitsuji:2013ura,Cisterna:2015uya}.
Scalar--Gauss--Bonnet (sGB) theories with an interaction of the form
$\xi(\phi)R_{\rm GB}^2$, where $R_{\rm GB}^2$ is the GB invariant, provide
another loophole.  If $\xi(\phi)$ is nonlinear in $\phi$, this interaction
breaks the continuous shift symmetry.  By contrast, for a linear coupling
$\xi(\phi)=\xi_0+\xi_1\phi$, the interaction is shift symmetric up to a
topological boundary term, but the associated radial shift current $J^r$
contains a term independent of $\phi'$.  The proportionality assumption used
in the Hui--Nicolis theorem is therefore violated.  Such theories admit
asymptotically Minkowski hairy BHs with dilatonic or shift-symmetric linear
couplings, as well as spontaneously scalarized solutions
\cite{Kanti:1995vq,Torii:1996yi,Kanti:1997br,Chen:2006ge,Guo:2008hf,
Pani:2009wy,Kleihaus:2011tg,Sotiriou:2013qea,Ayzenberg:2014aka,
Sotiriou:2014pfa,Maselli:2015tta,Kleihaus:2015aje,Doneva:2017bvd,
Silva:2017uqg,Antoniou:2017acq,Minamitsuji:2018xde}.
Another loophole retains a stationary metric while allowing the scalar to
depend linearly on the time coordinate $t$, $\phi=q_t t+\Psi(r)$, where
$q_t\ne0$ is constant and $\Psi(r)$ depends only on the areal radius $r$
\cite{Babichev:2013cya,Kobayashi:2014eva,Babichev:2015rva,
Babichev:2016fbg}.  We focus instead on the strictly static profile
$\phi=\phi(r)$.

The existence of a background solution is only the first viability test.  The
odd- and even-parity perturbations of static, spherically symmetric Horndeski
backgrounds were formulated in
Refs.~\cite{Kobayashi:2012kh,Kobayashi:2014wsa,Kase:2021mix,Kase:2023mho},
yielding high-frequency conditions for the absence of ghosts and radial or
angular Laplacian instabilities.  These conditions were applied first to
shift-symmetric theories with a static scalar \cite{Minamitsuji:2022mlv} and
then to the full $G_i(\phi,X)$ class \cite{Minamitsuji:2022vbi}.  At a
nonextremal horizon $r=r_s$, the inverse radial metric component
$h=g^{rr}$ behaves as $h\propto r-r_s$.  Since $X=-h\phi'^2/2$, a finite
nonzero value $X_s\equiv X(r_s)$ requires
$\phi'\propto(r-r_s)^{-1/2}$.  Except for special simultaneous degeneracies
requiring a separate dominant-balance analysis, one generically finds either
a divergent radial propagation speed or mutually incompatible no-ghost and
radial/angular gradient-stability conditions
\cite{Minamitsuji:2022mlv,Minamitsuji:2022vbi}. Thus, static hairy BHs on the
$X_s\ne0$ branch are generically unstable arbitrarily close to the horizon,
excluding the known nonminimal-derivative-coupling solutions as stable
candidates \cite{Kolyvaris:2011fk,Rinaldi:2012vy,Anabalon:2013oea,
Minamitsuji:2013ura,Cisterna:2015uya}.  The same instability was recently shown
to obstruct globally regular Horndeski BHs, since this horizon branch is
eliminated before central regularity is imposed \cite{DeFelice:2026rbi}.

For the complementary branch $X_s=0$, regular positive-power functions of
$\phi$ and $X$ generically lead back to the asymptotically Minkowski,
hairless Schwarzschild branch \cite{Minamitsuji:2022mlv,Minamitsuji:2022vbi}.
A well-established exception is an sGB coupling, whose Horndeski
representation is nonanalytic in $X$ and contains a
logarithmic quintic interaction.  Weakly coupled sGB BHs, as well as suitable
scalarized branches, can satisfy the local odd- and even-parity stability
conditions outside the horizon
\cite{Silva:2018qhn,Blazquez-Salcedo:2018jnn,
Blazquez-Salcedo:2020rhf,Blazquez-Salcedo:2020caw,
Minamitsuji:2022vbi,Minamitsuji:2024twp}.
Hence, among presently known asymptotically Minkowski
Horndeski solutions with a static scalar, sGB interactions provide the main
class supporting scalar hair while satisfying all local high-frequency
stability conditions throughout the exterior.

This restrictive situation motivates the search for stable hairy BHs in
extensions of Horndeski theories.
In a $3+1$ Arnowitt--Deser--Misner (ADM) formulation
\cite{Arnowitt:1962hi}, GLPV theories are obtained by relaxing the two
relations characterizing the quartic and quintic Horndeski Lagrangians
\cite{Gleyzes:2014dya}.  Covariantly, these departures are encoded by the
beyond-Horndeski functions $F_4(\phi,X)$ and $F_5(\phi,X)$.
Although the resulting field equations can contain higher derivatives,
degeneracy can remove the would-be extra scalar mode.  For the combined
quartic--quintic theory to retain its primary constraint, the covariant
degeneracy directions of the two sectors must coincide.  This yields the
$F_4$--$F_5$ compatibility condition for a common local disformal map to
Horndeski, provided the map is regular and invertible.
Invertible disformal transformations, introduced by Bekenstein
\cite{Bekenstein:1992pj}, preserve the number of propagating degrees of freedom
and clarify the relation of GLPV theories to Horndeski and the broader DHOST
framework
\cite{Gleyzes:2014qga,Domenech:2015tca,Langlois:2015cwa,
Crisostomi:2016tcp,Crisostomi:2016czh,BenAchour:2016fzp};
the transformation properties of cosmological perturbations under such maps
have also been studied extensively
\cite{Minamitsuji:2014waa,Tsujikawa:2014uza,Watanabe:2015uqa,
Motohashi:2015dis,Domenech:2015hka,Tsujikawa:2015upa}.
Early studies of spherically symmetric configurations addressed central
regularity and Vainshtein screening
\cite{DeFelice:2015GLPV,Kase:2015GLPV}.  
Because GLPV interactions
modify both the BH background and the perturbation principal matrices, the
near-horizon instability result derived in Horndeski theories for the
$X(r_s)\ne0$ branch cannot be applied directly.

Several classes of spherically symmetric BH solutions have been explored in
beyond-Horndeski and DHOST theories.  For a strictly radial scalar profile
$\phi=\phi(r)$, asymptotically flat BH and compact-object solutions were
constructed in Refs.~\cite{Babichev:2017guv,Bakopoulos:2022bho}.  A distinct
class employs the linearly time-dependent ansatz $\phi=q_t t+\Psi(r)$ in
shift-symmetric quadratic beyond-Horndeski and DHOST theories, yielding stealth
and primary-hair solutions
\cite{Motohashi:2019sen,Takahashi:2020hso,Bakopoulos:2023psh,
Baake:2023bph,Bakopoulos:2023cpp}.  Because the two scalar ansatzes differ in
their horizon regularity and constraint structures, their stability must be
analyzed separately.  For general spherically symmetric backgrounds, previous
studies have developed odd-parity perturbation theory, master-variable and
initial-value formulations, and a complete set of high-frequency stability
criteria
\cite{Takahashi:2019oxz,Takahashi:2021bml,Nakashi:2022wdg,
Mironov:2024bho}.  Odd-parity (axial) and radial perturbations of recent
primary-hair solutions with $\phi=q_t t+\Psi(r)$ have also been analyzed
\cite{Charmousis:2025axial,Charmousis:2026radial}.
Nevertheless, a complete odd- and even-parity perturbation analysis of
non-shift-symmetric quartic--quintic GLPV theories with a strictly static
scalar, covering regions in which either $t$ or $r$ is timelike, has so far
been lacking.

In this paper, we perform a complete quadratic-order analysis of odd- and
even-parity perturbations of static, spherically symmetric configurations
with a radial scalar profile in general quartic--quintic GLPV theories,
including Horndeski as a subclass.  We derive the local no-ghost and
high-frequency gradient-stability conditions in regions where either $t$ or
$r$ is timelike.  On regular, nondegenerate branches, each multipole with 
$\ell\geq2$ contains one tensor mode in the odd sector and 
two coupled tensor--scalar modes in the even sector.
The odd dipole carries no local propagating degree of freedom, 
whereas the even monopole and dipole each carry one.
The tensor modes share the same
radial cone.  When $r$ is timelike, the squared radial speeds are the
reciprocals of their $t$-timelike counterparts, while generic antisymmetric
mixing produces nonstandard even-parity angular dispersion.  Away from
accidental prefactor zeros, requiring finite, nonzero, scale-independent
eikonal phase speeds for both angular branches selects the Horndeski-related
$F_4$--$F_5$ compatibility condition.

For finite nonzero $X_s$, every nontrivial branch of the known
exact quadratic GLPV BH is locally unstable or has a degenerate tensor cone
near a simple outer horizon.  Apart from identified exceptional branches,
this obstruction extends to general shift- and reflection-symmetric
quadratic GLPV theories and to the regular Horndeski-related
quartic--quintic branch.

For the complementary $X_s=0$ branch, we construct analytic power-law
beyond-Horndeski deformations of weakly coupled sGB BHs, with the linked
quintic function fixed by the same compatibility condition.  At sufficiently
small coupling, all local no-ghost and high-frequency gradient-stability
conditions, including those for the even monopole and dipole, hold throughout
the exterior.  For a linear GB coupling, we identify an interior scale below
which the background and stability expansions lose perturbative control.
This does not imply a physical instability.

This paper is organized as follows.  In Sec.~\ref{backgroundgeneralsec}, we
introduce the GLPV action and derive the background equations.
Section~\ref{oddsec} treats the odd-parity higher multipoles and dipole, while
Sec.~\ref{evensec} analyzes the even-parity $\ell\geq2$, monopole, and dipole
sectors where either $t$ or $r$ is timelike.  The unreduced even-parity
coefficients are collected in Appendix~\ref{AppEvenCoefficients}.  In
Sec.~\ref{horizonInstabilitySec}, we study three near-horizon settings with
$X_s\ne0$.  Section~\ref{sGBsec} treats the complementary weakly coupled sGB
branch, its compatible quartic--quintic deformation, exterior stability, and
the interior perturbative matching scale.  We summarize our conclusions in
Sec.~\ref{consec}.

\section{GLPV theories and black-hole backgrounds}
\label{backgroundgeneralsec}

%
\subsection{Action and conventions}

We consider general GLPV theories with the
action~\cite{Gleyzes:2014dya}
\be
S=\int {\rm d}^4 x \sqrt{-g}\,
\left({\cal L}_2+{\cal L}_3+{\cal L}_4+{\cal L}_5
+{\cal L}_4^{\rm bH}+{\cal L}_5^{\rm bH}\right),
\label{action}
\ee
where $g_{\mu\nu}$ is the spacetime metric and $g=\det(g_{\mu\nu})$.
The individual Lagrangians are
\ba
{\cal L}_2 &=& G_2(\phi,X),\notag\\
{\cal L}_3 &=&-G_3(\phi,X)\square\phi,\notag\\
{\cal L}_4 &=&G_4(\phi,X)R+G_{4,X}(\phi,X)
\left[(\square\phi)^2
-(\nabla_\mu\nabla_\nu\phi)(\nabla^\mu\nabla^\nu\phi)\right],\notag\\
{\cal L}_5 &=&G_5(\phi,X)G_{\mu\nu}\nabla^\mu\nabla^\nu\phi
-\frac{G_{5,X}(\phi,X)}{6}\left[(\square\phi)^3
-3\square\phi\,(\nabla_\mu\nabla_\nu\phi)
(\nabla^\mu\nabla^\nu\phi)
+2(\nabla_\mu\nabla^\nu\phi)(\nabla_\nu\nabla^\rho\phi)
(\nabla_\rho\nabla^\mu\phi)\right],\notag\\
{\cal L}_4^{\rm bH} &=&F_4(\phi,X)
\epsilon^{\mu\nu\rho\sigma}\epsilon_{\mu'\nu'\rho'\sigma}
\nabla^{\mu'}\phi\nabla_\mu\phi
\nabla^{\nu'}\nabla_\nu\phi
\nabla^{\rho'}\nabla_\rho\phi,\notag\\
{\cal L}_5^{\rm bH} &=&F_5(\phi,X)
\epsilon^{\mu\nu\rho\sigma}\epsilon_{\mu'\nu'\rho'\sigma'}
\nabla^{\mu'}\phi\nabla_\mu\phi
\nabla^{\nu'}\nabla_\nu\phi
\nabla^{\rho'}\nabla_\rho\phi
\nabla^{\sigma'}\nabla_\sigma\phi\,,
\label{bHLagrangians}
\ea
with
\be
 X=-\frac12g^{\mu\nu}\nabla_\mu\phi\nabla_\nu\phi\,.
\label{Xdefinition}
\ee
Here, $G_i(\phi,X)$ ($i=2,3,4,5$), $F_4(\phi,X)$, and $F_5(\phi,X)$
are arbitrary functions of $\phi$ and $X$.  Horndeski theories are recovered
by setting $F_4=F_5=0$.  We define
$\square\phi=\nabla_\mu\nabla^\mu\phi$, where $\nabla_\mu$ is the covariant
derivative operator.  The quantities $R$ and $G_{\mu\nu}$ denote the Ricci
scalar and Einstein tensor, respectively.  A comma in a function subscript
denotes partial differentiation with respect to the corresponding argument,
e.g., $G_{i,\phi}=\partial G_i/\partial\phi$ and
$G_{i,X}=\partial G_i/\partial X$.  We adopt the Levi--Civita tensor
conventions
$\epsilon_{0123}=+\sqrt{-g}$ and $\epsilon^{0123}=-1/\sqrt{-g}$, so that
$\epsilon^{\mu\nu\rho\sigma}\epsilon_{\alpha\beta\gamma\sigma}
=-\delta^{\mu\nu\rho}_{\alpha\beta\gamma}$, where
$\delta^{\mu\nu\rho}_{\alpha\beta\gamma}$ is the generalized Kronecker
delta.

The functions $F_4$ and $F_5$ are treated as independent in the background
and perturbation equations derived in
Secs.~\ref{backgroundgeneralsec}, \ref{oddsec}, and~\ref{evensec}.  The
quartic--quintic compatibility relation, introduced later in
Eq.~(\ref{GLPVcompatibility}), is imposed only in
Secs.~\ref{fullGLPVNoGoSubsec} and~\ref{sGBsec}.  It selects the regular
Horndeski-related subclass relevant to the near-horizon theorem in the
former and determines $F_5$ from $F_4$ in the latter.  This relation also
ensures that the quartic and quintic kinetic Hessians possess the same null
direction, so their sum retains the primary degeneracy constraint.  Thus,
the earlier equations with independent $F_4$ and $F_5$ are algebraically
general, whereas the compatible branch defines the fully covariantly
degenerate Horndeski-related subclass used in the subsequent physical
applications.
Shift symmetry is
imposed only in the applications considered in
Secs.~\ref{targetBHsubsec} and~\ref{generalGLPVNoGoSubsec}, for which
$G_i=G_i(X)$, $F_4=F_4(X)$, and $F_5=F_5(X)$.  The general equations
elsewhere retain explicit $\phi$ dependence and mixed $\phi$--$X$ derivatives.

We consider the spherically symmetric background ansatz
\be
{\rm d}s^2=-f(r){\rm d}t^2+h^{-1}(r){\rm d}r^2
+r^2{\rm d}\Omega_2^2,
\qquad \phi=\phi(r),
\label{metric}
\ee
where $f$, $h$, and $\phi$ are functions only of the areal radius $r$, and
${\rm d}\Omega_2^2$ is the line element on the unit two-sphere. For this
ansatz, $X=-h\phi'^2/2$, where a prime denotes differentiation with respect
to $r$.  In regions with $f,h>0$, the $t$ direction is timelike and the $r$
direction is spacelike, so a nontrivial real scalar profile has $X<0$.  For
$f,h<0$, their causal roles are reversed and $X>0$.  In deriving the
background and perturbation equations, we require only $f/h>0$ and do not
otherwise fix the common sign of $f$ and $h$.  The resulting equations
therefore apply to both causal configurations, irrespective of the number
and ordering of Killing horizons.  We define
\be
s\equiv\sqrt{\frac{f}{h}}>0\,,
\label{positiveSDefinition}
\ee
where the positive real branch of the square root is understood.

\subsection{Background equations}
\label{GLPVBackgroundSubsec}

For the spherically symmetric background ansatz in
Eq.~(\ref{metric}), the independent metric field equations can be chosen as
the $tt$, $rr$, and $\theta\theta$ components.  The
$\varphi\varphi$ component is equivalent to the $\theta\theta$ component by
spherical symmetry.  Denoting these three equations by
${\cal E}_{00}=0$, ${\cal E}_{11}=0$, and ${\cal E}_{22}=0$,
respectively, we obtain
\ba
{\cal E}_{00}&\equiv&
\left(C_1+\frac{C_2}{r}+\frac{C_3}{r^2}\right)\phi''
+\left(\frac{\phi'}{2h}C_1+\frac{C_4}{r}+\frac{C_5}{r^2}\right)h'
+C_6+\frac{C_7}{r}+\frac{C_8}{r^2}
+\frac{2}{r^2}\left({\cal Q}'-\frac{h'}{2h}{\cal Q}
-U_4\right)=0\,,
\label{EGLPV00}\\[-2pt]
{\cal E}_{11}&\equiv&
-\left(\frac{\phi'}{2h}C_1+\frac{C_4}{r}+\frac{C_5}{r^2}\right)\frac{hf'}{f}
+C_9-\frac{2\phi'}{r}C_1
-\frac{1}{r^2}\left[\frac{\phi'}{2h}C_2+(h-1)C_4\right]
-\frac{4}{r^2}V_4
-\frac{2f'}{r^2f}{\cal Q}_h=0\,,
\label{EGLPV11}\\[-2pt]
{\cal E}_{22}&\equiv&
\left[\left\{C_2+\frac{(2h-1)\phi'C_3+2hC_5}{h\phi'r}\right\}\frac{f'}{4f}
+C_1+\frac{C_2}{2r}\right]\phi''
+\frac{1}{4f}\left(2hC_4-\phi'C_2
+\frac{2hC_5-\phi'C_3}{r}\right)
\left(f''-\frac{f'^2}{2f}\right)
\notag\\
&&+\left[C_4+\frac{2h(2h+1)C_5-\phi'C_3}{2h^2r}\right]
\frac{f'h'}{4f}
+\left(\frac{C_7}{4}+\frac{C_{10}}{r}\right)\frac{f'}{f}
+\left(\frac{\phi'}{h}C_1+\frac{C_4}{r}\right)\frac{h'}{2}
+C_6+\frac{C_7}{2r}
\notag\\
&&+\frac{1}{r}\left[-\frac{f'}{f}U_4+{\cal Z}'
+\left(\frac{f'}{2f}-\frac{h'}{2h}\right){\cal Z}\right]=0\,.
\label{EGLPV22}
\ea
The ten combinations of the Horndeski functions appearing in
Eqs.~(\ref{EGLPV00})--(\ref{EGLPV22}) are
\ba
C_1&=&-h^2(G_{3,X}-2G_{4,\phi X})\phi'^2-2G_{4,\phi}h,
\notag\\
C_2&=&2h^3(2G_{4,XX}-G_{5,\phi X})\phi'^3
-4h^2(G_{4,X}-G_{5,\phi})\phi',
\notag\\
C_3&=&-h^4G_{5,XX}\phi'^4+h^2G_{5,X}(3h-1)\phi'^2,
\notag\\
C_4&=&h^2(2G_{4,XX}-G_{5,\phi X})\phi'^4
+h(3G_{5,\phi}-4G_{4,X})\phi'^2-2G_4,
\notag\\
C_5&=&-\frac12\left[G_{5,XX}h^3\phi'^5
-hG_{5,X}(5h-1)\phi'^3\right],
\notag\\
C_6&=&h(G_{3,\phi}-2G_{4,\phi\phi})\phi'^2+G_2,
\notag\\
C_7&=&-2h^2(2G_{4,\phi X}-G_{5,\phi\phi})\phi'^3
-4G_{4,\phi}h\phi',
\notag\\
C_8&=&G_{5,\phi X}h^3\phi'^4
-h(2G_{4,X}h-G_{5,\phi}h-G_{5,\phi})\phi'^2
-2G_4(h-1),
\notag\\
C_9&=&-h(G_{2,X}-G_{3,\phi})\phi'^2-G_2,
\notag\\
C_{10}&=&\frac12G_{5,\phi X}h^3\phi'^4
-\frac12h^2(2G_{4,X}-G_{5,\phi})\phi'^2-G_4h.
\label{Cicoefficients}
\ea
The beyond-Horndeski contributions are expressed in terms of the following
combinations:
\ba
U_4&\equiv&h^3\phi'^4F_4,\qquad
U_5\equiv h^4\phi'^5F_5,\qquad
U_{4,h}=h^2\phi'^4\left(3F_4+XF_{4,X}\right),\qquad
U_{5,h}=h^3\phi'^5\left(4F_5+XF_{5,X}\right), \notag\\
V_4&\equiv&hU_{4,h}-\frac12U_4
=h^3\phi'^4\left(\frac52F_4+XF_{4,X}\right),\qquad
V_5\equiv hU_{5,h}-\frac12U_5
=h^4\phi'^5\left(\frac72F_5+XF_{5,X}\right),\notag\\
{\cal Q}&\equiv&2rU_4+3U_5,\qquad
{\cal Q}_h\equiv2rV_4+3V_5,\qquad
{\cal Z}\equiv U_4\left(\frac{rf'}{f}+2\right)
+3\frac{f'}{f}U_5\,.
\label{GLPVMetricBuildingBlocks}
\ea
Here $U_{4,h}\equiv(\partial U_4/\partial h)_{\phi,\phi'}$ and
$U_{5,h}\equiv(\partial U_5/\partial h)_{\phi,\phi'}$.  Since
$X=-h\phi'^2/2$, these derivatives include the implicit $h$ 
dependence through
$(\partial X/\partial h)_{\phi,\phi'}=X/h$.  No additional
beyond-Horndeski building blocks involving derivatives with respect to
$\phi$ enter the metric equations.  By contrast, the subscript $h$ in
${\cal Q}_h=2rV_4+3V_5$ labels the combination generated by the metric
variation and does not denote $\partial{\cal Q}/\partial h$.  In deriving
Eq.~(\ref{EGLPV22}), we used the exact identity
$s'/s=f'/(2f)-h'/(2h)$.  Consequently,
Eqs.~(\ref{EGLPV00})--(\ref{GLPVMetricBuildingBlocks}) are manifestly real
for $f/h>0$ in both causal sign sectors, $f,h>0$ and $f,h<0$.

The complete scalar-field equation can be written as
\be
{\cal E}_{\phi}\equiv\frac{1}{r^2s}
\left(r^2sJ^r\right)'
+\frac{\partial{\cal P}}{\partial\phi}=0,
\label{EphiGLPV}
\ee
where $J^r$ is the radial scalar current.  Its beyond-Horndeski
contributions depend on
$U_{4,\phi'}\equiv(\partial U_4/\partial\phi')_{h,\phi}$ and
$U_{5,\phi'}\equiv(\partial U_5/\partial\phi')_{h,\phi}$, given by
\be
U_{4,\phi'}=h^3\phi'^3\left(4F_4+2XF_{4,X}\right),\qquad
U_{5,\phi'}=h^4\phi'^4\left(5F_5+2XF_{5,X}\right).
\label{GLPVScalarCurrentDerivatives}
\ee
These expressions include the implicit $\phi'$ dependence of $X$, with
$(\partial X/\partial\phi')_{h,\phi}=2X/\phi'$.  
The two quantities in Eq.~(\ref{EphiGLPV}) are
\ba
J^r&=&
\left(C_1+\frac{C_2}{r}+\frac{C_3}{r^2}\right)\frac{f'}{2f}
-\frac{C_6+C_9}{\phi'}+\frac{2C_1}{r}
+\frac{1}{r^2}\left[\frac{1+h}{2h}C_2
-\frac{C_4+C_8-2C_{10}}{\phi'}\right]
\notag\\
&&
+\frac{2}{r^2}\left(1+\frac{rf'}{f}\right)U_{4,\phi'}
+\frac{3f'}{r^2f}U_{5,\phi'}.
\label{JrGLPV}
\\[-2pt]
{\cal P}&=&
\left[C_1+\frac{1}{r^2}\left(\frac{C_3}{2h}-\frac{C_5}{\phi'}\right)\right]
\left(\phi''+\frac{\phi'h'}{2h}\right)
+\left[\frac{\phi'}{2}C_2-hC_4
+\frac{1}{2r}\left(\frac{\phi'}{2}C_3-hC_5\right)\right]
\frac{f'}{rf}
\notag\\
&&+C_6+\frac{1}{r^2}
\left(\frac{\phi'}{2}C_2-hC_4+C_8-2C_{10}\right)
-\frac{2}{r^2}\left(1+\frac{rf'}{f}\right)U_4
-\frac{3f'}{r^2f}U_5.
\label{PGLPV}
\ea
In evaluating $\partial{\cal P}/\partial\phi$, the quantities
$f$, $h$, $\phi'$, $f'$, $h'$, and $\phi''$ are held fixed.  The derivative
therefore acts only on the explicit $\phi$ dependence of the coefficients
$C_i$ and the functions $F_4(\phi,X)$ and $F_5(\phi,X)$ contained in $U_4$
and $U_5$.  
In the Horndeski limit, $F_4$ and $F_5$, together with all their
derivatives, vanish. 
Equations~(\ref{EGLPV00})--(\ref{PGLPV}) then reduce to the Horndeski
background equations of Refs.~\cite{Kobayashi:2012kh,Kobayashi:2014wsa}.

Only three of the four equations
${\cal E}_{00}=0$, ${\cal E}_{11}=0$, ${\cal E}_{22}=0$, and
${\cal E}_{\phi}=0$ are independent.  On a branch with $\phi'\ne0$, radial
diffeomorphism invariance yields the identity
\be
{\cal E}_{\phi}=-\frac{1}{\phi'}\left[
\frac{f'}{2f}{\cal E}_{00}
+\left({\cal E}_{11}\right)'
+\left(\frac{f'}{2f}+\frac{2}{r}\right){\cal E}_{11}
+\frac{2}{r}{\cal E}_{22}\right].
\label{backgroundBianchiIdentity}
\ee
This identity provides a nontrivial check of the terms arising from explicit
$\phi$ dependence.

\section{Odd-parity stability}
\label{oddsec}

In this section, we derive the quadratic action for odd-parity perturbations
and obtain the local no-ghost conditions and the principal-part conditions for
radial and angular gradient stability.  We consider the higher multipoles
$\ell\geq2$ and the dipole $\ell=1$ separately in the following subsections.
The local criteria derived in this and the next section are necessary but do
not constitute a proof of full mode stability: finite-wavelength and
low-frequency instabilities can depend on the full algebraic matrices (or
effective potentials in the one-field sectors), the background profile, and
the boundary conditions.

\subsection{Higher multipoles: $\ell\geq2$}
\label{oddellge2sec}

We decompose the metric into its background and perturbed parts as
$g_{\mu\nu}=\bar g_{\mu\nu}+\delta g_{\mu\nu}$, where
$\bar g_{\mu\nu}$ is the background metric in Eq.~(\ref{metric}) and
$\delta g_{\mu\nu}$ is its perturbation.  For $\ell\geq2$, we impose the
Regge--Wheeler gauge $\delta g_{ab}=0$.  The odd-parity perturbations, which
have parity $(-1)^{\ell+1}$, are then expressed as
\be
\delta g_{tt}=\delta g_{tr}=\delta g_{rr}=\delta g_{ab}=0,\qquad
\delta g_{ta}=\sum_{\ell}Q(t,r)E_a{}^b\nabla_bY_{\ell},\qquad
\delta g_{ra}=\sum_{\ell}W(t,r)E_a{}^b\nabla_bY_{\ell}\,,
\label{odddecompositiongeneral}
\ee
where $Q(t,r)$ and $W(t,r)$ are metric perturbation amplitudes depending on
$t$ and $r$.  Henceforth, we suppress the $\ell$ labels of all perturbation
amplitudes in both parity sectors.  The function
$Y_{\ell}\equiv Y_{\ell0}(\theta,\varphi)$ is the $m=0$ scalar spherical
harmonic on the unit two-sphere, normalized according to
\begin{equation}
\gamma^{ab}\nabla_a\nabla_bY_{\ell}
=-\ell(\ell+1)Y_{\ell},\qquad
\int {\rm d}\Omega_2\,Y_{\ell}Y_{\ell'}
=\delta_{\ell\ell'}.
\end{equation}
On a spherically symmetric background, modes with different $m$ decouple
and have identical quadratic actions, so we may set $m=0$ without loss of
generality.  This representative-mode convention is used in both parity
sectors, with the sum over $m$ restored when needed.
Here the indices $a,b$ run over $\theta,\varphi$, and $\gamma_{ab}$ and
$\nabla_a$ denote the metric on the unit two-sphere and its compatible
covariant derivative, respectively.  The tensor $E_{ab}$ is the
corresponding Levi--Civita tensor, with
$E_{\theta\varphi}=\sin\theta$, and
${\rm d}\Omega_2=\sin\theta\,{\rm d}\theta\,{\rm d}\varphi$.  Since
$\delta\phi$ has even parity, the odd sector contains no scalar-field
perturbation.  The auxiliary field $\chi$ introduced below is therefore a
metric master variable, not a scalar-field perturbation.

Consider the infinitesimal coordinate transformation
$x^\mu\to x^\mu-\zeta^\mu$, under which the metric perturbation transforms as
$\delta g_{\mu\nu}\to\delta g_{\mu\nu}
+\bar\nabla_\mu\zeta_\nu+\bar\nabla_\nu\zeta_\mu$, where $\bar\nabla_\mu$ is
the covariant derivative compatible with the background metric.  The
odd-parity gauge generator is given by
$\zeta_t=\zeta_r=0$ and
$\zeta_a=\Lambda(t,r)E_a{}^b\nabla_bY_{\ell}$, where $\Lambda(t,r)$ is an
arbitrary function.  Before imposing the Regge--Wheeler gauge, the
perturbation amplitudes transform as
\be
Q\to Q+\dot\Lambda,\qquad
W\to W+\Lambda'-\frac{2}{r}\Lambda\,,
\label{oddgaugetransformation}
\ee
where a dot denotes differentiation with respect to $t$.  For $\ell\geq2$,
the Regge--Wheeler gauge completely fixes this gauge freedom.  
In particular, the combination
\be
\dot W-Q'+\frac{2}{r}Q
\label{oddgaugeinvariantcombination}
\ee
is invariant under the transformation in
Eq.~(\ref{oddgaugetransformation}).

We expand the full action to second order in the odd-parity perturbations,
integrate over the angular coordinates, and use the background equations
(\ref{EGLPV00})--(\ref{EGLPV22}) to eliminate $G_2$, $G_{3,X}$, and
$f''$.  Defining the angular eigenvalue
\be
L\equiv\ell(\ell+1),
\label{oddLdefinition}
\ee
the resulting quadratic action on the background solution is
\ba
S_{\rm odd}^{(2)}&=&\sum_{\ell}L\int {\rm d}t\,{\rm d}r\,
{\cal L}_{\rm odd},\notag\\
{\cal L}_{\rm odd}&=&
\frac{1}{4s}\left\{
\mH\left(\dot W-Q'+\frac{2Q}{r}\right)^2
+\frac{L-2}{r^2}\left(
\frac{\mF}{h}Q^2-f\mG W^2\right)\right\},
\label{oddmastergeneral}
\ea
where
\begin{align}
\mH={}&2G_4+2h\phi'^2\left(G_{4,X}-\frac12G_{5,\phi}\right)
-\frac{h^2\phi'^3}{r}G_{5,X}
-2h^2\phi'^4F_4-\frac{6h^3\phi'^5}{r}F_5,
\label{mH} \\
\mF={}&2G_4+h\phi'^2G_{5,\phi}
-\frac12hh'\phi'^3G_{5,X}-h^2\phi'^2\phi''G_{5,X},
\label{mF} \\
\mG={}&2G_4+h\phi'^2\left(2G_{4,X}-G_{5,\phi}\right)
-\frac{h^2f'\phi'^3}{2f}G_{5,X}
-2h^2\phi'^4F_4-\frac{3h^3f'\phi'^5}{f}F_5.
\label{oddGLPVCoefficients}
\end{align}
Here and in what follows, $s$ denotes the positive quantity defined in
Eq.~(\ref{positiveSDefinition}).  The assumption $f/h>0$ ensures that it is real and implies that
$f$ and $h$ have a common sign, which we denote by
$\varsigma\equiv\operatorname{sgn}(f)=\operatorname{sgn}(h)=\pm1$.
No separate assumptions about the signs of $f$ and $h$ are imposed.
All the functions $G_i$, $F_4$, $F_5$, and their derivatives appearing in
this section are evaluated on the background at $(\phi(r),X(r))$.  
The coefficients satisfy two useful consistency checks. 
For $F_4=F_5=0$,
Eqs.~(\ref{mH})--(\ref{oddGLPVCoefficients}) reproduce the Horndeski
results of Ref.~\cite{Kobayashi:2012kh}.  For constant $G_4$ with all other
couplings set to zero, one finds $\mF=\mG=\mH=2G_4$, as expected in GR.

For the monopole $\ell=0$, the odd-parity vector harmonic vanishes
identically.  Equivalently, $L=0$ and hence $S_{\rm odd}^{(2)}=0$.  For the
dipole $\ell=1$, one has $L=2$, so the term proportional to $L-2$ in
Eq.~(\ref{oddmastergeneral}) vanishes.  This case is treated separately in
Sec.~\ref{oddell1sec}.

For $\ell\geq2$, the two metric amplitudes $Q$ and $W$ jointly
describe a single propagating odd-parity degree of freedom.  To derive an
action for this mode, we introduce an auxiliary field $\chi$ and write
Eq.~(\ref{oddmastergeneral}) in the dynamically equivalent form
\be
{\cal L}_{\rm odd}=
\frac{1}{4s}\left\{
\mH\left[2\chi\left(\dot W-Q'+\frac{2Q}{r}\right)-\chi^2\right]
+\frac{L-2}{r^2}\left(\frac{\mF}{h}Q^2-f\mG W^2\right)\right\}.
\label{oddauxiliaryaction}
\ee
Variation with respect to $\chi$ yields the algebraic equation
\be
\chi=\dot W-Q'+\frac{2Q}{r}.
\label{oddauxiliarydefinition}
\ee
Substitution of Eq.~(\ref{oddauxiliarydefinition}) into
Eq.~(\ref{oddauxiliaryaction}) recovers the original Lagrangian in
Eq.~(\ref{oddmastergeneral}).
Thus, Eq.~(\ref{oddauxiliaryaction}) also provides a first-order parent
formulation when $r$ is timelike.  On the nondegenerate branch used below,
eliminating $Q$ and $W$ requires no inversion of an $r$-derivative
operator, so the reduced and original formulations remain locally
equivalent for either common sign of $f$ and $h$.

Variation with respect to $W$ and $Q$ yields, respectively,
\be
(L-2)f\mG W+r^2\mH\dot\chi=0,
\label{oddWequation}
\ee
and
\be
\left[2(L-2)\mF Q+4rh\mH\chi
+2r^2h\mH\chi'
+r^2\left(\mH h'+2h\mH'\right)\chi\right]f
-r^2f'h\mH\chi=0.
\label{oddQequation}
\ee
In a nondegenerate region where $\mF\mG\mH\neq0$, we define
\be
\alpha_M\equiv-\frac{r^2h\mH}{\mF}
\left(\frac{\mH'}{\mH}
-\frac{f'}{2f}+\frac{h'}{2h}+\frac{2}{r}\right).
\label{oddalphaM}
\ee
Solving Eqs.~(\ref{oddWequation}) and~(\ref{oddQequation}) for $W$ and $Q$
then gives
\ba
W&=&-\frac{r^2\mH}{(L-2)f\mG}\dot\chi,
\notag\\
Q&=&\frac{1}{L-2}\left(
\alpha_M\chi-\frac{r^2h\mH}{\mF}\chi'\right).
\label{oddWQsolutions}
\ea
Substituting Eq.~(\ref{oddWQsolutions}) into
Eq.~(\ref{oddauxiliaryaction}) and integrating by parts gives the reduced
quadratic action for $\chi$,
\be
S_{\rm odd}^{(2)}=\sum_{\ell}L\int {\rm d}t\,{\rm d}r
\left(K_\chi\dot\chi^2+G_\chi\chi'^2+M_\chi\chi^2\right),
\label{oddreducedaction}
\ee
where
\be
K_\chi=\frac{r^2\mH^{2}}{4(L-2)f s \mG},
\qquad
G_\chi=-\frac{r^2h\mH^{2}}{4(L-2) s \mF},
\qquad
M_\chi=-\frac{\mH}{4(L-2)s}
\left(L-2+\alpha_M'-\frac{2\alpha_M}{r}\right).
\label{oddKGMcoefficients}
\ee
Although $F_{4,\phi}$, $F_{4,X}$, $F_{5,\phi}$, and $F_{5,X}$ do not
appear explicitly in Eqs.~(\ref{mH})--(\ref{oddGLPVCoefficients}), they
enter the $L$-independent part of $M_\chi$ through the radial derivatives
of $\mH$ contained in $\alpha_M$ and $\alpha_M'$.  They therefore affect
the effective potential but not the leading eikonal principal symbol.

We now derive the local high-frequency stability conditions without fixing
the individual signs of $f$ and $h$.  In the eikonal limit $\ell\gg1$, the
principal part of Eq.~(\ref{oddreducedaction}) can be written as
$S_{\rm odd,pr}^{(2)}
=\sum_{\ell}L\int {\rm d}t\,{\rm d}r\,{\cal L}_{\rm pr}$, where
\be
{\cal L}_{\rm pr}=
\frac{r^2\mH^2}{4(L-2)s}
\left(\frac{\dot\chi^2}{f\mG}-\frac{h\chi'^2}{\mF}
-\frac{L-2}{r^2\mH}\chi^2\right).
\label{Lpr}
\ee
For $\mH\neq0$ and $\ell\geq2$, the prefactor multiplying the parentheses
is positive.  To express the angular term in local orthonormal coordinates
after harmonic projection, we denote the gradient along the two angular
directions by
$\nabla_{\hat\Omega}$.  For each multipole, its eikonal contribution is
identified through
$(\nabla_{\hat\Omega}\chi)^2\equiv k_{\hat\Omega}^2\chi^2$, where
$k_{\hat\Omega}^2=(L-2)/r^2\simeq\ell^2/r^2$ is the physical angular wave
number squared.

For $\varsigma=+1$, the coordinate $t$ is timelike, and the locally
orthonormal derivatives in the $(t,r)$ plane are
$\partial_{\hat 0}=\partial_t/\sqrt{f}$ and
$\partial_{\hat 1}=\sqrt{h}\,\partial_r$. For $\varsigma=-1$, the
coordinate $r$ is timelike, and they are
$\partial_{\hat 0}=\sqrt{-h}\,\partial_r$ and
$\partial_{\hat 1}=\partial_t/\sqrt{-f}$.  Up to an overall positive
normalization, the principal Lagrangian is therefore
\be
{\cal L}_{\rm pr}\propto
\begin{cases}
(\partial_{\hat 0}\chi)^2/\mG-(\partial_{\hat 1}\chi)^2/\mF
-(\nabla_{\hat\Omega}\chi)^2/\mH, & \varsigma=+1,\\[2pt]
(\partial_{\hat 0}\chi)^2/\mF-(\partial_{\hat 1}\chi)^2/\mG
-(\nabla_{\hat\Omega}\chi)^2/\mH, & \varsigma=-1.
\end{cases}
\ee
The no-ghost condition read from the timelike kinetic coefficient of the
reduced master variable $\chi$ therefore depends on the common sign
$\varsigma$ of $f$ and $h$:
\be
\mG>0\quad({\rm for}~\varsigma=+1),\qquad
\mF>0\quad({\rm for}~\varsigma=-1).
\label{oddnoghostcondition}
\ee
At each spacetime point, $\partial_{\hat 0}$ denotes differentiation along
the timelike direction of a local orthonormal frame.  Such frame derivatives
need not define globally integrable coordinates, but they are sufficient to
determine the local principal normalization.  We define
$q_\chi^{(\pm)}$ relative to the reduced invariant measure
$r^2s\,{\rm d}t\,{\rm d}r$, including the harmonic factor $L$, by
\be
S_{{\rm odd,kin}}^{(2)}=
\frac{1}{2}\sum_{\ell}\int {\rm d}t\,{\rm d}r\,
r^2s\,
q_\chi^{(\pm)}\left(\partial_{\hat 0}\chi\right)^2.
\label{oddlocalkineticdefinition}
\ee
Using $\dot\chi^2=f(\partial_{\hat 0}\chi)^2$ for $f,h>0$ and
$\chi'^2=(\partial_{\hat 0}\chi)^2/(-h)$ for $f,h<0$,
Eqs.~(\ref{oddreducedaction}) and~(\ref{oddKGMcoefficients}) give
\be
\begin{aligned}
q_\chi^{(+)}&\equiv
\frac{2fLK_\chi}{r^2s}
=\frac{L}{L-2}\frac{h}{f}\frac{\mH^2}{2\mG},
&& f,h>0,\\[2pt]
q_\chi^{(-)}&\equiv
\frac{2LG_\chi}{(-h)r^2s}
=\frac{L}{L-2}\frac{h}{f}\frac{\mH^2}{2\mF},
&& f,h<0.
\end{aligned}
\label{oddlocalkineticnormalizations}
\ee
The factor of $2$ in Eq.~(\ref{oddlocalkineticnormalizations}) arises from
the conventional factor of $1/2$ in
Eq.~(\ref{oddlocalkineticdefinition}), while $L$ originates from the angular
norm of the odd vector harmonic and already appears in
Eq.~(\ref{oddreducedaction}).  Accordingly, the no-ghost conditions in
Eq.~(\ref{oddnoghostcondition}) are equivalent to
$q_\chi^{(+)}>0$ and $q_\chi^{(-)}>0$.  
For $f,h<0$, $\chi$ is the momentum-type master variable conjugate to the
original amplitude $Q$: the parent action exchanges the roles played by the
timelike kinetic and angular-gradient coefficients when $Q$ is eliminated.
Thus, in the original $Q$ description $\mH$ multiplies the $r$-time kinetic
square, whereas in the reduced $\chi$ description $\mF$ determines the
timelike kinetic normalization.  This does not change the complete local
high-frequency stability criterion below, which requires both $\mF>0$ and
$\mH>0$ (as well as $\mG>0$).
In the eikonal limit,
\be
q_\chi^{(+)}\longrightarrow
\frac{h}{f}\frac{\mH^2}{2\mG},\qquad
q_\chi^{(-)}\longrightarrow
\frac{h}{f}\frac{\mH^2}{2\mF}.
\label{oddlocalkineticeikonal}
\ee
Equations~(\ref{oddlocalkineticnormalizations}) and
(\ref{oddlocalkineticeikonal}) show explicitly that $\mH\to0$ is a
strong-coupling limit.  If $h/f$, $\mF$, and $\mG$ remain finite and
nonzero, the local timelike kinetic normalization and the radial-gradient
coefficient vanish as $\mH^2$.  As seen from Eq.~(\ref{Lpr}), the
angular-gradient coefficient instead vanishes linearly in $\mH$, because
the factor $1/\mH$ in the angular term cancels one power of the overall
$\mH^2$.  The quadratic principal action thus loses its nonzero
normalization at $\mH=0$, and canonical normalization becomes singular,
signaling strong coupling.

For $\mH\neq0$, hyperbolicity in the two-dimensional $(t,r)$ sector
requires
\be
\frac{\mG}{\mF}>0.
\label{oddradialspeed}
\ee
For $\varsigma=+1$, the ratio $\mG/\mF$ is the squared radial propagation
speed.  For $\varsigma=-1$, the $t$ direction is spacelike, and the
corresponding squared propagation speed is the reciprocal ratio
$\mF/\mG$.  Combining Eqs.~(\ref{oddnoghostcondition}) and
(\ref{oddradialspeed}) therefore yields $\mF>0$ and $\mG>0$ in both sign
sectors.  The absence of angular gradient instabilities, as determined from
Eq.~(\ref{Lpr}), further requires
\be
\frac{\mG}{\mH}>0\quad({\rm for}~\varsigma=+1),\qquad
\frac{\mF}{\mH}>0\quad({\rm for}~\varsigma=-1).
\label{oddangularspeed}
\ee
The complete local high-frequency stability conditions for the odd-parity
sector are therefore
\be
\mG>0,\qquad
\mF>0,\qquad
\mH>0.
\label{oddgeneralconditions}
\ee
Although the stability conditions are identical in the two sign sectors,
the causal roles of $t$ and $r$ are interchanged.  For $f,h>0$, consider a
local mode $\chi\propto\exp[\mathrm{i}(\omega t-kr)]$.  Its propagation speed in the
radial direction of a local orthonormal frame is
$c_r=\omega/(k\sqrt{fh})$.  For $f,h<0$, the radial direction is timelike,
and a mode $\chi\propto\exp[\mathrm{i}(\omega r-kt)]$ has proper frequency
$\sqrt{-h}\,\omega$ and wave number $k/\sqrt{-f}$ along the spacelike
$t$ direction.  We accordingly define the propagation speed in this
direction as $c_\parallel=\sqrt{fh}\,\omega/k$.  The squared propagation
speeds are
\be
\begin{aligned}
\left(c_{r,{\rm odd}}^{2(+)},c_{\Omega,{\rm odd}}^{2(+)}\right)
&=\left(\frac{\mG}{\mF},\frac{\mG}{\mH}\right),
&& f,h>0,\\
\left(c_{\parallel,{\rm odd}}^{2(-)},c_{\Omega,{\rm odd}}^{2(-)}\right)
&=\left(\frac{\mF}{\mG},\frac{\mF}{\mH}\right),
&& f,h<0.
\end{aligned}
\label{oddgeneralspeeds}
\ee
Thus, $c_{r,{\rm odd}}^{2(+)}$ describes radial propagation in a region
with $f,h>0$, whereas $c_{\parallel,{\rm odd}}^{2(-)}$ describes propagation
along the spacelike $t$ direction in a region with $f,h<0$.
Equation~(\ref{oddgeneralconditions}) excludes ghosts, a vanishing principal
normalization, and high-frequency gradient instabilities in all local
spatial directions.  At a Killing horizon, these conditions are understood
as one-sided limits or evaluated in a regular frame.

\subsection{Dipole: $\ell=1$}
\label{oddell1sec}

For the dipole mode, $\ell=1$ and hence $L=2$, so the terms proportional to
$L-2$ in Eq.~(\ref{oddmastergeneral}) vanish.  The odd-parity tensor
harmonic associated with $\delta g_{ab}$ also vanishes identically.
Consequently, $\delta g_{ab}=0$ holds automatically and does not fix the
dipole gauge freedom.  We use the residual freedom to set $W=0$.  Since the
variation with respect to $W$ yields a nontrivial constraint, we first vary
the unfixed quadratic action and impose this gauge afterward.

The gauge-invariant dipole combination is
\be
{\cal E}\equiv\frac{\mH}{s}
\left(Q'-\frac{2Q}{r}-\dot W\right).
\label{oddell1Edefinition}
\ee
Varying the unfixed quadratic action with respect to $W$ and $Q$ yields,
respectively,
\be
\dot{\cal E}=0,\qquad (r^2{\cal E})'=0.
\label{oddell1equations}
\ee
After deriving these constraints, we impose the gauge condition
\be
W=0.
\label{oddell1gauge}
\ee
Equations~(\ref{oddell1Edefinition}) and~(\ref{oddell1equations}) then give
\be
Q'-\frac{2Q}{r}=\frac{C}{r^2\mH} s,
\label{oddell1firstintegral}
\ee
where $C$ is constant in both $t$ and $r$.  The general solution is
\be
Q=r^2\left[C\int^r\frac{s(\bar r)}{\bar r^4\mH(\bar r)}
{\rm d}\bar{r}+C_0(t)\right].
\label{oddell1Qsolution}
\ee
The function $C_0(t)$ is the homogeneous solution of the radial equation.
Gauge transformations preserving $W=0$ satisfy
$\Lambda'-2\Lambda/r=0$ and hence take the form
$\Lambda(t,r)=\Lambda_0(t)r^2$, where $\Lambda_0(t)$ is arbitrary.
Since such a transformation shifts $Q$ by $\dot\Lambda_0 r^2$, the function
$C_0(t)$ is pure gauge.  The constant $C$ is therefore the only physical
dipole integration constant and parametrizes an infinitesimal rotation.

As a consistency check, after deriving the dipole constraints from the
unfixed action, we set $L=2$ and $W=0$ in
Eq.~(\ref{oddmastergeneral}).  The resulting Lagrangian is
$\left.{\cal L}_{\rm odd}\right|_{\ell=1,W=0}
=\mH(Q'-2Q/r)^2/(4s)$.  Substitution of
Eq.~(\ref{oddell1firstintegral}) gives
\be
{\cal L}_{\rm odd}^{(\ell=1)}=
\frac{C^2}{4r^4\mH}s\,.
\label{oddell1action}
\ee
Since $C$ is an integration constant rather than an independent field, this
on-shell Lagrangian contains no kinetic term.  The dipole sector therefore
has no local propagating degree of freedom and no associated ghost or
propagation speed.  For a nontrivial rotational perturbation with $C\neq0$,
and assuming the remaining background functions are finite and regular,
regularity requires $\mH\neq0$.  This requirement is already ensured by the
higher-multipole stability condition $\mH>0$ in
Eq.~(\ref{oddgeneralconditions}), so the dipole sector imposes no additional
local stability condition.

\section{Even-parity stability}
\label{evensec}

In this section, we derive the quadratic action for even-parity perturbations
and obtain the local no-ghost conditions and the principal-part conditions for
radial and angular gradient stability.  Using the scalar spherical harmonics
$Y_\ell$ defined in Sec.~\ref{oddellge2sec}, we decompose the even-parity
metric perturbation $h_{\mu\nu}\equiv\delta g_{\mu\nu}$ on the
background~(\ref{metric}) as
\begin{align}
h_{tt}&=\sum_{\ell}f(r)H_0(t,r)Y_{\ell}, &
h_{tr}&=\sum_{\ell}H_1(t,r)Y_{\ell}, &
h_{rr}&=\sum_{\ell}h^{-1}(r)H_2(t,r)Y_{\ell}, \notag\\
h_{ta}&=\sum_{\ell}h_0(t,r)\nabla_aY_{\ell}, &
h_{ra}&=\sum_{\ell}h_1(t,r)\nabla_aY_{\ell}, &
h_{ab}&=\sum_{\ell} \left[ K(t,r)g_{ab}Y_{\ell}
+r^2G(t,r)\nabla_a\nabla_bY_{\ell} \right],
\label{evendecompositiongeneral}
\end{align}
where $g_{ab}=r^2\gamma_{ab}$, with the angular indices $a,b$ running over
$\theta,\varphi$.  The scalar field is expanded as
\be
\phi=\bar{\phi}(r)+\sum_{\ell}\delta\phi(t,r)Y_{\ell}\,.
\label{evenscalardecomposition}
\ee
We henceforth omit the overbar from the background scalar.  For $\ell\geq2$,
we fix the three even-parity gauge freedoms by imposing the uniform-curvature
gauge \cite{Kobayashi:2014wsa,Kase:2021mix,Kase:2023mho}
\be
h_0=0\,,\qquad K=0\,,\qquad G=0\,,
\label{evenuniformcurvaturegauge}
\ee
The remaining perturbation variables are $H_0$, $H_1$, $H_2$, $h_1$, and
$\delta\phi$.

\subsection{Higher multipoles ($\ell\geq2$): quadratic action and field equations}
\label{evenSecondOrderActionSubsec}

We expand Eq.~(\ref{action}) to second order, integrate over the angles,
perform integrations by parts, and use the background
Eqs.~(\ref{EGLPV00})--(\ref{EphiGLPV}).  With
$L\equiv\ell(\ell+1)$ as defined in Eq.~(\ref{oddLdefinition}), the resulting
quadratic action takes the form
\be
S_{\rm even}^{(2)}=\sum_{\ell\geq2}\int {\rm d}t\,{\rm d}r\,
{\cal L}_{\rm even}^{(2)},
\label{evenReducedActionGeneral}
\ee
where
\begin{align}
{\cal L}_{\rm even}^{(2)}={}&
H_0\Bigl[a_1\delta\phi''+a_2\delta\phi'+a_3H_2'+La_4h_1'
+(a_5+La_6)\delta\phi+(a_7+La_8)H_2+La_9h_1\Bigr]\notag\\
&
+Lb_1H_1^2
+H_1\Bigl(b_2\dot{\delta\phi}'+b_3\dot{\delta\phi}+b_4\dot H_2+Lb_5\dot h_1\Bigr)
+c_1\dot{\delta\phi}\dot H_2
+H_2\Bigl[c_2\delta\phi'+(c_3+Lc_4)\delta\phi+Lc_5h_1\Bigr]
+c_6H_2^2\notag\\
&+Ld_1\dot h_1^{\,2}+Lh_1(d_2\delta\phi'+d_3\delta\phi)+Ld_4h_1^2
+e_1\dot{\delta\phi}^{\,2}+e_2\delta\phi'^{\,2}+(e_3+Le_4)\delta\phi^2.
\label{evenStotal}
\end{align}
The coefficients $a_1,\ldots,e_4$ are given explicitly in
Appendix~\ref{AppEvenCoefficients}, including all contributions from the
general functions $F_4(\phi,X)$ and $F_5(\phi,X)$; no shift symmetry is
assumed.  Whenever $s$ appears in these coefficients, it denotes the positive
quantity defined in Eq.~(\ref{positiveSDefinition}).  Hence, the action
requires only $f/h>0$, with no separate assumption that $f$ and $h$ are
individually positive.

Before eliminating the nondynamical fields, we record the Euler--Lagrange
equations for the five perturbation variables.  These equations provide the
starting point for the constraint reduction below. Variation with respect
to $H_0$ yields
\begin{align}
a_1\delta\phi''+a_2\delta\phi'+a_3H_2'+La_4h_1'
+(a_5+La_6)\delta\phi+(a_7+La_8)H_2+La_9h_1=0\,.
\label{evenH0equation}
\end{align}
Since $H_1$ appears without derivatives, its equation of motion is algebraic,
\begin{align}
2Lb_1H_1+b_2\dot{\delta\phi}'+b_3\dot{\delta\phi}
+b_4\dot H_2+Lb_5\dot h_1=0\,.
\label{evenH1equation}
\end{align}
Variation with respect to $H_2$ and $h_1$ gives, respectively,
\begin{align}
&-c_1\ddot{\delta\phi}-b_4\dot H_1+c_2\delta\phi'
+(c_3+Lc_4)\delta\phi+Lc_5h_1+2c_6H_2
-a_3H_0'+(a_7-a_3'+La_8)H_0=0,
\label{evenH2equation}\\
&-2d_1\ddot h_1+d_2\delta\phi'+d_3\delta\phi+2d_4h_1
-a_4H_0'+(a_9-a_4')H_0-b_5\dot H_1+c_5H_2=0 .
\label{evenh1equation}
\end{align}
Finally, varying the scalar perturbation gives
\begin{align}
&-2e_1\ddot{\delta\phi}-2e_2\delta\phi''-2e_2'\delta\phi'
+2(e_3+Le_4)\delta\phi
+a_1H_0''+(2a_1'-a_2)H_0'
+(a_1''-a_2'+a_5+La_6)H_0\notag\\
&+b_2\dot H_1'+(b_2'-b_3)\dot H_1-c_1\ddot H_2-c_2H_2'
+(c_3-c_2'+Lc_4)H_2
-Ld_2h_1'+L(d_3-d_2')h_1=0.
\label{evenscalarequation}
\end{align}
%

\subsection{Higher multipoles ($\ell\geq2$): stability conditions}
\label{evenStabilitySubsec}

For the higher multipoles $\ell\geq2$, where
$L=\ell(\ell+1)\geq6$, we follow
Refs.~\cite{Kobayashi:2014wsa,Kase:2023mho} and reduce the action to its two
dynamical variables by imposing the $H_0$ constraint and eliminating the
auxiliary fields, while retaining all GLPV contributions. This procedure
applies locally on the regular branch with
$\phi'\ne0$, $fh\ne0$, $a_3\ne0$, $a_4/f\ne0$, and
${\cal D}_{\rm ev}\ne0$, and with every further denominator introduced below
nonzero. If any of these quantities vanishes, the constraint system can
change rank, and the corresponding branch must instead be analyzed directly
using Eqs.~(\ref{evenH0equation})--(\ref{evenscalarequation}). The conditions
at a Killing horizon are understood as one-sided limits from a region with
fixed signs of $f$ and $h$, or equivalently in a horizon-regular frame; the
coordinate expressions containing $f$ or $h$ in denominators are not to be
evaluated by direct substitution at the horizon.
To write the reduced action compactly, we introduce the following background
combinations:
\ba
\vartheta_1&\equiv& \frac{fh\phi'^3}{rs}
\left(F_4r+3F_5h\phi'\right),\qquad
{\cal Z}_{\rm ev} \equiv \frac{a_4}{h}+\phi'\vartheta_1,\qquad
\Delta_{\rm ev}\equiv\frac1r-\frac{f'}{2f}
 +\frac{L{\cal Z}_{\rm ev}}{2a_3},\qquad
{\cal D}_{\rm ev}
\equiv La_4\Delta_{\rm ev},\nonumber\\
{\cal R}_{\rm ev} &\equiv& f{\cal D}_{\rm ev}, \qquad
\Sigma_{\rm ev} \equiv \frac{{\cal R}_{\rm ev}'}{{\cal R}_{\rm ev}}
 =\frac{f'}f+\frac{a_4'}{a_4}+\frac{\Delta_{\rm ev}'}{\Delta_{\rm ev}},
\qquad
{\cal B}_{\rm ev}\equiv a_5+La_6,\qquad
{\cal C}_{\rm ev} \equiv a_2-a_1'
 +\frac{La_1{\cal Z}_{\rm ev}}{2a_3},\nonumber\\
c_{\rm ev}\ &\equiv&-\frac{a_1}{fh}+\frac{2r\vartheta_1}{f},\qquad
\beta_{\rm ev}\equiv\frac{a_1}{a_3}, \qquad
{\cal N}_{\rm ev}\equiv a_1'-a_2+L\vartheta_1,\qquad
{\cal M}_{\rm ev} \equiv {\cal N}_{\rm ev}
 +\frac{Lra_4\vartheta_1}{a_3}-\frac{La_1a_4}{2a_3h}.
\label{evenReductionDefinitions}
\ea

Since $H_0$ appears linearly and without derivatives in the quadratic action,
it acts as a Lagrange multiplier.  Its equation of motion therefore imposes a
constraint relating $H_2$, $h_1$, and $\delta\phi$.  To recast this constraint
as a first-order radial relation, we introduce
\be
\psi\equiv H_2+\frac{La_4}{a_3}h_1+\frac{a_1}{a_3}\delta\phi'.
\label{evenpsidefinition}
\ee
Since $\psi$ contains $\delta\phi'$, this is a derivative-dependent change of
variables.  When $f,h>0$, the radial coordinate is spacelike, so the standard
constraint reduction applies directly.  When $f,h<0$, however, $r$ is
timelike and $\delta\phi'$ is a time derivative; the equivalence of the
reduced and original formulations is then not automatic.  We establish this
equivalence separately in Sec.~\ref{evenInteriorStabilitySubsubsec} using the
first-order parent action in Eq.~(\ref{evenInteriorParentAction}).

The coefficient relations $a_7=a_3'$ and
$a_8=-a_4/(2h)-\phi'\vartheta_1/2$ allow the $H_0$ constraint to be
written as
\be
a_3\psi'+\left(a_3'-\frac{L}{2}{\cal Z}_{\rm ev}\right)\psi
 +{\cal C}_{\rm ev}\delta\phi'+{\cal B}_{\rm ev}\delta\phi
 +{\cal D}_{\rm ev}h_1=0.
\label{evenH0constraintpsi}
\ee
On the regular branch $a_3{\cal D}_{\rm ev}\ne0$,
Eq.~(\ref{evenH0constraintpsi}) is algebraic in $h_1$, while
Eq.~(\ref{evenpsidefinition}) is algebraic in $H_2$.  Solving them gives
\ba
h_1&=& -\frac{1}{{\cal D}_{\rm ev}}\left[
a_3\psi'+\left(a_3'-\frac{L}{2}{\cal Z}_{\rm ev}\right)\psi
+{\cal C}_{\rm ev}\delta\phi'+{\cal B}_{\rm ev}\delta\phi\right],
\label{evenh1H2solutions0}\\
H_2&=&\psi-\frac{La_4}{a_3}h_1-\beta_{\rm ev}\delta\phi'.
\label{evenh1H2solutions}
\ea
The remaining auxiliary perturbation $H_1$ enters the action algebraically
and can therefore be eliminated. The coefficients in its equation obey the
exact relations
\be
b_2-b_4\beta_{\rm ev}=0, \qquad
Lb_5-\frac{La_4}{a_3}b_4=\frac{La_4}{f},\qquad
b_3=\frac{2{\cal N}_{\rm ev}}{f}, \qquad
b_3+\frac{La_4}{a_3}c_{\rm ev}=\frac{2{\cal M}_{\rm ev}}{f}.
\label{evenReductionIdentities}
\ee
On the regular branch $La_4\ne0$, the $H_1$ equation then gives
\be
H_1=\frac{2a_3}{La_4}\dot\psi
-\frac{2{\cal N}_{\rm ev}}{La_4}\dot{\delta\phi}
-\dot h_1.
\label{evenH1solution}
\ee
We then substitute Eqs.~(\ref{evenh1H2solutions})
and~(\ref{evenH1solution}) into the original quadratic
action~(\ref{evenStotal}). The definitions in
Eq.~(\ref{evenReductionDefinitions}), together with
$a_3=-\phi'a_1/2-ra_4$, imply
${\cal C}_{\rm ev}+{\cal M}_{\rm ev}=0$.
Consequently, the mixed radial--time derivative terms combine into
$(2a_3{\cal M}_{\rm ev}/{\cal R}_{\rm ev})
(\dot\psi\dot{\delta\phi})'$. After one integration by parts in $r$, they
give
$-(2a_3{\cal M}_{\rm ev}/{\cal R}_{\rm ev})'
\dot\psi\dot{\delta\phi}$. Since
$\dot{\vec{\cal X}}^{\rm T}{\bm K}\dot{\vec{\cal X}}$ contains the cross
term $2K_{12}\dot\psi\dot{\delta\phi}$, the corresponding contribution to
$K_{12}$ is
$-(a_3{\cal M}_{\rm ev}/{\cal R}_{\rm ev})'$.
Performing the remaining integrations by parts and omitting boundary terms,
we obtain the reduced action
\be
S_{\rm even}^{(2)}=
\sum_{\ell\geq2}\int {\rm d}t\,{\rm d}r\,
\left[\dot{\vec{\cal X}}^{{\rm T}} \matrixsym{K}
\dot{\vec{\cal X}}
 +(\vec{\cal X}')^{{\rm T}}\matrixsym{G}\vec{\cal X}'
 +\vec{\cal X}^{{\rm T}}\matrixsym{Q}\vec{\cal X}'
 +\vec{\cal X}^{{\rm T}}\matrixsym{M}\vec{\cal X}\right]\,,
\label{evenReducedTwoFieldAction}
\ee
where the two-component dynamical field is
\be
\vec{\cal X}\equiv (\psi,\delta\phi)^{\rm T}\,.
\ee
The matrices ${\bm K}$, ${\bm G}$, and ${\bm M}$ are symmetric. The symmetric
part of the single-radial-derivative mixing matrix can be integrated by parts
and absorbed into ${\bm M}$. Thus, without loss of generality, ${\bm Q}$ can
be chosen to be antisymmetric and contains only the irreducible antisymmetric
mixing.

Using the exact identities among the GLPV coefficients given in
Appendix~\ref{AppEvenCoefficients}, the components of the kinetic matrix
${\bm K}$ take the form
\begin{align}
K_{11}={}&-\frac{a_3^2}{{\cal R}_{\rm ev}}
\left(\frac{a_4'}{a_4}+\frac{\Delta_{\rm ev}'}{\Delta_{\rm ev}}+\frac2r\right),
\label{evenK11}\\
K_{12}={}&\frac{2a_3{\cal N}_{\rm ev}}{fLa_4}+\frac{c_{\rm ev}}2
+\frac{a_3{\cal M}_{\rm ev}\Sigma_{\rm ev}-a_3{\cal M}_{\rm ev}'
-a_3{\cal B}_{\rm ev}-L{\cal Z}_{\rm ev}{\cal M}_{\rm ev}/2}
{{\cal R}_{\rm ev}},
\label{evenK12}\\
K_{22}={}&e_1-\frac{2{\cal N}_{\rm ev}^2}{fLa_4}
+\frac{2{\cal B}_{\rm ev}{\cal M}_{\rm ev}
-{\cal C}_{\rm ev}'{\cal M}_{\rm ev}-{\cal C}_{\rm ev}{\cal M}_{\rm ev}'
+{\cal C}_{\rm ev}{\cal M}_{\rm ev}\Sigma_{\rm ev}}{{\cal R}_{\rm ev}}
+\frac12(\beta_{\rm ev}'c_{\rm ev}+\beta_{\rm ev}c_{\rm ev}').
\label{evenK22}
\end{align}
To write the remaining matrices compactly, we introduce the following
two-component vectors:
\begin{align}
{\bm e}_{\delta\phi}&\equiv\begin{pmatrix}0\\1\end{pmatrix},&
{\bm e}_{\psi}&\equiv\begin{pmatrix}1\\0\end{pmatrix},&
{\bm p}_{\rm ev}&\equiv\begin{pmatrix}a_3\\{\cal C}_{\rm ev}\end{pmatrix},&
{\bm s}_{\rm ev}&\equiv
\begin{pmatrix}a_3'-L{\cal Z}_{\rm ev}/2\\{\cal B}_{\rm ev}\end{pmatrix},\notag\\
{\bm u}_{\rm ev}&\equiv-\frac{{\bm p}_{\rm ev}}{{\cal D}_{\rm ev}},&
{\bm v}_{\rm ev}&\equiv-\frac{{\bm s}_{\rm ev}}{{\cal D}_{\rm ev}},&
{\bm w}_{\rm ev}&\equiv\frac{La_4}{a_3{\cal D}_{\rm ev}}{\bm p}_{\rm ev}
-\beta_{\rm ev}{\bm e}_{\delta\phi},&
{\bm z}_{\rm ev}&\equiv{\bm e}_{\psi}
+\frac{La_4}{a_3{\cal D}_{\rm ev}}{\bm s}_{\rm ev}.
\label{evenSpatialVectors}
\end{align}
In this notation, the solutions in 
Eqs.~(\ref{evenh1H2solutions0}) and 
(\ref{evenh1H2solutions}) are
\be
h_1={\bm u}_{\rm ev}^{{\rm T}}\vec{\cal X}'
+{\bm v}_{\rm ev}^{{\rm T}}\vec{\cal X},
\qquad
H_2={\bm w}_{\rm ev}^{{\rm T}}\vec{\cal X}'
+{\bm z}_{\rm ev}^{{\rm T}}\vec{\cal X}.
\ee
For any two column vectors ${\bm x}$ and ${\bm y}$, we denote their
symmetrized outer product by
\be
\operatorname{sym}({\bm x},{\bm y})\equiv
\frac{{\bm x}{\bm y}^{{\rm T}}+{\bm y}{\bm x}^{{\rm T}}}2\,.
\ee
Substituting these expressions for $h_1$ and $H_2$ into the
radial-derivative sector of Eq.~(\ref{evenStotal}) and collecting the terms
quadratic in $\vec{\cal X}'$ gives the symmetric radial-gradient matrix
\be
{\bm G}=c_2\operatorname{sym}({\bm w}_{\rm ev},{\bm e}_{\delta\phi})
+Lc_5\operatorname{sym}({\bm w}_{\rm ev},{\bm u}_{\rm ev})
+c_6{\bm w}_{\rm ev}{\bm w}_{\rm ev}^{{\rm T}}
+Ld_2\operatorname{sym}({\bm u}_{\rm ev},{\bm e}_{\delta\phi})
+Ld_4{\bm u}_{\rm ev}{\bm u}_{\rm ev}^{{\rm T}}
+e_2{\bm e}_{\delta\phi}{\bm e}_{\delta\phi}^{{\rm T}}.
\label{evenGmatrix}
\ee
Before the final integration by parts in $r$, the terms containing one radial
derivative and those containing no derivatives can be written as
$\vec{\cal X}^{\rm T}{\bm J}\vec{\cal X}'$ and
$\vec{\cal X}^{\rm T}{\bm M}_0\vec{\cal X}$, respectively, where
\begin{align}
{\bm J}={}&c_2{\bm z}_{\rm ev}{\bm e}_{\delta\phi}^{{\rm T}}
+(c_3+Lc_4){\bm e}_{\delta\phi}{\bm w}_{\rm ev}^{{\rm T}}
+Lc_5({\bm v}_{\rm ev}{\bm w}_{\rm ev}^{{\rm T}}
+{\bm z}_{\rm ev}{\bm u}_{\rm ev}^{{\rm T}})
+2c_6{\bm z}_{\rm ev}{\bm w}_{\rm ev}^{{\rm T}}\notag\\
&
+Ld_2{\bm v}_{\rm ev}{\bm e}_{\delta\phi}^{{\rm T}}
+Ld_3{\bm e}_{\delta\phi}{\bm u}_{\rm ev}^{{\rm T}}
+2Ld_4{\bm v}_{\rm ev}{\bm u}_{\rm ev}^{{\rm T}},
\label{evenJmatrix}\\
{\bm M}_0={}&(c_3+Lc_4)\operatorname{sym}({\bm z}_{\rm ev},{\bm e}_{\delta\phi})
+Lc_5\operatorname{sym}({\bm z}_{\rm ev},{\bm v}_{\rm ev})
+c_6{\bm z}_{\rm ev}{\bm z}_{\rm ev}^{{\rm T}}\notag\\
&
+Ld_3\operatorname{sym}({\bm v}_{\rm ev},{\bm e}_{\delta\phi})
+Ld_4{\bm v}_{\rm ev}{\bm v}_{\rm ev}^{{\rm T}}
+(e_3+Le_4){\bm e}_{\delta\phi}{\bm e}_{\delta\phi}^{{\rm T}}.
\label{evenM0matrix}
\end{align}
The antisymmetric part of ${\bm J}$ gives the derivative-mixing matrix
${\bm Q}$. For its symmetric part
${\bm J}_{\rm S}=({\bm J}+{\bm J}^{\rm T})/2$, integration by parts gives,
up to a total radial derivative,
$\vec{\cal X}^{\rm T}{\bm J}_{\rm S}\vec{\cal X}'
=-\vec{\cal X}^{\rm T}{\bm J}_{\rm S}'\vec{\cal X}/2$.
It therefore contributes to the algebraic matrix ${\bm M}$, and the final
matrices are
\be
{\bm Q}=\frac{{\bm J}-{\bm J}^{{\rm T}}}2,
\qquad
{\bm M}={\bm M}_0-\frac12
\left(\frac{{\bm J}+{\bm J}^{{\rm T}}}2\right)'.
\label{evenQMmatrices}
\ee
Equations~(\ref{evenK11})--(\ref{evenQMmatrices}) specify all four matrices in
the reduced two-field action~(\ref{evenReducedTwoFieldAction}). In this form,
no individual field factor carries more than one derivative, consistently
with the GLPV constraint structure.

\subsubsection{Region with timelike $t$: $f,h>0$}
\label{evenExteriorStabilitySubsubsec}

For $f>0$ and $h>0$, the coordinate $t$ is timelike. The Hessian of the
reduced Lagrangian with respect to $\dot\psi$ and $\dot{\delta\phi}$ is
$2{\bm K}$. Hence, on the nondegenerate two-field branch, the no-ghost
condition is the positive definiteness of ${\bm K}$. Writing
${\bm K}=(K_{ij})$, Sylvester's criterion gives
\be
K_{11}>0,\qquad
\Delta_K\equiv\det{\bm K}=K_{11}K_{22}-K_{12}^2>0.
\label{evenNoGhostCondition}
\ee
Once $K_{11}>0$ is imposed, the determinant condition $\Delta_K>0$ is
equivalent to the positivity of the Schur complement,
$K_{22}-K_{12}^2/K_{11}>0$.

Both principal minors admit a compact factorization in full GLPV theories.
Using the positive quantity $s$ defined in
Eq.~(\ref{positiveSDefinition}), we introduce
\begin{align}
\mu_{\rm ev}&\equiv-\frac{4a_3}{hs}
=\frac{2(\phi'a_1+2ra_4)}{hs},\qquad
\mH_\vartheta\equiv\frac{2{\cal Z}_{\rm ev}}s
=\mH+\frac{2\phi'\vartheta_1}{s},\notag\\
{\cal P}_{1}&\equiv\frac1s
\left(\frac{s r^2\mH\mH_\vartheta}{\mu_{\rm ev}}\right)',\qquad
{\cal P}_{2}\equiv
\frac{h}{f}(rf'-2f)\mu_{\rm ev},\notag\\
\Theta_{\rm ev}&\equiv{\cal P}_{2}+2rL\mH_\vartheta,\qquad
\Delta_{\rm ev}=-\frac{\Theta_{\rm ev}}{2rh\mu_{\rm ev}}.
\label{evenGLPVPDefinitions}
\end{align}
The last relation in Eq.~(\ref{evenGLPVPDefinitions}) is an on-shell
re-expression of $\Delta_{\rm ev}$, which was defined in
Eq.~(\ref{evenReductionDefinitions}), rather than an independent definition.
The difference of the background metric equations yields the on-shell
identity
\be
\left[\frac{r\mH(rf'-2f)}s\right]'=-2s\mF .
\label{evenPOnShellIdentity}
\ee
Substituting ${\cal C}_{\rm ev}=-{\cal M}_{\rm ev}$,
${\cal B}_{\rm ev}=-{\cal N}_{\rm ev}'+La_6$, and
${\cal M}_{\rm ev}={\cal N}_{\rm ev}
+Lf a_4c_{\rm ev}/(2a_3)$ into Eqs.~(\ref{evenK11})--(\ref{evenK22}),
and then using Eq.~(\ref{evenPOnShellIdentity}), we obtain
\begin{align}
K_{11}&=\frac{s h^2\mu_{\rm ev}^4
\left(L{\cal P}_{1}-\mF\right)}
{2fL\mH^2\Theta_{\rm ev}^2},\label{K11con}\\
\Delta_K&=\frac{(L-2)h\mu_{\rm ev}^4\mF}
{4fL\mH^2\phi'^2\Theta_{\rm ev}^2}
\left(2{\cal P}_{1}-\mF\right).
\label{evenKineticPFactorizations}
\end{align}
On the regular branch with $f,h>0$, assuming $\mF>0$ and $L\geq 6$, the two
no-ghost inequalities in Eq.~(\ref{evenNoGhostCondition}) reduce to the single
condition
\be
\mK\equiv 2{\cal P}_{1}-\mF>0\,.
\label{evenKDefinition}
\ee
Indeed, this inequality also guarantees $K_{11}>0$ through
Eq.~(\ref{K11con}). In the Horndeski limit, where $\vartheta_1\to0$ and
$\mH_\vartheta\to\mH$, ${\cal P}_1$ reduces to the corresponding quantity
defined in Ref.~\cite{Kase:2023mho}.

We next analyze radial-gradient stability. Writing
${\bm u}_{\rm ev}=(u_1,u_2)^{{\rm T}}$ and
${\bm w}_{\rm ev}=(w_1,w_2)^{{\rm T}}$, the independent components of the
symmetric radial-gradient matrix ${\bm G}$ in Eq.~(\ref{evenGmatrix}) are
\begin{align}
G_{11}={}&Lc_5w_1u_1+c_6w_1^2+Ld_4u_1^2, \label{evenGcomponents0}\\
G_{12}={}&\frac12\left[c_2w_1+Lc_5(w_1u_2+w_2u_1)
+2c_6w_1w_2+Ld_2u_1+2Ld_4u_1u_2\right],\label{evenGcomponents1}\\
G_{22}={}&c_2w_2+Lc_5w_2u_2+c_6w_2^2
+Ld_2u_2+Ld_4u_2^2+e_2.
\label{evenGcomponents}
\end{align}
In the radial eikonal limit, the background coefficients can be treated as
locally constant. We use the plane-wave ansatz
$\vec{\cal X}(t,r)=\vec{\cal X}_0\exp[\mathrm{i}(\omega t-kr)]$, where
$\vec{\cal X}_0$ is a constant nonvanishing amplitude vector. In a local
orthonormal frame, the frequency and radial wave number are
$\omega_{\rm loc}=\omega/\sqrt{f}$ and $k_{\rm loc}=k\sqrt{h}$,
respectively, so the locally measured radial propagation speed is
$c_r\equiv\omega_{\rm loc}/k_{\rm loc}=\omega/(k\sqrt{fh})$.
The principal part of the field equations admits a nontrivial solution for
$\vec{\cal X}_0$ only if
\be
\det\left(fh\,c_r^2{\bm K}+{\bm G}\right)=0.
\label{evenRadialDispersion}
\ee
We define the two combinations
\begin{align}
\Delta_G&\equiv\det{\bm G}
=G_{11}G_{22}-G_{12}^2,\label{evenRadialInvariants0}\\
T_r&\equiv K_{11}G_{22}+K_{22}G_{11}
-2K_{12}G_{12}.
\label{evenRadialInvariants}
\end{align}
Solving Eq.~(\ref{evenRadialDispersion}) for $c_r^2$ gives the two squared
radial propagation speeds
\be
c_{r\pm,{\rm even}}^{2(+)}
=\frac{-T_r\pm\sqrt{T_r^2-4\Delta_K\Delta_G}}
{2fh\,\Delta_K}.
\label{evenRadialSpeeds}
\ee
The corresponding eigenmodes are generally mixtures of the scalar and tensor
perturbations. Since ${\bm K}$ and ${\bm G}$ are not, in general,
simultaneously diagonal, neither matrix alone identifies the tensor branch.
The two branches can instead be separated by factorizing the full radial
characteristic polynomial. For this purpose, we rewrite
Eq.~(\ref{evenPOnShellIdentity}) using $a_4=f\mH/(2s)$ and combine it with
the coefficient identity for $\vartheta_2$:
\begin{align}
(2f-rf')a_4'={}&
\left(rf''-\frac{rf'^2}{f}+2f'-\frac{2f}{r}\right)a_4
+\frac{fs}{r}\mF,\notag\\
\vartheta_2={}&\frac{hf'}{f\phi'}\vartheta_1
+\frac{fh^2\phi'^2F_4}{s}
\left(\frac{1}{r}-\frac{f'}{2f}\right).
\label{evenTensorRadialRelations}
\end{align}
The first relation in Eq.~(\ref{evenTensorRadialRelations}) is therefore
equivalent to Eq.~(\ref{evenPOnShellIdentity}), whereas the second follows
directly from the coefficient definitions. The dependence on $F_4$ and $F_5$
is already incorporated in $\mH$, and hence in $a_4=f\mH/(2s)$, as well as in
the remaining background coefficients. Combining
Eq.~(\ref{evenTensorRadialRelations}) with the coefficient identities in
Appendix~\ref{AppEvenCoefficients} yields the exact on-shell identity
\be
\mF^2\Delta_G+fh\,\mF\mG T_r
+f^2h^2\mG^2\Delta_K=0.
\label{evenTensorRadialIdentity}
\ee
Consequently, the radial characteristic polynomial factorizes as
\be
\det\left(fh\,c_r^2{\bm K}+{\bm G}\right)
=f^2h^2\Delta_K
\left(c_r^2-\frac{\mG}{\mF}\right)
\left(c_r^2-\frac{\mF\Delta_G}
{f^2h^2\mG\Delta_K}\right).
\label{evenRadialFactorization}
\ee
The two branches are therefore identified unambiguously as
\be
c_{r,{\rm grav}}^{2(+)}=\frac{\mG}{\mF}
=c_{r,{\rm odd}}^{2(+)},\qquad
c_{r,{\rm sc}}^{2(+)}=
\frac{\mF\Delta_G}{f^2h^2\mG\Delta_K}.
\label{evenRadialModeSpeeds}
\ee
Thus, the even- and odd-parity tensor modes propagate on the same radial
characteristic cone,
$c_{r,{\rm grav}}^{2(+)}=c_{r,{\rm odd}}^{2(+)}$, even in the presence of
both $F_4$ and $F_5$. Although neither $F_4$ nor $F_5$ enters $\mF$
explicitly, both contribute to $\mG$ and $\mH$ and hence affect the scalar
radial speed and the scalar--tensor mixing in the even-parity sector. In the
Horndeski limit, the two radial propagation
speeds reduce to those obtained in Ref.~\cite{Kobayashi:2014wsa}.

Under the odd-parity stability conditions $\mF>0$ and $\mG>0$, the squared
tensor radial speed $c_{r,{\rm grav}}^{2(+)}$ in
Eq.~(\ref{evenRadialModeSpeeds}) is positive. On the regular branch with
$f,h>0$, once the even-parity no-ghost condition $\Delta_K>0$ is imposed,
the squared scalar radial speed $c_{r,{\rm sc}}^{2(+)}$ in
Eq.~(\ref{evenRadialModeSpeeds}) is positive if and only if $\Delta_G>0$.
Hence, in addition to the odd-parity conditions, the required even-parity
no-ghost and radial-gradient stability conditions are
\be
\Delta_K>0,\qquad \Delta_G>0.
\label{evenRadialStability}
\ee
Together with $\mF>0$ and $\mG>0$, the inequalities in
Eq.~(\ref{evenRadialStability}) also fix the sign of $T_r$. Specifically,
Eq.~(\ref{evenTensorRadialIdentity}) gives
\be
T_r=-\frac{\mF^2\Delta_G+f^2h^2\mG^2\Delta_K}
{fh\,\mF\mG}<0,
\ee
where the strict inequality follows because
$f,h,\mF,\mG,\Delta_K,\Delta_G>0$. Thus, no additional condition on $T_r$ is
required. Since ${\bm K}$ is positive definite, $T_r<0$ together with
$\Delta_G>0$ is equivalent to the negative definiteness of ${\bm G}$, namely
$G_{11}<0$ and $\det{\bm G}=\Delta_G>0$.

We finally consider angular propagation in the eikonal limit, $L\gg1$, with
vanishing radial wave number. The term involving ${\bm Q}$ contains a radial
derivative and therefore does not contribute in this limit. Locally, the full
perturbation, including its spherical-harmonic dependence, can be represented
by an angular WKB factor
$\exp[\mathrm{i}(\omega t-\sqrt{L}\,\theta)]$; the reduced amplitudes
$\vec{\cal X}(t,r)$ themselves remain independent of $\theta$. The frequency
and angular wave number measured in a local orthonormal frame are
$\omega/\sqrt{f}$ and $k_\Omega=\sqrt{L}/r$, respectively. The corresponding
angular propagation speed is
$c_\Omega=\omega/(\sqrt{f}\,k_\Omega)=r\omega/(\sqrt{fL})$.
At leading order in large $L$, the principal equations yield
\be
\det\left(f c_\Omega^2\widetilde{\bm K}
+r^2\widetilde{\bm M}\right)=0.
\label{evenAngularDispersion}
\ee
The two reduced fields scale differently in the large-$L$ limit. To isolate
finite coefficients, we define
\begin{align}
\widetilde K_{11}&\equiv\lim_{L\to\infty}L^2K_{11},&
\widetilde K_{12}&\equiv\lim_{L\to\infty}LK_{12},&
\widetilde K_{22}&\equiv\lim_{L\to\infty}K_{22},\notag\\
\widetilde M_{11}&\equiv\lim_{L\to\infty}LM_{11},&
\widetilde M_{12}&\equiv\lim_{L\to\infty}M_{12},&
\widetilde M_{22}&\equiv\lim_{L\to\infty}\frac{M_{22}}L.
\label{evenLargeLdefinitions}
\end{align}
Defining the $L$-independent combination $\bar a_2\equiv a_2-L\vartheta_1$,
the terms proportional to $L\vartheta_1$ cancel exactly in both $b_3$ and
$e_1$, leaving them independent of $L$. The remaining large-$L$ expressions
can be written compactly in terms of
\begin{align}
\kappa_\infty&\equiv\frac{a_4}{f},&
\lambda_\infty&\equiv\frac{a_4}{a_3},&
{\cal D}_{\infty1}&\equiv a_9-a_4',&
{\cal D}_{\infty2}&\equiv-a_8\lambda_\infty,\notag\\
{\cal C}_\infty&\equiv\vartheta_1-a_8\beta_{\rm ev},&
{\cal E}_\infty&\equiv\lambda_\infty c_{\rm ev},&
{\cal N}_{\infty1}&\equiv a_8b_3+a_7{\cal E}_\infty+a_6b_4,&
{\cal N}_{\infty2}&\equiv a_8{\cal E}_\infty.
\label{evenLargeLAuxiliary}
\end{align}
On the regular branch characterized by ${\cal D}_{\infty 2}\neq0$, a direct
large-$L$ expansion of the kinetic matrix gives
\begin{align}
\widetilde K_{11}={}&b_4\left(\frac{a_7}{{\cal D}_{\infty2}}
-\frac{a_8{\cal D}_{\infty1}}{{\cal D}_{\infty2}^2}\right)
-\frac12\left(\frac{a_3b_4}{{\cal D}_{\infty2}}\right)',
\label{evenLargeLKcomponents0}\\
\widetilde K_{12}={}&\frac12\left[-\frac{b_4b_3}{\kappa_\infty}
+\frac{{\cal N}_{\infty1}}{{\cal D}_{\infty2}}
-\frac{{\cal N}_{\infty2}{\cal D}_{\infty1}}{{\cal D}_{\infty2}^2}
-\left(\frac{a_3{\cal E}_\infty}{{\cal D}_{\infty2}}\right)'\right],
\label{evenLargeLKcomponents1}\\
\widetilde K_{22}={}&e_1+\frac{a_6{\cal E}_\infty}{{\cal D}_{\infty2}}
-\frac12\left(\frac{{\cal C}_\infty{\cal E}_\infty}
{{\cal D}_{\infty2}}-\beta_{\rm ev}c_{\rm ev}\right)'.
\label{evenLargeLKcomponents}
\end{align}
For the leading angular terms, it is useful to introduce
\begin{align}
\nu_\infty&\equiv-\frac{a_8}{{\cal D}_{\infty2}},&
\tau_\infty&\equiv\frac{a_4}{{\cal D}_{\infty2}},&
\zeta_\infty&\equiv\lambda_\infty
\left(\frac{a_7}{{\cal D}_{\infty2}}
-\frac{a_8{\cal D}_{\infty1}}{{\cal D}_{\infty2}^2}\right),\notag\\
\tau_0&\equiv\frac{\lambda_\infty{\cal C}_\infty}
{{\cal D}_{\infty2}}-\beta_{\rm ev}=-\frac{\vartheta_1}{a_8},&
\zeta_0&\equiv\frac{\lambda_\infty a_6}{{\cal D}_{\infty2}}
=-\frac{a_6}{a_8}.
\label{evenLargeLVectors}
\end{align}
After a final integration by parts in $r$, the components of the leading-order
algebraic matrix are
\begin{align}
\widetilde M_{11}={}&c_5\zeta_\infty\nu_\infty+d_4\nu_\infty^2
-\frac12(c_5\nu_\infty\tau_\infty)',\label{evenLargeLMcomponents0}\\
\widetilde M_{12}={}&\frac12(c_4\zeta_\infty+c_5\nu_\infty\zeta_0
+d_3\nu_\infty)
-\frac14(c_5\nu_\infty\tau_0+d_2\nu_\infty
+c_4\tau_\infty)',\label{evenLargeLMcomponents1}\\
\widetilde M_{22}={}&e_4+c_4\zeta_0-\frac12(c_4\tau_0)'.
\label{evenLargeLMcomponents}
\end{align}
The reduction of Eq.~(\ref{evenAngularDispersion}) to the invariants and speed
formulas below assumes the nondegenerate large-$L$ branch with
$\det\widetilde{\bm K}\ne0$; on the no-ghost branch,
$\widetilde{\bm K}$ is positive definite. If
$\det\widetilde{\bm K}=0$, the leading kinetic matrix is degenerate and the
large-$L$ balance must instead be obtained from
Eq.~(\ref{evenReducedTwoFieldAction}) with the relevant subleading terms
retained.

We then introduce the combinations
\begin{align}
\Delta_{\widetilde K}&\equiv\det\widetilde{\bm K},&
\Delta_{\widetilde M}&\equiv\det\widetilde{\bm M},&
T_\Omega^{(+)}&\equiv
\widetilde K_{11}\widetilde M_{22}
+\widetilde K_{22}\widetilde M_{11}
-2\widetilde K_{12}\widetilde M_{12},\notag\\
B_1^{(+)}&\equiv\frac{r^2T_\Omega^{(+)}}
{2f\Delta_{\widetilde K}},&
B_2^{(+)}&\equiv\frac{r^4\Delta_{\widetilde M}}
{f^2\Delta_{\widetilde K}}.
\label{evenAngularBdefinitions}
\end{align}
Solving the angular characteristic equation gives
\be
c_{\Omega\pm,{\rm even}}^{2(+)}
=-B_1^{(+)}
\pm\sqrt{\left[B_1^{(+)}\right]^2-B_2^{(+)}}.
\label{evenAngularSpeeds}
\ee
Both squared speeds are real and positive if and only if
\be
B_1^{(+)}<0,\qquad
0<B_2^{(+)}\leq\left[B_1^{(+)}\right]^2.
\label{evenAngularStability}
\ee
For positive-definite $\widetilde{\bm K}$, these conditions are equivalent to
$T_\Omega^{(+)}<0$ and $\Delta_{\widetilde M}>0$, 
supplemented by the hyperbolicity bound
$B_2^{(+)}\leq\left[B_1^{(+)}\right]^2$. In the Horndeski limit
$F_4=F_5=0$, one has $\vartheta_1=\vartheta_2=\tau_0=0$, and $B_1^{(+)}$ and
$B_2^{(+)}$ reduce to the corresponding quantities obtained in
Ref.~\cite{Kase:2023mho}. Equation~(\ref{evenAngularSpeeds}) then reproduces
Eq.~(95) of that reference exactly.

\subsubsection{Region with timelike $r$: $f,h<0$}
\label{evenInteriorStabilitySubsubsec}

For $f,h<0$, the radial coordinate $r$ is timelike, whereas $t$ is spacelike.
Consequently, ${\bm G}$ rather than ${\bm K}$ is the kinetic matrix. The field
redefinition in Eq.~(\ref{evenpsidefinition}) contains $\delta\phi'$, which is
now a time derivative, so its equivalence to the original formulation must be
established separately. We therefore introduce the parent action
\be
S_{\rm par}^{(2)}=S_{\rm unred}^{(2)}
+\sum_{\ell\geq2}\int{\rm d}t\,{\rm d}r\,\lambda_\psi
\left(\psi-H_2-\frac{La_4}{a_3}h_1-\beta_{\rm ev}\delta\phi'\right).
\label{evenInteriorParentAction}
\ee
Here, $S_{\rm unred}^{(2)}$ denotes the action in
Eqs.~(\ref{evenReducedActionGeneral}) and (\ref{evenStotal}) before the
constraints are eliminated. Variation with respect to $\lambda_\psi$ imposes
Eq.~(\ref{evenpsidefinition}); substituting this constraint into the parent
action recovers $S_{\rm unred}^{(2)}$. On the regular branch
\be
\phi'\ne0,\qquad a_3\ne0,\qquad \frac{a_4}{f}\ne0,
\qquad {\cal D}_{\rm ev}\ne0,
\ee
the $H_0$ constraint, the definition enforced by $\lambda_\psi$, and the
$H_1$, $H_2$, and $h_1$ equations can be solved successively and locally for
$h_1$, $H_2$, $H_1$, and then $H_0$ and $\lambda_\psi$. Eliminating these
variables in either order requires no inverse $r$-derivative operator and gives
the same
two-field action~(\ref{evenReducedTwoFieldAction}). Thus, the parent,
original, and reduced formulations have the same local constraint rank and
are locally equivalent. Taking
$r$ as the evolution coordinate, the velocity Hessian of the reduced action
is $2{\bm G}$. The term involving the antisymmetric matrix ${\bm Q}$ is only
linear in the $r$ derivatives and does not contribute to this Hessian.
Therefore, the absence of ghosts requires ${\bm G}$ to be positive definite.
Since ${\bm G}$ is a real symmetric $2\times2$ matrix, Sylvester's criterion
gives
\be
G_{11}>0,\qquad \Delta_G={\rm det}\,{\bm G}>0.
\label{evenInteriorNoGhostCondition}
\ee
To analyze gradients along the spacelike $t$ direction, we use the same
algebraic definitions as in Eq.~(\ref{evenGLPVPDefinitions}), evaluated in the
region $f,h<0$:
\ba
\mu_{\rm ev}^{(-)} &\equiv& -\frac{4a_3}{hs},
\qquad \qquad \qquad \quad
\mH_\vartheta^{(-)} \equiv\mH+\frac{2\phi'\vartheta_1}{s},\notag\\
{\cal P}_{1}^{(-)} &\equiv& \frac1s
\left(\frac{s r^2\mH\mH_\vartheta^{(-)}}{\mu_{\rm ev}^{(-)}}\right)',\qquad
{\cal P}_{2}^{(-)}\equiv
\frac{h}{f}(rf'-2f)\mu_{\rm ev}^{(-)},\qquad
\Theta_{\rm ev}^{(-)}\equiv{\cal P}_{2}^{(-)}
+2rL\mH_\vartheta^{(-)}.
\label{evenInteriorPDefinitions}
\ea
For $f,h<0$, the ratio $f/h$ remains positive, and we retain the convention
$s=\sqrt{f/h}>0$. The superscript $(-)$ labels the region and does not
introduce an additional sign in, for example,
$\mu_{\rm ev}^{(-)}=-4a_3/(hs)$. Whenever $\sqrt{fh}$ occurs, the consistent
algebraic branch is $\sqrt{fh}=-hs=-f/s$. The quantities in
Eq.~(\ref{evenInteriorPDefinitions}) are therefore real. The exact coefficient
identities yield
\begin{align}
K_{11}&=\frac{s h^2(\mu_{\rm ev}^{(-)})^4
\left(L{\cal P}_{1}^{(-)}-\mF\right)}
{2fL\mH^2(\Theta_{\rm ev}^{(-)})^2},\\
\Delta_K&=\frac{(L-2)h(\mu_{\rm ev}^{(-)})^4\mF}
{4fL\mH^2\phi'^2(\Theta_{\rm ev}^{(-)})^2}
\left(2{\cal P}_{1}^{(-)}-\mF\right).
\label{evenInteriorKFactorizations}
\end{align}
These factorizations do not determine the no-ghost conditions for $f,h<0$,
because the kinetic matrix is ${\bm G}$ rather than ${\bm K}$. In particular,
${\bm G}$ also depends on the coefficients $c_2$, $c_4$, $e_2$, and
$\vartheta_2$. Instead, Eq.~(\ref{evenInteriorKFactorizations}) organizes the
gradient-stability conditions for propagation along the spacelike $t$
direction.

We next consider the principal characteristics in the $(t,r)$ plane. For a
local Fourier mode
$\vec{\cal X}=\vec{\cal X}_0\exp[\mathrm{i}(\omega r-kt)]$, the frequency
measured with respect to proper time along the timelike $r$ direction is
$\sqrt{-h}\,\omega$, whereas the physical wave number along the spacelike
$t$ direction is $k/\sqrt{-f}$. The corresponding propagation speed is
$c_\parallel=\sqrt{fh}\,\omega/k$. The principal part of the perturbation
equations then gives
\be
\det\left(fh\,{\bm K}+c_\parallel^2{\bm G}\right)=0,
\qquad
c_{\parallel\pm,{\rm even}}^{2(-)}
=
\frac{fh\left[-T_r\pm\sqrt{T_r^2-4\Delta_K\Delta_G}\right]}
{2\Delta_G}\,,
\label{evenInteriorRadialSpeeds}
\ee
where $T_r$ is the mixed invariant defined in
Eq.~(\ref{evenRadialInvariants}). Interchanging the timelike and spacelike
directions makes the characteristic polynomial for $f,h<0$ reciprocal to its
$f,h>0$ counterpart. More explicitly, defining
$P_+(z_+)\equiv\det(fh\,z_+{\bm K}+{\bm G})$, we have
\be
P_-(z_-)\equiv\det(fh\,{\bm K}+z_-{\bm G})
=z_-^2P_+(1/z_-).
\label{evenReciprocalRadialPolynomial}
\ee
Using the exact factorization in Eq.~(\ref{evenRadialFactorization}), 
the two squared propagation speeds are
\be
c_{\parallel,{\rm grav}}^{2(-)}
=\frac{\mF}{\mG}
=c_{\parallel,{\rm odd}}^{2(-)},
\qquad
c_{\parallel,{\rm sc}}^{2(-)}
=
\frac{f^2h^2\mG\Delta_K}{\mF\Delta_G}.
\label{evenInteriorRadialModeSpeeds}
\ee
Thus, the odd- and even-parity tensor characteristics continue to coincide
for $f,h<0$. The reciprocal forms of both propagation speeds
relative to their $f,h>0$ counterparts arise solely from the interchange
of the timelike and spacelike directions in the $(t,r)$ plane.

Under the odd-parity stability conditions $\mF>0$ and $\mG>0$, the
squared tensor propagation speed in
Eq.~(\ref{evenInteriorRadialModeSpeeds}) is positive. The even-parity
no-ghost conditions in Eq.~(\ref{evenInteriorNoGhostCondition}) imply
$\Delta_G>0$, so the squared scalar propagation speed is positive if and
only if $\Delta_K>0$. Hence, the additional even-parity stability condition
against gradients along the spacelike $t$ direction is
\be
\Delta_K>0.
\label{evenInteriorRadialStability}
\ee
For $f,h<0$ and $L\geq6$, the exact factorization in
Eq.~(\ref{evenInteriorKFactorizations}) shows that this condition is
equivalent to
\be
\mK^{(-)}\equiv2{\cal P}_{1}^{(-)}-\mF>0.
\ee
Moreover, $\mK^{(-)}>0$ implies
$L{\cal P}_{1}^{(-)}-\mF>0$ for $\mF>0$ and $L\geq6$, and hence
$K_{11}<0$. Together with $\Delta_K>0$, this confirms that
${\bm K}$ is negative definite. Thus, $\mK^{(-)}>0$ is the compact
condition for stability against gradients along the spacelike $t$
direction, whereas the independent no-ghost conditions remain those in
Eq.~(\ref{evenInteriorNoGhostCondition}).

Angular propagation has a qualitatively different principal structure when
$f,h<0$. Since $r$ is timelike, an $r$ derivative is a time derivative, so
the term involving the antisymmetric matrix ${\bm Q}$ contributes to the
principal symbol and cannot be neglected. To characterize this contribution,
we define
\begin{align}
U_{\infty1}&\equiv-\frac{a_3}{{\cal D}_{\infty2}},&
U_{\infty2}&\equiv-\frac{{\cal C}_\infty}{{\cal D}_{\infty2}},&
W_{\infty1}&\equiv\frac{a_4}{{\cal D}_{\infty2}},&
W_{\infty2}&\equiv-\frac{\vartheta_1}{a_8}.
\end{align}
The large-$L$ kinetic matrix for evolution along the timelike $r$ direction
is
\begin{equation}
\widetilde{\bm G}\equiv
\lim_{L\to\infty}
\operatorname{diag}(L,1)\,{\bm G}\,\operatorname{diag}(L,1),
\end{equation}
whose components are
\begin{align}
\widetilde G_{11}={}&c_5W_{\infty1}U_{\infty1}
+c_6W_{\infty1}^2,\label{evenInteriorLargeLG0} \\
\widetilde G_{12}={}&\frac12\left[
c_2W_{\infty1}
+c_5(W_{\infty1}U_{\infty2}+W_{\infty2}U_{\infty1})
+2c_6W_{\infty1}W_{\infty2}
+d_2U_{\infty1}\right],\label{evenInteriorLargeLG1} \\
\widetilde G_{22}={}&c_2W_{\infty2}
+c_5W_{\infty2}U_{\infty2}
+c_6W_{\infty2}^2
+d_2U_{\infty2}+e_2.
\label{evenInteriorLargeLG}
\end{align}
The same large-$L$ expansion gives the leading antisymmetric coefficient
\begin{align}
\Upsilon_\infty
&\equiv\lim_{L\to\infty}Q_{12}
=\frac{c_5\vartheta_1-d_2a_8-c_4a_4}
{2{\cal D}_{\infty2}}\notag\\
&=\frac{1}{2{\cal D}_{\infty2}}
\left\{
\frac{a_4\phi'}{h}\vartheta_2
-\vartheta_1\left[
\frac{hs}{2r}\mG
+\frac{f'a_4}{2f}
\right]
\right\}.
\label{evenInteriorQleading}
\end{align}
Using Eq.~(\ref{evenTensorRadialRelations}), $a_4=f\mH/(2s)$, and the
definitions of $\mG$ and $\mH$, this coefficient factorizes as
\be
\Upsilon_\infty=
\frac{fh^3\phi'^4}{2r{\cal D}_{\infty2}}
\left(\frac{f'}{2f}-\frac1r\right)
\left[
3F_5\left(G_4-2XG_{4,X}+XG_{5,\phi}\right)
-XF_4G_{5,X}
\right].
\label{evenInteriorUpsilonFactorization}
\ee
The square bracket vanishes under the quartic--quintic Horndeski-related
compatibility condition
\be
XG_{5,X}F_4
=3F_5\left(G_4-2XG_{4,X}+XG_{5,\phi}\right).
\label{GLPVcompatibility}
\ee
This condition aligns the covariant degeneracy directions of the quartic and
quintic sectors, so that the combined kinetic Hessian retains the primary
constraint that removes the additional higher-derivative mode
\cite{Crisostomi:2016tcp,BenAchour:2016fzp}. Equivalently, the two sectors
admit the same disformal map to Horndeski theories
\cite{Crisostomi:2016tcp,Mironov:2022ffa}. 
Equation~(\ref{GLPVcompatibility}) is therefore a nontrivial restriction on
the otherwise independent functions $F_4$ and $F_5$: although the quartic
and quintic sectors are separately degenerate, their sum retains covariant
degeneracy only when their kinetic null directions coincide. Imposing this
condition sets the square bracket in
Eq.~(\ref{evenInteriorUpsilonFactorization}) to zero and hence gives
$\Upsilon_\infty=0$ on the regular large-$L$ branch. The compatibility
condition is automatically satisfied in the Horndeski limit $F_4=F_5=0$
and in the quartic subclass $G_{5,X}=F_5=0$. Even when the compatibility
condition is not satisfied, $\Upsilon_\infty$ can vanish accidentally at
special background loci, such as $\phi'=0$ or $f'/(2f)=1/r$ in
Eq.~(\ref{evenInteriorUpsilonFactorization}); such zeros do not imply
covariant degeneracy.

We now isolate from the reduced even-parity
Lagrangian~(\ref{evenReducedTwoFieldAction}) the local principal terms that
govern large-$L$ angular propagation for $f,h<0$. We denote them by
${\cal L}_{\Omega,-}^{(2)}$, where $\Omega$ refers to angular propagation and
the minus sign labels the branch $f,h<0$. These terms are obtained by setting
the momentum along the spacelike $t$ direction to zero and retaining the
kinetic terms with derivatives along the timelike $r$ direction, together
with the leading angular-gradient terms.

Let ${\bm R}_L=\operatorname{diag}(L,1)$,
$\vec{\cal X}={\bm R}_L\vec{\cal Y}$, and
${\bm\epsilon}_2=\begin{pmatrix}0&1\\-1&0\end{pmatrix}$.
The definitions of $\widetilde{\bm G}$ and $\widetilde{\bm M}$, together
with Eq.~(\ref{evenInteriorQleading}), imply
\be
{\bm R}_L{\bm G}{\bm R}_L
=\widetilde{\bm G}+o(1),\qquad
{\bm R}_L{\bm Q}{\bm R}_L
=L\Upsilon_\infty{\bm\epsilon}_2+o(L),\qquad
{\bm R}_L{\bm M}{\bm R}_L
=L\widetilde{\bm M}+o(L).
\label{evenInteriorLargeLMatrixScaling}
\ee
Here, $o(1)$ denotes terms that vanish as $L\to\infty$, whereas $o(L)$
denotes terms whose ratio to $L$ vanishes in this limit, with the local
background quantities held fixed. Since ${\bm Q}$ is antisymmetric, the
off-diagonal rescaling produces the factor
$({\bm R}_L)_{11}({\bm R}_L)_{22}=L$ in the second relation, while
$\lim_{L\to\infty}Q_{12}=\Upsilon_\infty$. The third relation is equivalent
to the componentwise definitions of $\widetilde{\bm M}$ in
Eq.~(\ref{evenLargeLdefinitions}). The factor of $L$ reflects the leading
angular-gradient contribution from the spherical-harmonic decomposition.

Freezing the background coefficients locally, we therefore obtain
\be
{\cal L}_{\Omega,-}^{(2)}=
(\vec{\cal Y}')^{{\rm T}}\widetilde{\bm G}\vec{\cal Y}'
+L\Upsilon_\infty\vec{\cal Y}^{{\rm T}}
{\bm\epsilon}_2\vec{\cal Y}'
+L\vec{\cal Y}^{{\rm T}}\widetilde{\bm M}\vec{\cal Y}
+o(L).
\label{evenInteriorAngularPrincipalAction}
\ee
Although the first term has no explicit factor of $L$, it contains the
highest derivatives with respect to the timelike coordinate $r$ and must be
retained. Its relative importance depends on the mode frequency, whose
large-$L$ scaling must be derived from the dispersion relation rather than
assumed in advance.

As in the $f,h>0$ case, we consider a local mode
$\vec{\cal Y}=\vec{\cal Y}_0\exp(-\mathrm{i}\varpi r)$, where $\varpi$ is
the coordinate frequency conjugate to the timelike coordinate $r$. The
locally measured frequency is $\sqrt{-h}\,\varpi$. Substituting this ansatz
into the Euler--Lagrange equations derived from
Eq.~(\ref{evenInteriorAngularPrincipalAction}) and requiring a nontrivial
amplitude $\vec{\cal Y}_0$ gives
\be
\det\left(
\varpi^2\widetilde{\bm G}
-\mathrm{i}\varpi L\Upsilon_\infty{\bm\epsilon}_2
+L\widetilde{\bm M}
\right)=0.
\label{evenInteriorAngularDispersion}
\ee
We restrict the following discussion to the nondegenerate large-$L$
no-ghost branch, on which $\widetilde{\bm G}$ is positive definite. The
degenerate branch with $\det\widetilde{\bm G}=0$ is not considered further
here. On that branch, the leading kinetic matrix for $r$-time evolution is
degenerate, and Eq.~(\ref{evenInteriorAngularDispersion}) need not capture
the dominant large-$L$ balance. Its stability must instead be analyzed
directly from Eq.~(\ref{evenReducedTwoFieldAction}), with the appropriate
subleading terms retained. On the nondegenerate branch, we define
\begin{align}
\Delta_{\widetilde G}&\equiv\det\widetilde{\bm G},&
T_\Omega^{(-)}&\equiv\widetilde G_{11}\widetilde M_{22}
+\widetilde G_{22}\widetilde M_{11}
-2\widetilde G_{12}\widetilde M_{12},\notag\\
B_1^{(-)}(L)&\equiv
\frac{(-h)r^2[T_\Omega^{(-)}-L\Upsilon_\infty^2]}
{2\Delta_{\widetilde G}},&
B_2^{(-)}&\equiv
\frac{h^2r^4\Delta_{\widetilde M}}{\Delta_{\widetilde G}}.
\label{evenInteriorAngularInvariants}
\end{align}
Although the characteristic matrix in
Eq.~(\ref{evenInteriorAngularDispersion}) contains a term linear in
$\varpi$, its determinant is even in $\varpi$: $\widetilde{\bm G}$ and
$\widetilde{\bm M}$ are symmetric, whereas ${\bm\epsilon}_2$ is
antisymmetric. Defining $c_\Omega^2=(-h)r^2\varpi^2/L$, the resulting
frozen-coefficient large-$L$ characteristic equation is quadratic in
$c_\Omega^2$, with formal roots
\be
c_{\Omega\pm,{\rm even}}^{2(-)}=
-B_1^{(-)}(L)
\pm\sqrt{\left[B_1^{(-)}(L)\right]^2-B_2^{(-)}}.
\label{evenInteriorAngularSpeeds}
\ee
Both roots are real and positive if and only if
\be
B_1^{(-)}(L)<0,\qquad
0<B_2^{(-)}\leq\left[B_1^{(-)}(L)\right]^2.
\label{evenInteriorAngularStability}
\ee
For $\Upsilon_\infty\ne0$ and fixed finite background coefficients, the
no-ghost condition $\Delta_{\widetilde G}>0$, together with $-h>0$, gives
$B_1^{(-)}(L)=-(-h)r^2L\Upsilon_\infty^2/
(2\Delta_{\widetilde G})+O(1)<0$ for sufficiently large $L$. Moreover,
$B_2^{(-)}=O(1)$ and
$\operatorname{sgn}B_2^{(-)}=\operatorname{sgn}\Delta_{\widetilde M}$,
whereas $[B_1^{(-)}(L)]^2=O(L^2)$. The upper bound in
Eq.~(\ref{evenInteriorAngularStability}) is therefore automatically
satisfied at sufficiently large $L$, and the remaining positivity condition
for the frozen-coefficient roots is $\Delta_{\widetilde M}>0$. On this branch,
the two roots have the asymptotic forms
\be
c_{\Omega+,{\rm even}}^{2(-)}
=
\frac{(-h)r^2L\Upsilon_\infty^2}
{\Delta_{\widetilde G}}
+O(1),
\qquad
c_{\Omega-,{\rm even}}^{2(-)}
=
\frac{(-h)r^2\Delta_{\widetilde M}}
{L\Upsilon_\infty^2}
+O(L^{-2}).
\label{evenInteriorAngularAsymptotics}
\ee
The first expression in Eq.~(\ref{evenInteriorAngularAsymptotics}) is the
plus root, which grows as $L$. The second is the minus root, which decreases
as $L^{-1}$ and tends to zero.
Since $c_\Omega^2=(-h)r^2\varpi^2/L$, their coordinate frequencies scale as
$\varpi_+^2=O(L^2)$ and $\varpi_-^2=O(1)$, respectively, instead of both
obeying the standard angular-eikonal scaling $\varpi^2=O(L)$.

The plus branch is a genuine high-frequency solution, with
$\varpi_+=O(L)$. 
For this branch, the local large-$L$ expansion obtained by treating the
background coefficients as constant at a fixed radius is asymptotically
self-consistent: the kinetic and antisymmetric-mixing terms scale as
$O(L^2)$, while the terms involving radial derivatives of the background
coefficients are subleading.
We introduce the locally measured angular wave
number and frequency, $k_\Omega=\sqrt{L}/r$ and
$\Omega=\sqrt{-h}\,\varpi$, and denote the phase and group velocities of the
plus branch by $v_{{\rm ph},+}$ and $v_{{\rm gr},+}$, respectively. 
The plus-branch relation in
Eq.~(\ref{evenInteriorAngularAsymptotics}) then gives
\be
 \Omega_+^2=
 \frac{(-h)r^4\Upsilon_\infty^2}{\Delta_{\widetilde G}}
 k_\Omega^4+O(k_\Omega^2),
 \qquad
 v_{{\rm ph},+}=\frac{\Omega_+}{k_\Omega},\qquad
 v_{{\rm gr},+}=\frac{{\rm d}\Omega_+}{{\rm d}k_\Omega}
 =2v_{{\rm ph},+}+O(k_\Omega^{-1}).
\label{evenInteriorHighBranchPhysicalDispersion}
\ee
For $\Upsilon_\infty\ne0$, Eq.~(\ref{evenInteriorHighBranchPhysicalDispersion})
gives $\Omega_+\propto k_\Omega^2$. Consequently, both
$v_{{\rm ph},+}$ and $v_{{\rm gr},+}$ grow linearly with $k_\Omega$ and
diverge in the formal limit $k_\Omega\to\infty$. The plus branch therefore
has no finite, scale-independent angular propagation cone in the standard
relativistic eikonal limit.

By contrast, although the minus branch has an angular wave number
$k_\Omega=O(\sqrt{L})$, its frequency remains finite,
$\varpi_-=O(1)$. It is therefore not a standard high-frequency eikonal
branch, and the background-derivative terms omitted in deriving
Eq.~(\ref{evenInteriorAngularDispersion}) can contribute at the same order
to its dispersion relation. 
Consequently,
Eq.~(\ref{evenInteriorAngularStability}) characterizes the two roots of the
frozen-coefficient problem but, for $\Upsilon_\infty\ne0$, does not provide
a uniform eikonal stability criterion for the plus and minus angular
branches in Eq.~(\ref{evenInteriorAngularSpeeds}). 
The large-$L$
expansion is also nonuniform near $\Upsilon_\infty=0$. Taking $L\to\infty$
at fixed $\Upsilon_\infty\ne0$ produces the $L^{\pm1}$ hierarchy in
Eq.~(\ref{evenInteriorAngularAsymptotics}), whereas setting
$\Upsilon_\infty=0$ first removes the antisymmetric mixing responsible for
this hierarchy and, on the regular nondegenerate branch, gives the standard
$O(1)$ eikonal scaling for both roots. Thus, the limits $L\to\infty$ and
$\Upsilon_\infty\to0$ do not commute.

Away from accidental zeros of the prefactors in
Eq.~(\ref{evenInteriorUpsilonFactorization}), requiring both angular branches
to have finite, nonzero phase speeds with the standard scaling
$\varpi^2=O(L)$ imposes $\Upsilon_\infty=0$. On the regular branch, the
background-independent compatibility condition~(\ref{GLPVcompatibility})
enforces this requirement, as explained after
Eq.~(\ref{evenInteriorUpsilonFactorization}). On the resulting regular-eikonal
branch, the $L\Upsilon_\infty^2$ term in $B_1^{(-)}(L)$
vanishes, and both angular speeds have finite eikonal limits provided the
remaining invariants are finite. Their reality and positivity still require
Eq.~(\ref{evenInteriorAngularStability}), whereas the $L^{\pm1}$ hierarchy
found in Eq.~(\ref{evenInteriorAngularAsymptotics}) does not arise, as in
Horndeski theories. If the action is regarded instead as a finite-cutoff
effective theory on a background with a preferred scalar-gradient direction,
however, Lorentz-invariant ultraviolet scaling need not apply, and the formal
limit $k_\Omega\to\infty$ may lie outside the theory's domain of validity.

For comparison with the Horndeski limit, let
$\widetilde{\bm G}_{\rm H}$ and $\widetilde{\bm M}_{\rm H}$ denote the
Horndeski limits of the corresponding matrices and set
$\Delta_{\widetilde G,{\rm H}}=\det\widetilde{\bm G}_{\rm H}$.  The
Horndeski invariants for $f,h<0$ are
\begin{align}
B_{1,{\rm H}}^{(-)}&\equiv
\frac{(-h)r^2}{2\Delta_{\widetilde G,{\rm H}}}
\left(\widetilde G_{11}^{\rm H}\widetilde M_{22}^{\rm H}
+\widetilde G_{22}^{\rm H}\widetilde M_{11}^{\rm H}
-2\widetilde G_{12}^{\rm H}\widetilde M_{12}^{\rm H}\right),
\label{evenInteriorHorndeskiB1}\\
B_{2,{\rm H}}^{(-)}&\equiv
\frac{h^2r^4\det\widetilde{\bm M}_{\rm H}}
{\Delta_{\widetilde G,{\rm H}}}.
\label{evenInteriorHorndeskiB2}
\end{align}
The two squared propagation speeds are
\be
\left.c_{\Omega\pm,{\rm even}}^{2(-)}\right|_{F_4=F_5=0}
=-B_{1,{\rm H}}^{(-)}
\pm\sqrt{\left(B_{1,{\rm H}}^{(-)}\right)^2-B_{2,{\rm H}}^{(-)}}.
\label{evenInteriorHorndeskiAngularSpeeds}
\ee
If $B_{1,{\rm H}}^{(-)}$ and $B_{2,{\rm H}}^{(-)}$ are finite and satisfy
$B_{1,{\rm H}}^{(-)}<0$ and
$0<B_{2,{\rm H}}^{(-)}\leq \left(B_{1,{\rm H}}^{(-)}\right)^2$,
both squared speeds are finite and strictly positive. 
The interior quantities $B_{1,{\rm H}}^{(-)}$ and
$B_{2,{\rm H}}^{(-)}$ do not generally coincide with their exterior
Horndeski counterparts in Ref.~\cite{Kase:2023mho}. This is because the
matrix governing the timelike principal part of the angular characteristic
problem is $\widetilde{\bm G}$ when $r$ is timelike, whereas it is
$\widetilde{\bm K}$ when $t$ is timelike. The reciprocal
identity~(\ref{evenReciprocalRadialPolynomial}) concerns only the radial
principal symbol in the $(t,r)$ plane and therefore does not relate the
interior and exterior angular invariants.

\subsection{Monopole: $\ell=0$}
\label{evenMonopoleSubsec}

For the monopole mode $\ell=0$, the spherical harmonic $Y_0$ is constant on
the two-sphere and satisfies
$\nabla_aY_0=\nabla_a\nabla_bY_0=0$. Hence, in the decomposition
(\ref{evendecompositiongeneral}), the terms multiplying $h_0$, $h_1$, and
$G$ vanish identically, and these amplitudes are absent. Among the gauge
conditions in Eq.~(\ref{evenuniformcurvaturegauge}), only $K=0$ remains
nontrivial and can be imposed using the radial monopole gauge freedom. We
therefore restrict the ungauge-fixed harmonic action directly to $\ell=0$
and impose $K=0$. The resulting action is algebraically identical to the
formal $L=0$ specialization of Eq.~(\ref{evenStotal}). For every coefficient
used below, we define
$q_i^{(0)}\equiv\left.q_i\right|_{L=0}$ with
$q_i\in\{a_i,c_i,e_i\}$. We then introduce
\be
\Pi_0\equiv a_1^{(0)}\delta\phi'
+\biggl[a_2^{(0)}-\left(a_1^{(0)}\right)'\biggr]\delta\phi
+a_3^{(0)}H_2.
\label{evenMonopolePi}
\ee

Using the exact coefficient identities collected in
Appendix~\ref{AppEvenCoefficients}, all terms containing $H_0$ or $H_1$
combine into
\be
H_0\Pi_0'-\frac{2}{f}H_1\dot\Pi_0.
\label{evenMonopoleConstraints}
\ee
Thus, $H_0$ and $H_1$ are Lagrange multipliers imposing
$\Pi_0'=0$ and $\dot\Pi_0=0$, respectively. On any connected region, these
constraints imply $\Pi_0=C_{\rm mon}$, where $C_{\rm mon}$ is a spacetime
constant fixed by the boundary conditions. Since it carries no local
dynamical degree of freedom, we set $C_{\rm mon}=0$ in the local stability
analysis. On the generic branch $a_3^{(0)}\ne0$, the constraint $\Pi_0=0$
then determines $H_2$ algebraically. Defining
\be
\lambda_0\equiv\frac{a_1^{(0)}}{a_3^{(0)}},\qquad
\nu_0\equiv
\frac{a_2^{(0)}-\left(a_1^{(0)}\right)'}{a_3^{(0)}},
\qquad
H_2=-\lambda_0\delta\phi'-\nu_0\delta\phi,
\label{evenMonopoleH2}
\ee
the monopole action reduces to an action for $\delta\phi$ alone.

Substituting Eq.~(\ref{evenMonopoleH2}) into the monopole action and integrating by parts with respect to $r$, we obtain the exact reduced action for the remaining scalar perturbation,
\be
S_0^{(2)}=\int {\rm d}t\,{\rm d}r
\left( {\cal K}_0 \dot{\delta\phi}^2
+{\cal G}_0 \delta\phi'^2
+{\cal U}_0 \delta\phi^2 \right),
\label{evenMonopoleAction}
\ee
where
\begin{align}
{\cal K}_0={}&e_1^{(0)}-c_1^{(0)}\nu_0
+\frac12(c_1^{(0)}\lambda_0)',\\
{\cal G}_0={}&e_2^{(0)}-c_2^{(0)}\lambda_0+c_6^{(0)}\lambda_0^2,\\
{\cal U}_0={}&e_3^{(0)}-c_3^{(0)}\nu_0+c_6^{(0)}\nu_0^2
-\frac12[-c_2^{(0)}\nu_0-c_3^{(0)}\lambda_0
+2c_6^{(0)}\lambda_0\nu_0]'.
\label{evenMonopoleCoefficients}
\end{align}
Thus, on the generic branch $a_3^{(0)}\ne0$, the monopole sector is
described by the single scalar perturbation $\delta\phi$. The kinetic and
gradient terms must be identified separately according to whether $t$ or $r$
is timelike. The coefficient ${\cal U}_0$ affects the low-frequency spectrum;
its profile and the boundary conditions must be analyzed separately to exclude
tachyonic or global mode instabilities.

\subsubsection{Region with timelike $t$: $f,h>0$}
\label{evenMonopoleExteriorSubsubsec}

For $f,h>0$, $t$ is timelike and $r$ is spacelike. Accordingly,
${\cal K}_0$ is the kinetic coefficient, whereas ${\cal G}_0$ controls the
radial gradient. The absence of ghosts and radial Laplacian instabilities requires
\be
{\cal K}_0>0,\qquad
c_{r,0}^{2(+)}=-\frac{{\cal G}_0}{fh{\cal K}_0}>0
\quad\Longleftrightarrow\quad {\cal G}_0<0.
\label{evenMonopoleExteriorStability}
\ee
No independent angular propagation condition arises in the monopole sector.

\subsubsection{Region with timelike $r$: $f,h<0$}
\label{evenMonopoleInteriorSubsubsec}

For $f,h<0$, $r$ is timelike and $t$ is spacelike. Hence,
${\cal G}_0$ is the kinetic coefficient for evolution 
along $r$, whereas ${\cal K}_0$ controls the gradient along the spacelike $t$ direction. 
The absence of ghosts and Laplacian instabilities requires
\be
{\cal G}_0>0,\qquad
c_{\parallel,0}^{2(-)}=-\frac{fh{\cal K}_0}{{\cal G}_0}>0
\quad\Longleftrightarrow\quad {\cal K}_0<0.
\label{evenMonopoleInteriorStability}
\ee
Here, $c_{\parallel,0}^{(-)}$ denotes the physical 
propagation speed along the spacelike $t$ direction. 
The monopole sector therefore propagates a single scalar 
mode in both regions whenever the corresponding kinetic and
gradient conditions are satisfied.

\subsection{Dipole: $\ell=1$}
\label{evenDipoleSubsec}

For $\ell=1$, the traceless even-parity tensor harmonic vanishes:
$\nabla_a\nabla_bY_1=-\gamma_{ab}Y_1$ is pure trace. Consequently, the
amplitudes $K$ and $G$ appear only through the combination $K-G$.  On a
background with
$\phi'\ne0$, the even-parity gauge freedom can be fixed completely, following
Ref.~\cite{Kase:2023mho}, by imposing
\be
h_0=0,\qquad K=G,\qquad \delta\phi=0.
\label{evenDipoleGauge}
\ee
In this gauge, the single local dynamical degree of freedom is represented by
the metric perturbation $\psi$.  A direct reduction of the dipole action
coincides with the result obtained by restricting the exact two-field reduced
action to $\delta\phi=0$ and subsequently setting $L=2$.  The resulting action
is
\be
S_1^{(2)}=\int {\rm d}t\,{\rm d}r
\left({\cal K}_1\dot\psi^{\,2}
+{\cal G}_1\psi'{}^2
+{\cal U}_1\psi^2\right),
\label{evenDipoleAction}
\ee
where
\be
{\cal K}_1\equiv\left.K_{11}\right|_{L=2},\qquad
{\cal G}_1\equiv\left.G_{11}\right|_{L=2},\qquad
{\cal U}_1\equiv\left.M_{11}\right|_{L=2}.
\label{evenDipoleCoefficients}
\ee
The antisymmetry of ${\bm Q}$ implies $Q_{11}=0$, 
so no one-field first-derivative mixing remains.  
The two-field determinant conditions obtained 
for $\ell\geq2$ do not apply to the dipole sector,
because the second perturbation is removed 
by gauge fixing before the physical kinetic 
and gradient coefficients are evaluated.
The coefficient ${\cal U}_1$ controls the remaining low-frequency potential
term and must be included, together with the boundary conditions, in any
global mode-stability analysis.

\subsubsection{Region with timelike $t$: $f,h>0$}
\label{evenDipoleExteriorSubsubsec}

For $f,h>0$, the absence of ghosts and radial Laplacian
instabilities requires
\be
{\cal K}_1>0,\qquad
c_{r,1}^{2(+)}
=-\frac{{\cal G}_1}{fh{\cal K}_1}>0
\quad\Longleftrightarrow\quad
{\cal G}_1<0.
\label{evenDipoleExteriorStability}
\ee
\subsubsection{Region with timelike $r$: $f,h<0$}
\label{evenDipoleInteriorSubsubsec}

For $f,h<0$, $r$ is timelike and therefore serves as the evolution
coordinate.  The absence of ghosts requires ${\cal G}_1>0$, while stability
against gradients along the spacelike $t$ direction requires
\be
{\cal G}_1>0,\qquad
c_{\parallel,1}^{2(-)}
=-\frac{fh{\cal K}_1}{{\cal G}_1}>0
\quad\Longleftrightarrow\quad
{\cal K}_1<0.
\label{evenDipoleInteriorStability}
\ee
Thus, the regular dipole sector contains one local propagating degree of freedom in both sign sectors.  If $\phi'=0$, the gauge choice~(\ref{evenDipoleGauge}) is not admissible. 
Similarly, if a denominator entering the generic constraint reduction vanishes, the reduced action above is no longer valid.  Such exceptional branches must be examined
directly using the original constraint equations.  
Since $\ell=1$ is fixed, the dipole sector has no independent angular-eikonal stability condition.

\section{Near-horizon instability}
\label{horizonInstabilitySec}

We examine the local perturbative stability of static hairy BHs in a
sufficiently small one-sided neighborhood just outside a simple outer
horizon in three settings.  First, we analyze an exact quadratic-GLPV BH
solution.  Second, we extend the analysis to general shift- and
reflection-symmetric quadratic GLPV theories.  Third, we consider the
regular Horndeski-related quartic--quintic branch satisfying the
compatibility condition and admitting a locally invertible disformal map.
In each case, instability means the violation of at least one local
no-ghost or high-frequency gradient-stability condition.  
The scope of each generic argument is stated explicitly, together with any
degenerate or fine-tuned branches that it does not cover.

\subsection{Exact quadratic-GLPV black hole}
\label{targetBHsubsec}

The exact BH solution constructed in Ref.~\cite{Bakopoulos:2022bho}
belongs to the class of shift- and reflection-symmetric quartic GLPV
theories.  It is characterized by a parameter $\eta_{\rm BH}$ and a positive length scale $\lambda_{\rm BH}>0$; the latter convention fixes the branch of the
arctangent in the asymptotic expansion. 
The subscript ``BH'' distinguishes these model parameters from the
angular multipole $\ell$ and the scalar kinetic normalization introduced in
Sec.~\ref{sGBsec}.

Throughout this subsection, we set the reduced Planck mass to unity, $\Mpl=1$. To adopt the standard Einstein--Hilbert normalization $G_4(0)=1/2$, we multiply the entire action of Ref.~\cite{Bakopoulos:2022bho}, where $G_4(0)=1$, by $1/2$. This overall rescaling leaves the background equations and their solutions unchanged. The resulting coupling functions are
\ba
G_2(X)&=&-\frac{4\eta_{\rm BH}}{\lambda_{\rm BH}^2}X^2,\qquad
G_4(X)=\frac12-2\eta_{\rm BH}\left(X+X^2\right),\notag\\
F_4(X)&=&\frac{\eta_{\rm BH}}{2}\left(\frac{1}{X}+3\right),
\qquad G_3=G_5=F_5=0\,.
\label{targetTheory}
\ea

Shift symmetry implies $\partial{\cal P}/\partial\phi=0$.  The scalar-field
equation~(\ref{EphiGLPV}) therefore admits the first integral
\be
r^2sJ^r=C_J,
\label{targetCurrentIntegral}
\ee
where $C_J$ is the integration constant associated with the scalar charge.
For the static configuration, the squared norm of the current is
\be
J_\mu J^\mu=g_{rr}(J^r)^2
=\frac{1}{h}\left(\frac{C_J}{r^2s}\right)^2
=\frac{C_J^2}{r^4f}.
\label{targetCurrentNorm}
\ee
Since $f\to0$ at a regular BH horizon, finiteness of the current norm in
Eq.~(\ref{targetCurrentNorm}) requires $C_J=0$.  It follows that
$J^r=0$ throughout the connected exterior region.  Thus, the
zero-scalar-charge branch is selected by horizon regularity rather than
imposed as an independent assumption.
Substituting Eq.~(\ref{targetTheory}) into the current
(\ref{JrGLPV}) gives
\be
J^r=\frac{4\eta_{\rm BH}h\phi'}{r^2\lambda_{\rm BH}^2}
\left[h\phi'^2(r^2+\lambda_{\rm BH}^2)
-\lambda_{\rm BH}^2\right] .
\label{targetCurrentExplicit}
\ee
On the nontrivial hairy branch with $\eta_{\rm BH}\phi'\ne0$,
$J^r=0$ therefore fixes the scalar profile algebraically as
\be
\phi'^2=\frac{\lambda_{\rm BH}^2}
{h(r^2+\lambda_{\rm BH}^2)}
=\frac{1}{h[1+(r/\lambda_{\rm BH})^2]},\qquad
X=-\frac{\lambda_{\rm BH}^2}
{2(r^2+\lambda_{\rm BH}^2)} .
\label{targetScalarProfile}
\ee

We now turn to the metric equations.  Solving Eqs.~(\ref{EGLPV00}) and
(\ref{EGLPV11}) for $h'$ and $f'$, respectively, and combining the results
gives
\be
\frac{f'}{f}-\frac{h'}{h}
=\frac{4\eta_{\rm BH}\phi'^2}{r\lambda_{\rm BH}^2}
\left[h\phi'^2(r^2+\lambda_{\rm BH}^2)
-\lambda_{\rm BH}^2\right]=0\,,
\label{targetMetricRatio}
\ee
where the last equality follows from Eq.~(\ref{targetScalarProfile}).
Thus, $f=C_fh$, where $C_f$ is constant.  Asymptotic flatness,
$f/h\to1$, fixes $C_f=1$, and hence
\be
f=h .
\label{targetFEqualsH}
\ee
Substituting Eqs.~(\ref{targetScalarProfile}) and
(\ref{targetFEqualsH}) into either metric equation gives the remaining
first-order equation
\be
(rh)'=1+\frac{\eta_{\rm BH}\lambda_{\rm BH}^2}
{r^2+\lambda_{\rm BH}^2}=1-2\eta_{\rm BH} X\,.
\label{targetMetricFirstOrder}
\ee
Integrating Eq.~(\ref{targetMetricFirstOrder}) gives
\be
rh=r+\eta_{\rm BH}\lambda_{\rm BH}
\arctan\!\left(\frac{r}{\lambda_{\rm BH}}\right)-2M,
\label{targetMetricIntegral}
\ee
where $M$ is an integration constant.  Together with
Eq.~(\ref{targetFEqualsH}), this gives
\be
f(r)=h(r)=1+\eta_{\rm BH}
\frac{\arctan(r/\lambda_{\rm BH})}
{r/\lambda_{\rm BH}}-\frac{2M}{r}\,.
\label{targetExactSolution}
\ee
At large $r$, Eq.~(\ref{targetExactSolution}) behaves as
$f=h=1-2M_{\rm ADM}/r+{\cal O}(r^{-2})$, where
$M_{\rm ADM}=M-\pi\eta_{\rm BH}\lambda_{\rm BH}/4$.

For later convenience, we introduce
\be
y\equiv\frac{r}{\lambda_{\rm BH}},\qquad
u\equiv1+y^2 .
\label{targetyuDefinitions}
\ee
Let $r_s$ denote the radius of a simple outer event horizon, at which
$f(r_s)=h(r_s)=0$.  We define
$f_{s,1}\equiv f'(r_s)$ and $h_{s,1}\equiv h'(r_s)$.  Since
$f=h$, these coefficients satisfy $f_{s,1}=h_{s,1}>0$.  We also define
$y_s\equiv y(r_s)$ and $u_s\equiv u(r_s)$.  Equation~(\ref{targetScalarProfile})
then gives $X=-1/(2u)$.  
Hence, the horizon value of the scalar kinetic term is
\be
X_s\equiv X(r_s)=-\frac{1}{2u_s}<0\,,
\label{targetXs}
\ee
which is finite and nonzero. 
Let $\delr\equiv r-r_s$ and
$\phi_s\equiv\phi(r_s)$.  Since $h$ vanishes linearly at a simple outer
horizon, the relation $X=-h\phi'^2/2$ with $X_s\ne0$ requires
$\phi'\propto\delr^{-1/2}$.  We therefore define $p_s$ by
$\phi'=p_s/\sqrt{\delr}+{\cal O}(\delr^{1/2})$.  The near-horizon expansion
can then be written as
\be
\begin{aligned}
f&=f_{s,1}\delr+{\cal O}(\delr^2),\qquad
h=h_{s,1}\delr+{\cal O}(\delr^2),\qquad
X=X_s+{\cal O}(\delr),\\
\phi&=\phi_s+2p_s\delr^{1/2}+{\cal O}(\delr^{3/2}),
\qquad p_s^2=-\frac{2X_s}{h_{s,1}}.
\end{aligned}
\label{targetHorizonExpansion}
\ee
The relation for $p_s^2$ follows directly from $X=-h\phi'^2/2$.  For the
exact solution considered here, $f=h$ implies $f_{s,1}=h_{s,1}$, while
$X_s=-1/(2u_s)$ gives $p_s^2=1/(h_{s,1}u_s)$.  Thus, although $\phi'$ diverges
as $\delr^{-1/2}$ in static coordinates, $X$ remains finite at the horizon,
and all the coupling functions in Eq.~(\ref{targetTheory}) are analytic in a
neighborhood of the nonzero value $X_s$.

We now examine the local perturbative stability of this solution near the
outer horizon.  For the coupling functions in Eq.~(\ref{targetTheory}), the
odd-parity coefficients reduce to
\be
\mG=\mH=1,\qquad
\mF=1+\eta_{\rm BH}\frac{1+2y^2}{u^2}\,.
\label{targetOddCoefficients}
\ee
Since $\mG=\mH=1$, the local high-frequency odd-parity stability conditions
in the exterior reduce to $\mF(r)>0$.  The associated squared radial and
angular propagation speeds are
\be
c_{r,{\rm odd}}^{2(+)}
=\left(1+\eta_{\rm BH}\frac{1+2y^2}{u^2}\right)^{-1},
\qquad
c_{\Omega,{\rm odd}}^{2(+)}=1\,.
\label{targetOddSpeeds}
\ee
We define the horizon value of $\mF$ as
$\mF_s\equiv\mF(r_s)=1+\eta_{\rm BH}(1+2y_s^2)/u_s^2$.
For $\mF_s\ne0$, continuity ensures that $\mF(r)$ has the same sign as
$\mF_s$ in a sufficiently small one-sided neighborhood exterior to the
outer horizon.  Hence, $\mF_s<0$ violates the odd-parity stability condition
there, whereas for $\mF_s>0$ the odd-parity sector satisfies all its local
high-frequency stability conditions in that neighborhood.  If $\mF_s=0$,
Eq.~(\ref{targetOddCoefficients}) fixes
$\eta_{\rm BH}=-u_s^2/(1+2y_s^2)<0$, and a near-horizon expansion gives
\ba
\mF &=&-\frac{4\eta_{\rm BH}y_s^3}
{\lambda_{\rm BH}u_s^3}\delr+{\cal O}(\delr^2),\\
c_{r,{\rm odd}}^{2(+)}
&=& -\frac{\lambda_{\rm BH}u_s^3}
{4\eta_{\rm BH}y_s^3}\frac{1}{\delr}+{\cal O}(1)
\longrightarrow+\infty\qquad(\delr\to0^+)\,.
\label{targetFzeroHorizonBehavior}
\ea
Thus, this branch does not admit a finite, nondegenerate radial tensor
characteristic cone at the horizon and is excluded as degenerate.  We
therefore restrict the remainder of this subsection to $\mF_s>0$.

The relevant even-parity quantities can also be evaluated exactly, without
invoking a near-horizon expansion.  In particular, we obtain the relations
\be
\mA\equiv\mH_\vartheta
=1+\eta_{\rm BH}\frac{1-2y^2}{u^2},
\qquad
G_{4,X}+2XG_{4,XX}+8XF_4+4X^2F_{4,X}=0\,.
\label{targetAidentity}
\ee
Substituting them into Eqs.~(\ref{evencoeffa}) and
(\ref{evenGLPVPDefinitions}) gives
\be
a_1=0,\qquad a_3=-\frac{rh}{2},\qquad a_4=\frac h2,\qquad
\mu_{\rm ev}=2r,\qquad
{\cal P}_{1}=\left(\frac{r\mA}{2}\right)' .
\label{targetEvenSimplifications}
\ee
The kinetic-determinant factorization in
Eq.~(\ref{evenKineticPFactorizations}) then identifies the second
even-parity no-ghost factor as
\be
\mK\equiv2{\cal P}_{1}-\mF
=-\frac{12\eta_{\rm BH} y^2}{u^3}.
\label{targetKexact}
\ee
Since $y^2/u^3>0$ throughout the exterior, Eq.~(\ref{targetKexact}) gives
$\mK<0$ for $\eta_{\rm BH}>0$.  Hence, the even-parity sector contains a
ghost on this branch.

It remains to analyze $\eta_{\rm BH}<0$, for which $\mK>0$.  Evaluating the
large-$L$ matrices in
Eqs.~(\ref{evenLargeLKcomponents0})--(\ref{evenLargeLKcomponents}) and
Eqs.~(\ref{evenLargeLMcomponents0})--(\ref{evenLargeLMcomponents}) for the
theory defined by Eq.~(\ref{targetTheory}), and combining the result with
Eq.~(\ref{evenKineticPFactorizations}), gives the exact determinant identity
\be
\mF\mK B_2^{(+)}
=-\frac{\mF\Xi_\Omega^2}{hr^2}\,{\cal Q}_\Omega\,.
\ee
Here, $B_2^{(+)}$ is normalized as in
Eq.~(\ref{evenAngularBdefinitions}), and the auxiliary quantities are
defined by
\be
\Xi_\Omega\equiv\frac{r\mA}{2},
\qquad
{\cal Q}_\Omega
\equiv r^4\left(\frac{{\cal P}_4}{r^3}\right)'
+2+\frac{[\mF{\cal P}_4-(rh'-2h)]^2}{4h\mF}\,,
\qquad
{\cal P}_4\equiv\frac{rh'+2h}{\mA}\,.
\label{targetAngularFactorization}
\ee
We first consider the branch with $\mA_s\equiv\mA(r_s)\ne0$, 
retaining the odd-parity condition $\mF_s>0$ imposed above. 
The first two terms in ${\cal Q}_\Omega$ remain
finite at $r=r_s$, whereas the quantity in square brackets in its last term
behaves as
\be
\mF{\cal P}_4-(rh'-2h)
=r_s h_{s,1}\left(\frac{\mF_s}{\mA_s}-1\right)
+{\cal O}(\delr).
\label{targetAngularSquareHorizon}
\ee
Moreover,
\be
\mF_s-\mA_s=\frac{4\eta_{\rm BH} y_s^2}{u_s^2}\,.
\label{targetFminusA}
\ee
Since $\eta_{\rm BH}<0$ and $y_s>0$, Eq.~(\ref{targetFminusA}) implies
$\mF_s-\mA_s\ne0$.  Together with $\mA_s\ne0$,
Eq.~(\ref{targetAngularSquareHorizon}) then shows that
$\mF{\cal P}_4-(rh'-2h)$ has a nonzero horizon limit.  Since
$h=h_{s,1}\delr+{\cal O}(\delr^2)$ and $\mF_s>0$, the last term produces
the leading $\delr^{-1}$ divergence of ${\cal Q}_\Omega$.
Combining these results with the determinant identity above gives
\be
\mF\mK B_2^{(+)}
=-\frac{r_s^2(\mF_s-\mA_s)^2}
{16\delr^2}
+{\cal O}(\delr^{-1})
=-\frac{\eta_{\rm BH}^2r_s^2y_s^4}
{u_s^4\delr^2}
+{\cal O}(\delr^{-1})<0\,.
\label{targetNegativeSquare}
\ee
On the branch under consideration, odd-parity stability gives $\mF>0$, 
while $\eta_{\rm BH}<0$ gives $\mK>0$.  Hence, $\mF\mK>0$ throughout 
a sufficiently small one-sided neighborhood exterior to the outer horizon.  
It then follows from Eq.~(\ref{targetNegativeSquare}) that $B_2^{(+)}<0$ 
there, signaling a local high-frequency angular Laplacian instability.

The branch $\mA_s=0$ also fails to restore stability.  For
$\eta_{\rm BH}>0$, the even-parity ghost identified above is already
present.  For $\eta_{\rm BH}<0$, the condition $\mA_s=0$ requires
$y_s^2<1/2$ and gives
$\mF_s=-4y_s^2/(1-2y_s^2)<0$, so the odd-parity stability condition is
violated.  Consequently, for every $\eta_{\rm BH}\ne0$, the solution either
violates a local no-ghost or high-frequency gradient-stability condition
arbitrarily close to a simple outer horizon, or belongs to the degenerate
$\mF_s=0$ branch, whose radial propagation speed diverges according to
Eq.~(\ref{targetFzeroHorizonBehavior}).

\subsection{Quadratic GLPV theories}
\label{generalGLPVNoGoSubsec}

The near-horizon obstruction found in Sec.~\ref{targetBHsubsec} is not
restricted to the particular coupling functions in Eq.~(\ref{targetTheory}).
We now extend the analysis to the general shift- and reflection-symmetric
quadratic GLPV class defined by
\be
G_2=G_2(X),\qquad G_4=G_4(X),\qquad F_4=F_4(X),
\qquad G_3=G_5=F_5=0,
\label{generalQuadraticGLPVClass}
\ee
which contains the theory in Eq.~(\ref{targetTheory}) as a special case.
For this class, the three background functions governing odd-parity stability reduce to
\be
 \mF=2G_4,\qquad
 \mG=\mH=2\left(G_4-2XG_{4,X}-4X^2F_4\right).
\label{oddquarticcheckpoint}
\ee

To analyze the even-parity sector, we further introduce the combinations
\begin{align}
\mA \equiv\mH+8X^2F_4=2(G_4-2XG_{4,X}),\qquad
{\cal T}\equiv G_{4,X}+2XG_{4,XX}+8XF_4+4X^2F_{4,X},
\qquad \mathfrak q\equiv-4X{\cal T}.
\label{generalQuadraticGLPVCombinations}
\end{align}
For this class, the coefficient relations collected in
Appendix~\ref{AppEvenCoefficients} yield the exact identities
\be
a_1=\frac{2fhr\phi'}s\,{\cal T},\qquad
a_3=-\frac{shr}{2}(\mH+\mathfrak q),\qquad
\mu_{\rm ev}=2r(\mH+\mathfrak q),\qquad
\mH_\vartheta=\mA,\qquad
{\cal P}_{1}
=\frac1s\left[
\frac{sr\mH\mA}{2(\mH+\mathfrak q)}\right]'.
\label{generalQuadraticGLPVCoefficientIdentities}
\ee
The large-$L$ angular determinant admits a corresponding exact factorization. To express it compactly, we define
\be
\Xi_\Omega \equiv
\frac{r\mH\mA}{2(\mH+\mathfrak q)},\qquad
{\cal Q}_\Omega \equiv
r^4s\left(\frac{{\cal P}_4}{sr^3}\right)'
+\frac{2f\mG}{h\mH^2}
+\frac{[\mF{\cal P}_4-(rf'-2f)]^2}{4f\mF},\qquad 
{\cal P}_4\equiv
\frac{(\mH+\mathfrak q)(rf'+2f)}{\mH\mA}\,,
\label{generalQuadraticGLPVAngularAuxiliaries}
\ee
which generalize the expressions in Eq.~(\ref{targetAngularFactorization}).

Although the individual components of the matrix $\widetilde{\bm M}$ given
in Eqs.~(\ref{evenLargeLMcomponents0})--(\ref{evenLargeLMcomponents}) are
lengthy, its determinant takes a remarkably compact form.  Substituting
Eq.~(\ref{generalQuadraticGLPVCoefficientIdentities}) into these components
and using the on-shell background identity in
Eq.~(\ref{evenPOnShellIdentity}), we obtain
\begin{equation}
\det\widetilde{\bm M}
=-\frac{h\mF(\mH+\mathfrak q)^2}
{4r^2\phi'^2}\,{\cal Q}_\Omega .
\end{equation}
Likewise, taking the large-$L$ limit of the kinetic-determinant factorization in Eq.~(\ref{evenKineticPFactorizations}) and using
$\mu_{\rm ev}=2r(\mH+\mathfrak q)$ and $\mH_\vartheta=\mA$, we find
\begin{equation}
\det\widetilde{\bm K}
=\frac{hr^2(\mH+\mathfrak q)^4}
{f\mH^2\phi'^2\mA^2}\,\mF\mK .
\label{generalQuadraticGLPVKineticDeterminant}
\end{equation}
Combining these determinant identities with the definition of
$B_2^{(+)}$ in Eq.~(\ref{evenAngularBdefinitions}) yields
\be
\mF\mK B_2^{(+)}
=-\frac{\mF\Xi_\Omega^2}{fr^2}\,{\cal Q}_\Omega .
\label{generalQuadraticGLPVAngularFactorization}
\ee
The compact expression for $\det\widetilde{\bm M}$ is the crucial algebraic result, as it isolates ${\cal Q}_\Omega$ as the single combination controlling the sign of the angular determinant before normalization by the kinetic matrix. All three identities are exact for arbitrary on-shell backgrounds within the class defined by Eq.~(\ref{generalQuadraticGLPVClass}); their derivation requires neither a near-horizon expansion nor any relation specific to the Horndeski limit.

We adopt the simple outer-horizon expansion introduced in
Eq.~(\ref{targetHorizonExpansion}), now with $X_s\ne0$ and without imposing
$f_{s,1}=h_{s,1}$.  We further assume that the functions entering
Eq.~(\ref{generalQuadraticGLPVCombinations}) are finite and analytic in a
neighborhood of $X_s$.  Here, a subscript $s$ denotes evaluation at $r_s$.
Since $\mG=\mH$ in this class, either $\mF_s<0$ or $\mH_s<0$ violates 
an odd-parity stability condition by continuity. At the boundary $\mG_s=0$ 
(and hence $\mH_s=0$), the odd-parity kinetic normalization vanishes, 
rendering the principal symbol degenerate. The remaining
boundary case $\mF_s=0$ with $\mH_s>0$ is also nonregular.  If $\mF>0$
immediately outside the horizon, Eq.~(\ref{oddgeneralspeeds}) gives
$c_{r,{\rm odd}}^{2(+)}=\mG/\mF\to+\infty$ as $r\to r_s^+$; if $\mF<0$
there, the odd-parity stability condition is violated.  Thus, $\mF_s=0$
does not admit a finite, nondegenerate radial tensor cone.  We therefore
work on the regular, nondegenerate branch characterized by
\be
\mF_s>0,\qquad \mH_s>0,\qquad
\mA_s\ne0,\qquad \mH_s+\mathfrak q_s\ne0,
\label{generalQuadraticGLPVRegularBranch}
\ee
On this branch, the first two terms defining ${\cal Q}_\Omega$ in Eq.~(\ref{generalQuadraticGLPVAngularAuxiliaries}) remain 
finite in the limit $r\to r_s^+$. 
To isolate the potentially singular contribution, we define
the horizon mismatch
\be
{\cal C}_s\equiv
\mF_s(\mH_s+\mathfrak q_s)-\mH_s\mA_s .
\label{generalQuadraticGLPVMismatch}
\ee
The leading near-horizon behavior is then
\begin{align}
\mF{\cal P}_4-(rf'-2f)
&=\frac{r_s f_{s,1}{\cal C}_s}{\mH_s\mA_s}
+{\cal O}(\delr),\notag\\
{\cal Q}_\Omega
&=\frac{r_s^2f_{s,1}{\cal C}_s^2}
{4\mF_s\mH_s^2\mA_s^2}\frac1{\delr}
+{\cal O}(1).
\label{generalQuadraticGLPVQExpansion}
\end{align}
Combining Eqs.~(\ref{generalQuadraticGLPVAngularFactorization}) and
(\ref{generalQuadraticGLPVQExpansion}) gives the universal negative-square behavior
\be
\mF\mK B_2^{(+)}
=-\frac{r_s^2{\cal C}_s^2}
{16(\mH_s+\mathfrak q_s)^2\delr^2}
+{\cal O}(\delr^{-1})<0
\qquad(\delr\to0^+,\ {\cal C}_s\ne0).
\label{generalQuadraticGLPVNegativeSquare}
\ee
On the branch defined by Eq.~(\ref{generalQuadraticGLPVRegularBranch}),
continuity gives $\mF>0$ and $\mH>0$ sufficiently close to the horizon.  If
$\mK<0$, the even-parity sector contains a ghost, while $\mK=0$ corresponds
to a degenerate kinetic matrix.  For $\mK>0$,
Eq.~(\ref{generalQuadraticGLPVNegativeSquare}) implies $B_2^{(+)}<0$
sufficiently close to the horizon, signaling a local high-frequency angular
Laplacian instability.  Thus, on the regular, nondegenerate branch, a static
BH with $X_s\ne0$ and generic ${\cal C}_s\ne0$ necessarily violates at least
one local no-ghost or high-frequency gradient-stability condition arbitrarily
close to its outer horizon.

As a consistency check of the normalization and overall sign, consider the Horndeski limit $F_4=0$. In this limit, 
$\mA=\mH$, and
\be
{\cal C}_s=-16X_s^2
\left(G_{4,X}^2+G_4G_{4,XX}\right)_s,\qquad
\mH_s+\mathfrak q_s
=2\left(G_4-4XG_{4,X}-4X^2G_{4,XX}\right)_s\,.
\ee
Substituting these expressions into Eq.~(\ref{generalQuadraticGLPVNegativeSquare}) yields
\be
\mF\mK B_2^{(+)}
=-\frac{4X_s^4
\left(G_{4,X}^2+G_4G_{4,XX}\right)_s^2}
{\left(G_4-4XG_{4,X}-4X^2G_{4,XX}\right)_s^2}
\frac{r_s^2}{\delr^2}+{\cal O}(\delr^{-1}).
\label{HorndeskiReflectionNegativeSquare}
\ee
This result agrees exactly with Eq.~(5.13) of Ref.~\cite{Minamitsuji:2022mlv}, where $B_2$ in that reference corresponds to $B_2^{(+)}$ in our notation.

The cases $\mF_s=0$ and $\mH_s=0$ have already been excluded as degenerate odd-parity branches. The qualifier \emph{generic} excludes the exceptional loci ${\cal C}_s=0$, $\mA_s=0$, and $\mH_s+\mathfrak q_s=0$. At ${\cal C}_s=0$, the leading term in Eq.~(\ref{generalQuadraticGLPVNegativeSquare}) vanishes, whereas either of the other two conditions renders the elimination of the nondynamical perturbations singular. These cases require a higher-order near-horizon expansion or an analysis of the unreduced equations and are not covered by the preceding result.
The factorization in Eq.~(\ref{generalQuadraticGLPVNegativeSquare}) applies only to the class defined by Eq.~(\ref{generalQuadraticGLPVClass}). Although Eq.~(\ref{evenKineticPFactorizations}) remains valid with explicit $\phi$ dependence and nonvanishing $G_3$, $G_5$, and $F_5$, it does not imply the angular factorization in Eq.~(\ref{generalQuadraticGLPVAngularFactorization}). 
We therefore analyze the general quartic--quintic case separately in the next subsection.

\subsection{Regular Horndeski-related quartic--quintic branch}
\label{fullGLPVNoGoSubsec}

We now allow all the functions $G_i$ $(i=2,\ldots,5)$, $F_4$, and $F_5$ to have general dependence on both $\phi$ and $X$. As shown in Sec.~\ref{evenInteriorStabilitySubsubsec}, away from accidental zeros of the background prefactors, requiring both angular modes to have finite, nonzero eikonal phase speeds selects the Horndeski-related $F_4$--$F_5$ compatibility condition~(\ref{GLPVcompatibility}). This condition aligns the quartic and quintic covariant degeneracy directions and permits a single local disformal transformation that removes both beyond-Horndeski interactions \cite{Crisostomi:2016tcp,BenAchour:2016fzp,Mironov:2022ffa}. 
We focus on the branch for which this transformation is regular and locally invertible in a one-sided neighborhood exterior to the outer horizon; this map underlies the near-horizon argument below.

We introduce the three combinations
\be
 {\cal U}\equiv G_4-2XG_{4,X}+XG_{5,\phi},
 \qquad 
 {\cal V}\equiv {\cal U}-4X^2F_4,
 \qquad
 {\cal Q}_5 \equiv G_{5,X}-12XF_5\,.
\label{fullGLPVHorizonCombinations}
\ee
Expanding Eqs.~(\ref{mH})--(\ref{oddGLPVCoefficients}) near the horizon gives
\be
 \mF=2(G_4-XG_{5,\phi})_s+{\cal O}(\delr^{1/2}),\qquad
 \mG=-\frac{h_{s,1}^2p_s^3}{2\sqrt{\delr}}
 ({\cal Q}_5)_s+{\cal O}(1),\qquad
 \mH=2{\cal V}_s+{\cal O}(\delr^{1/2})\,.
\label{fullGLPVOddHorizonExpansion}
\ee
Here and below, the subscript $s$ indicates that the corresponding quantity
is evaluated at the horizon $r=r_s$.

For $({\cal Q}_5)_s\ne0$, Eq.~(\ref{fullGLPVOddHorizonExpansion}) shows that $\mG$ diverges as $\delr^{-1/2}$. If $\mF_s=2(G_4-XG_{5,\phi})_s\ne0$, the common odd- and even-parity tensor radial speed squared, $\mG/\mF$, diverges at the same rate. If $\mF_s=0$, then $\mF={\cal O}(\delr^{1/2})$; the odd-parity principal symbol is degenerate at the horizon, and $|\mG/\mF|$ diverges at least as $\delr^{-1}$ wherever $\mF\ne0$ immediately outside it. Thus, necessary conditions for a finite, nondegenerate tensor cone are $({\cal Q}_5)_s=0$ and $\mF_s\ne0$.

If $({\cal Q}_5)_s=0$ but $F_{5,s}\ne0$, then
$G_{5,X,s}=12X_sF_{5,s}$.  Equation~(\ref{GLPVcompatibility}) consequently
gives ${\cal U}_s=4X_s^2F_{4,s}$, so ${\cal V}_s=0$ and, by
Eq.~(\ref{fullGLPVOddHorizonExpansion}), $\mH_s=0$.  This branch has a
degenerate odd-parity angular principal symbol.  The remaining candidate
branch on which a finite, nondegenerate tensor cone may be realized must
therefore satisfy the necessary conditions
\be
 F_{5,s}=G_{5,X,s}=0,\qquad
 \mF_s\ne0,\qquad {\cal V}_s\ne0.
\label{fullGLPVFiniteConeBranch}
\ee

To keep both the disformal map and the even-parity constraint reduction
regular, we further require
\be
{\cal U}_s\ne0,\qquad
\mu_{{\rm ev},s}\mH_{\vartheta,s}\ne0.
\label{fullGLPVRegularDisformalBranch}
\ee
Together with ${\cal V}_s\ne0$ in
Eq.~(\ref{fullGLPVFiniteConeBranch}), the first condition ensures that
${\cal U}_s/{\cal V}_s$, identified below with the field-space Jacobian,
is finite and nonzero. The second condition keeps the standard large-$L$
constraint reduction nondegenerate. As seen from
Eq.~(\ref{evenGLPVPDefinitions}), $\mu_{\rm ev}=0$ makes the generic
reduction singular, whereas $\mH_\vartheta=0$ removes the leading large-$L$
term from $\Theta_{\rm ev}$. These are regularity assumptions rather than
additional stability conditions.

By continuity, ${\cal U}$, ${\cal V}$, $\mu_{\rm ev}$, and
$\mH_\vartheta$ remain nonzero throughout a sufficiently small one-sided
neighborhood exterior to the horizon. In this neighborhood, we introduce
the pure disformal transformation
\be
\bar g_{\mu\nu}=g_{\mu\nu}
+D(\phi,X)\nabla_\mu\phi\nabla_\nu\phi .
\label{fullGLPVDisformalTransformation}
\ee
An overbar denotes a quantity in the transformed frame. Under this
transformation, the scalar kinetic term and the quartic beyond-Horndeski
function become \cite{Gleyzes:2014qga,Mironov:2022ffa}
\be
\bar X=\frac{X}{1-2DX},\qquad
\bar F_4(\phi,\bar X)=\frac{(1-2DX)^{5/2}}{{\cal J}}
\left(F_4-\frac12{\cal V}D_{,X}\right),
\label{fullGLPVQuarticDisformalRule}
\ee
where
\be
{\cal J}\equiv1+2X^2D_{,X}.
\label{Jacobian}
\ee
The quantity ${\cal J}$ is the determinant of the linearized map
$\delta g_{\mu\nu}\mapsto\delta\bar g_{\mu\nu}$ on the space of independent
symmetric metric components, evaluated at fixed scalar configuration.
Holding the scalar fixed in this field-space derivative does not amount to
imposing the perturbative gauge condition $\delta\phi=0$. The same factor
also enters the kinetic-variable Jacobian,
$\partial\bar X/\partial X
={\cal J}/(1-2DX)^2$.

In the neighborhood where ${\cal V}\ne0$, suppose that the two
invertibility factors $1-2DX$ and ${\cal J}$ are nonzero. The quartic
transformation law then gives
\be
\bar F_4=0
\quad\Longleftrightarrow\quad
D_{,X}=\frac{2F_4}{{\cal V}}.
\label{fullGLPVQuarticHorndeskiCondition}
\ee
On the same invertible patch, the quintic transformation law gives
\cite{Gleyzes:2014qga,Mironov:2022ffa}
\be
\bar F_5=0
\quad\Longleftrightarrow\quad
F_5-\frac{X{\cal Q}_5}{6}D_{,X}=0.
\label{fullGLPVQuinticHorndeskiCondition}
\ee
Using the definitions in Eq.~(\ref{fullGLPVHorizonCombinations}), the
compatibility condition~(\ref{GLPVcompatibility}) can be rewritten as
\be
X{\cal Q}_5F_4=3F_5{\cal V}.
\ee
For $X{\cal Q}_5\ne0$, this identity gives
\be
\frac{6F_5}{X{\cal Q}_5}
=\frac{2F_4}{{\cal V}}.
\ee
Hence, the value of $D_{,X}$ that sets $\bar F_4=0$ also satisfies
Eq.~(\ref{fullGLPVQuinticHorndeskiCondition}) and therefore sets
$\bar F_5=0$. At a simultaneous zero ${\cal Q}_5=F_5=0$, 
the finite value of $D_{,X}$ fixed by Eq.~(\ref{fullGLPVQuarticHorndeskiCondition}) 
satisfies Eq.~(\ref{fullGLPVQuinticHorndeskiCondition}) identically. Thus, the same
transformation removes both beyond-Horndeski interactions.
Substituting Eq.~(\ref{fullGLPVQuarticHorndeskiCondition}) into
Eq.~(\ref{Jacobian}) gives
\be
{\cal J}=1+2X^2D_{,X}=\frac{{\cal U}}{{\cal V}},
\label{fullGLPVDisformalJacobian}
\ee
independently of the $\phi$-dependent integration function in $D$.
Equations~(\ref{fullGLPVFiniteConeBranch}) and
(\ref{fullGLPVRegularDisformalBranch}), together with continuity, ensure
that ${\cal J}$ remains finite and nonzero throughout the chosen one-sided
exterior neighborhood.

Equation~(\ref{fullGLPVQuarticHorndeskiCondition}) determines the
$X$ dependence of $D$ but leaves an arbitrary integration function
$D_0(\phi)$. Since $\phi'\ne0$ throughout the exterior neighborhood,
$\phi$ is locally monotonic and the background trajectory can be written as
$X=X_{\rm bg}(\phi)$. A convenient form of the general local solution is
\be
D(\phi,X)=D_0(\phi)+\int_{X_{\rm bg}(\phi)}^X
\frac{2F_4(\phi,Y)}{{\cal V}(\phi,Y)}\,{\rm d}Y.
\label{fullGLPVDisformalIntegral}
\ee
Before fixing $D_0(\phi)$, we linearize the transformation for arbitrary
$D$. On the static, spherically symmetric background $\phi=\phi(r)$, where
$\nabla_\mu\phi=\phi'\delta_\mu^{\,r}$ and $g^{rr}=h$, the variation of
$X=-g^{\mu\nu}\nabla_\mu\phi\nabla_\nu\phi/2$ is
\be
\delta X
=\frac12\nabla^\mu\phi\nabla^\nu\phi\,\delta g_{\mu\nu}
-\nabla^\mu\phi\nabla_\mu\delta\phi
=-Xh\,\delta g_{rr}-h\phi'\delta\phi'.
\label{fullGLPVDeltaX}
\ee
The linearized disformal transformation is
\begin{equation}
\delta\bar g_{\mu\nu}
=\delta g_{\mu\nu}
+\delta D\,\nabla_\mu\phi\nabla_\nu\phi
+2D\nabla_{(\mu}\phi\nabla_{\nu)}\delta\phi,
\qquad
\delta D=D_{,\phi}\delta\phi+D_{,X}\delta X.
\label{fullGLPVGeneralLinearDisformalMap}
\end{equation}
The last term introduces additional derivative mixing whenever the
background value of $D$ is nonzero.

For the present local argument, we use the freedom in $D_0(\phi)$ to choose
$D_0(\phi)=0$, which gives
\be
D\bigl(\phi,X_{\rm bg}(\phi)\bigr)=0.
\label{fullGLPVDisformalBackgroundChoice}
\ee
This normalization changes neither $D_{,X}$ nor the conditions
$\bar F_4=\bar F_5=0$. On the background, it implies $\bar g_{\mu\nu}
=g_{\mu\nu}$ and $1-2DX=1$, so the two frames have the same areal radius
and horizon data. Together with ${\cal J}\ne0$, these relations make the
transformation locally invertible. On the background, this choice also eliminates 
the final term proportional
to $D$ in Eq.~(\ref{fullGLPVGeneralLinearDisformalMap}).

We next establish the property of this map needed for the instability
argument: it preserves the sign of $\mF\mK B_2^{(+)}$.  The even-parity
action was derived in the
uniform-curvature gauge~(\ref{evenuniformcurvaturegauge}), which already
fixes all three scalar gauge functions.  A pure disformal transformation
acts only on the metric; the scalar is unchanged, so $\bar\phi=\phi$ and
$\delta\bar\phi=\delta\phi$.  With the background normalization
(\ref{fullGLPVDisformalBackgroundChoice}), substitution of
Eq.~(\ref{fullGLPVDeltaX}) into 
Eq.~(\ref{fullGLPVGeneralLinearDisformalMap}) yields
\be
\begin{aligned}
 \delta\bar g_{rr}={}&{\cal J}\,\delta g_{rr}
 -h\phi'^3D_{,X}\delta\phi'
 +\phi'^2D_{,\phi}\delta\phi,\\
 \delta\bar g_{\mu\nu}={}&\delta g_{\mu\nu}\quad
 [\,(\mu,\nu)\ne(r,r)\,].
\end{aligned}
\label{fullGLPVLinearDisformalMap}
\ee
Here, the derivatives of $D$ are evaluated on the background.  In particular,
$h_0$, $K$, and $G$ are unchanged, so the uniform-curvature gauge is preserved
in the transformed frame.  
To extract the leading large-$L$ angular principal symbol, we freeze the
background coefficients locally and set the radial wave number to zero,
$k_r=0$. In the corresponding Fourier representation,
$\delta\phi'\mapsto {\rm i}k_r\delta\phi$, so the term proportional to
$\delta\phi'$ does not contribute at principal order. The induced algebraic
map from $(\delta g_{rr},\delta\phi)$ to
$(\delta\bar g_{rr},\delta\bar\phi)$ is then triangular, with determinant
${\cal J}$. Since ${\cal J}\ne0$ throughout the chosen neighborhood, this
map is nonsingular, as required for the congruence relation below.

Let $\widetilde{\bm M}_{\rm H}$ denote the reduced angular principal matrix
in the transformed Horndeski frame.  The reduced dynamical pairs obtained
after constraint elimination in the GLPV and Horndeski frames are
$\vec{\cal X}=(\psi,\delta\phi)^{\rm T}$ and
$\vec{\cal X}_{\rm H}=(\psi_{\rm H},\delta\phi)^{\rm T}$, respectively,
where $\psi_{\rm H}$ is defined by the barred-frame analogue of
Eq.~(\ref{evenpsidefinition}).  These reduced pairs should not be confused
with the pre-reduction pair $(\delta g_{rr},\delta\phi)$ appearing above.
The nonsingular angular field redefinition established above, together with
the regularity of both constraint reductions, induces a relation
$\vec{\cal X}_{\rm H}={\bm R}_\Omega\vec{\cal X}$, where
${\bm R}_\Omega$ is real, finite, and nonsingular.  Substituting this
relation into the Horndeski-frame quadratic form
$\vec{\cal X}_{\rm H}^{\rm T}\widetilde{\bm M}_{\rm H}
\vec{\cal X}_{\rm H}$ and matching it to
$\vec{\cal X}^{\rm T}\widetilde{\bm M}\vec{\cal X}$ gives
\be
 \widetilde{\bm M}
 ={\bm R}_\Omega^{\rm T}\widetilde{\bm M}_{\rm H}
 {\bm R}_\Omega,\qquad
 \det\widetilde{\bm M}
 =(\det{\bm R}_\Omega)^2\det\widetilde{\bm M}_{\rm H},\qquad
 \operatorname{sgn}\det\widetilde{\bm M}
 =\operatorname{sgn}\det\widetilde{\bm M}_{\rm H}.
\label{fullGLPVAngularCongruence}
\ee
The large-$L$ limit of Eq.~(\ref{evenKineticPFactorizations}) 
gives the exact identity
\be
 \det\widetilde{\bm K}={\cal W}_K\mF\mK,
 \qquad
 {\cal W}_K\equiv
 \frac{h\mu_{\rm ev}^4}
 {16f\mH^2\phi'^2r^2\mH_\vartheta^2}>0 .
\label{fullGLPVLargeLKineticDeterminant}
\ee
The strict positivity follows in the one-sided exterior neighborhood
from $f,h>0$, $\phi'^2>0$, and the nonvanishing factors specified in
Eqs.~(\ref{fullGLPVFiniteConeBranch}) and
(\ref{fullGLPVRegularDisformalBranch}).  The identity (\ref{fullGLPVLargeLKineticDeterminant}) 
holds for general quartic--quintic GLPV theories on the regular large-$L$ branch and does not
rely on the compatibility condition~(\ref{GLPVcompatibility}).
In the quadratic subclass of Sec.~\ref{generalGLPVNoGoSubsec}, the relations
$\mu_{\rm ev}=2r(\mH+\mathfrak q)$ and $\mH_\vartheta=\mA$ reduce
Eq.~(\ref{fullGLPVLargeLKineticDeterminant}) exactly to
Eq.~(\ref{generalQuadraticGLPVKineticDeterminant}).
The same identity, with bars on all theory-dependent quantities, holds in the
Horndeski frame and has $\bar{\cal W}_K>0$.  Combining these identities with
Eq.~(\ref{evenAngularBdefinitions}) gives
\be
 \mF\mK B_2^{(+)}
 =\frac{r^4}{f^2{\cal W}_K}
 \det\widetilde{\bm M},\qquad
 \bar{\mF}\bar{\mK}\bar B_2^{(+)}
 =\frac{r^4}{f^2\bar{\cal W}_K}
 \det\widetilde{\bm M}_{\rm H}.
\label{fullGLPVProductMassDeterminant}
\ee
Since both normalization factors are positive,
Eq.~(\ref{fullGLPVAngularCongruence}) immediately yields
\be
 \operatorname{sgn}\!\left(\mF\mK B_2^{(+)}\right)
 =\operatorname{sgn}\det\widetilde{\bm M}
 =\operatorname{sgn}\det\widetilde{\bm M}_{\rm H}
 =\operatorname{sgn}\!\left(
 \bar{\mF}\bar{\mK}\bar B_2^{(+)}\right).
\label{fullGLPVProductSignInvariance}
\ee

We now apply the corresponding near-horizon results in the Horndeski frame.
The background normalization~(\ref{fullGLPVDisformalBackgroundChoice})
implies $\bar X_s=X_s$ and leaves $f_{s,1}$, $h_{s,1}$, and $r_s$
unchanged. All barred couplings and their derivatives appearing below are
therefore evaluated at
$(\phi_s,\bar X_s)=(\phi_s,X_s)$. The regular quintic transformation,
together with Eq.~(\ref{fullGLPVFiniteConeBranch}), also gives
$\bar G_{5,\bar X,s}=0$, as required for a finite tensor cone in the
Horndeski frame.

The radial characteristic associated with the even-parity scalar degree of
freedom imposes an additional independent condition. On a regular,
nondegenerate horizon branch with $\bar\zeta_r\ne0$,
Refs.~\cite{Minamitsuji:2022vbi,DeFelice:2026rbi} give
\begin{align}
\bar c_{r,{\rm sc}}^{2(+)}={}&
\frac{2h_{s,1}X_s\bar\kappa_r}
{\bar\zeta_r\delr}+{\cal O}(1),\\
\bar\kappa_r\equiv{}&
X_sr_s^2\left(2X_s\bar G_{3,\bar X\bar X}
-\bar G_{3,\bar X}\right)
+r_s^2\left(3\bar G_{4,\phi}
-4X_s^2\bar G_{4,\phi\bar X\bar X}\right)
+2X_s^2\bar G_{5,\bar X\bar X}.
\end{align}
Here, $\bar\zeta_r$ is the denominator coefficient in the scalar radial
characteristic. For finite nonzero $\bar\zeta_r$, finiteness of
$\bar c_{r,{\rm sc}}^{2(+)}$ at the horizon requires
\be
\bar\kappa_r=0.
\ee
If instead $\bar\kappa_r\ne0$, the speed diverges as $\delr^{-1}$.
The requirement $\bar\kappa_r=0$ is not implied by
Eqs.~(\ref{fullGLPVFiniteConeBranch}) and
(\ref{fullGLPVRegularDisformalBranch}), which instead ensure regularity of
the tensor cone and disformal map and nondegeneracy of the constraint
reduction. When $\bar\zeta_r=0$, the dominant near-horizon balance changes,
and the scalar radial characteristic must be analyzed separately.

After imposing $\bar\kappa_r=0$, the near-horizon angular result in
Eq.~(3.23) of Ref.~\cite{Minamitsuji:2022vbi} gives, for
$\bar\kappa_{\rm H}\bar\zeta_{\rm H}\ne0$,
\be
\bar{\mF}\bar{\mK}\bar B_2^{(+)}
=-\frac{4h_{s,1}^2X_s^4r_s^4\bar\kappa_{\rm H}^2}
{\bar\zeta_{\rm H}^2\delr^2}
+{\cal O}(\delr^{-1})<0
\qquad (\delr\to0^+),
\label{fullGLPVHorndeskiNegativeSquare}
\ee
where
\begin{align}
\bar\kappa_{\rm H}\equiv{}&
\bar G_4\bar G_{4,\bar X\bar X}+\bar G_{4,\bar X}^2
-\bar G_{5,\phi\bar X}
(\bar G_4-X_s\bar G_{4,\bar X})
-\bar G_{5,\phi}
(2\bar G_{4,\bar X}+X_s\bar G_{4,\bar X\bar X}
-\bar G_{5,\phi}),
\label{fullGLPVBarredKappa}\\
\bar\zeta_{\rm H}\equiv{}&
2\bar G_{3,\phi\bar X}X_s^2r_s^2
+r_s\bigl(
\bar G_4h_{s,1}
-2\bar G_{4,\phi\phi}X_sr_s
-4\bar G_{4,\bar X}X_sh_{s,1}
-4\bar G_{4,\bar X\bar X}X_s^2h_{s,1}
-4\bar G_{4,\phi\phi\bar X}X_s^2r_s
\bigr)
\notag\\
&+X_s\bigl[
3\bar G_{5,\phi}h_{s,1}r_s
+2\bar G_{5,\phi\bar X}X_s(1+h_{s,1}r_s)
\bigr].
\label{fullGLPVBarredZeta}
\end{align}
Equation~(\ref{fullGLPVProductSignInvariance}) shows that the negative sign
in Eq.~(\ref{fullGLPVHorndeskiNegativeSquare}) is invariant under the
disformal transformation and hence also holds in the GLPV frame. 
Therefore, there exists $\varepsilon_s>0$ such that
\be
0<r-r_s<\varepsilon_s
\quad\Longrightarrow\quad
\mF\mK B_2^{(+)}<0.
\label{fullGLPVStrictNegativeProduct}
\ee
On the odd-parity-stable branch, $\mF>0$. If $\mK>0$ also holds,
Eq.~(\ref{fullGLPVStrictNegativeProduct}) implies $B_2^{(+)}<0$, signaling
a local high-frequency even-parity angular Laplacian instability. Otherwise,
the odd-parity sector is unstable or degenerate, or the even-parity sector
contains a ghost for $\mK<0$ or has a degenerate kinetic matrix for
$\mK=0$.

We have therefore established a generic local near-horizon obstruction in
the regular Horndeski-related subclass. If $({\cal Q}_5)_s\ne0$, the tensor
cone is singular, whereas $\mF_s=0$ or $\mH_s=0$ makes the odd-parity
principal symbol degenerate. On the remaining regular branch,
$\bar\kappa_r\ne0$ yields a divergent scalar radial propagation speed. 
For $\bar\kappa_r=0$ and
$\bar\kappa_{\rm H}\bar\zeta_{\rm H}\ne0$,
Eq.~(\ref{fullGLPVStrictNegativeProduct}) implies that at least one local
no-ghost or high-frequency gradient-stability condition must be violated.

This argument does not cover the exceptional loci ${\cal U}_s=0$,
$\mu_{{\rm ev},s}\mH_{\vartheta,s}=0$, $\bar\zeta_r=0$, or
$\bar\kappa_r=0$ with
$\bar\kappa_{\rm H}\bar\zeta_{\rm H}=0$. Each requires a separate
dominant-balance analysis, and no conclusion about its stability follows
from the present result.

\section{Beyond-Horndeski deformations of scalar--Gauss--Bonnet black holes}
\label{sGBsec}

Finally, we investigate the local high-frequency stability of sGB BHs with
beyond-Horndeski deformations. Unlike the backgrounds considered in
Sec.~\ref{horizonInstabilitySec}, these solutions satisfy $X_s=0$ and
therefore lie outside the assumptions of the near-horizon obstruction
derived there. Starting from the perturbatively controlled sGB background
of Refs.~\cite{Sotiriou:2014pfa,Minamitsuji:2022vbi}, we introduce a beyond-Horndeski
deformation satisfying the compatibility
condition~(\ref{GLPVcompatibility}). For the sGB Horndeski functions, this
condition uniquely fixes $F_5$ in terms of $F_4$. We then construct the
weak-coupling background expansion, derive sufficient conditions for local
high-frequency stability throughout the exterior, and identify the scale
at which the interior expansion ceases to be perturbatively controlled.

\subsection{Perturbative backgrounds}
\label{GBbackgroundsubsec}

The perturbative construction below describes the asymptotically flat solution
outside the outer horizon and its horizon-regular continuation into the region
with $f,h<0$, where $r$ is timelike.  Since the background equations require only
$f/h>0$, they apply in both sign sectors wherever the small-$\alpha$
expansion remains under perturbative control.  No statement is made about
the deeper interior where this expansion breaks down. 
In Sec.~\ref{GBinteriorcriticalsubsec}, we examine this continuation and
estimate the associated effective-field-theory (EFT) matching scale at which
perturbative control is lost.

We consider the class of theories described by the action
\be
 S=\int {\rm d}^4x\sqrt{-g}\left[
 \frac{\Mpl^2}{2}R+\eta X+\alpha\xi(\phi)R_{\rm GB}^2
 +{\cal L}_4^{\rm bH}+{\cal L}_5^{\rm bH}\right]\,,
\label{GBactionMain}
\ee
where ${\cal L}_4^{\rm bH}$ and ${\cal L}_5^{\rm bH}$ are defined in
Eq.~(\ref{bHLagrangians}).  We assume $\eta>0$.  The dimensionless parameter
$\alpha$, satisfying $|\alpha|\ll1$, is introduced as a bookkeeping
parameter for the perturbative expansion.  
The GB invariant is defined by
\be
 R_{\rm GB}^2\equiv R^2-4R_{\mu\nu}R^{\mu\nu}
 +R_{\mu\nu\rho\sigma}R^{\mu\nu\rho\sigma}\,,
\ee
where $R$, $R_{\mu\nu}$, and $R_{\mu\nu\rho\sigma}$
denote the Ricci scalar, Ricci tensor, and Riemann tensor,
respectively.  We define the asymptotic scalar value by
$\phi_0\equiv\lim_{r\to\infty}\phi(r)$, assume that $\xi(\phi)$ is analytic
in a neighborhood of $\phi_0$, and introduce
\be
 \xi^{(j)}_0\equiv
 \left.\frac{{\rm d}^{j}\xi}{{\rm d}\phi^{j}}\right|_{\phi=\phi_0}\,,
 \qquad
 \xi_1\equiv\xi^{(1)}_0\ne0\,.
\label{GBxiDerivatives}
\ee
Here $j$ is a nonnegative integer.  At a general field value, we use
$\xi^{(j)}(\phi)\equiv{\rm d}^{j}\xi/{\rm d}\phi^{j}$ and
$\xi_{,\phi}\equiv\xi^{(1)}(\phi)$.  The condition
$\xi_1\ne0$ selects the weakly hairy branch whose scalar profile is
sourced at first order in $\alpha$.  When $\xi_1=0$, the scalar field
is not sourced at this order; scalarized BH solutions, if they exist, belong
to a distinct nonperturbative branch and are not captured by the expansion
considered here.

Up to total derivatives, the sGB interaction admits an equivalent Horndeski
representation in terms of the following functions
\cite{Kobayashi:2011nu}:
\ba
G_2^{\rm GB}&=&
8\alpha\xi^{(4)}(\phi)X^2(3-\ln|X|),
\qquad
G_3^{\rm GB}=
4\alpha\xi^{(3)}(\phi)X(7-3\ln|X|),
\notag\\
G_4^{\rm GB}&=&
4\alpha\xi^{(2)}(\phi)X(2-\ln|X|),
\qquad
G_5^{\rm GB}=
-4\alpha\xi^{(1)}(\phi)\ln|X| \,.
\label{GBHorndeskiEmbedding}
\ea
After including the canonical scalar kinetic term and the Einstein--Hilbert
term, the Horndeski-sector functions used below are
$G_2=\eta X+G_2^{\rm GB}$, $G_3=G_3^{\rm GB}$,
$G_4=\Mpl^2/2+G_4^{\rm GB}$, and $G_5=G_5^{\rm GB}$.
Here and below, a fixed reference scale $X_\star$ is implicit in $\ln|X|$;
the apparent nonanalyticity at $X=0$ cancels from the sGB background equations
and does not obstruct the small-$\alpha$ expansion
\cite{Minamitsuji:2022vbi}.
For these functions, the two Horndeski combinations entering the
compatibility condition~(\ref{GLPVcompatibility}) reduce identically to
\be
 G_4-2XG_{4,X}+XG_{5,\phi}
 =\frac{\Mpl^2}{2},
 \qquad
 XG_{5,X}=-4\alpha\xi^{(1)}(\phi).
\label{GBcompatibilityCombinations}
\ee
Imposing the compatibility condition~(\ref{GLPVcompatibility}) on the full
theory in Eq.~(\ref{GBactionMain}) and using the identities in
Eq.~(\ref{GBcompatibilityCombinations}) uniquely determines the quintic
beyond-Horndeski function in terms of the quartic one:
\be
 F_5(\phi,X)
 =-\frac{8\alpha\xi^{(1)}(\phi)}{3\Mpl^2}F_4(\phi,X).
\label{GBcompatibleF5}
\ee
Although $G_{5,X}\propto X^{-1}$ in the Horndeski representation, the
combination $XG_{5,X}$ entering the compatibility condition and the
resulting expression for $F_5$ in Eq.~(\ref{GBcompatibleF5}) remain finite
as $X\to0$ for the regular $F_4$ considered below.

To make the weak-coupling orders explicit, we parameterize the
beyond-Horndeski functions as
\be
F_4=\alpha{\cal A}_{\rm bH}(\phi,X),\qquad
F_5=-\frac{8\alpha^2\xi^{(1)}(\phi)}{3\Mpl^2}
{\cal A}_{\rm bH}(\phi,X)\,,
\label{GBbeyondHorndeskiInteractions}
\ee
and adopt the simple ansatz
\be
{\cal A}_{\rm bH}(\phi,X)=\mu_\phi\phi^{m_\phi}+\mu_X X^n,
\ee
where $m_\phi$ and $n$ are positive integers. The constants $\mu_\phi$ and
$\mu_X$ have the dimensions required for ${\cal A}_{\rm bH}$ to have the
same dimension as $F_4$ and are held fixed as $\alpha\to0$. The explicit
prefactors in $F_4$ and $F_5$ are therefore $\alpha$ and $\alpha^2$,
respectively; their background orders also depend on the component of
${\cal A}_{\rm bH}$.
We impose Eq.~(\ref{GBcompatibleF5}) as a functional identity in
$(\phi,X)$, rather than only along the BH background. For nonlinear
$\xi(\phi)$, the additional factor $\xi^{(1)}(\phi)$ implies that $F_5$
does not generally retain the two-term monomial form of
${\cal A}_{\rm bH}$. Moreover, wherever $\xi^{(1)}(\phi)\ne0$, a
nonvanishing quartic deformation necessarily entails a nonvanishing
quintic partner.

Let $M$ denote the Schwarzschild mass parameter. The zeroth-order geometry
has a simple outer event horizon at $r_s\equiv2M$. 
Introducing the dimensionless radius $\rho$ and the Schwarzschild factor
$q_s(\rho)$ as
\be
\rho\equiv\frac{r}{M},\qquad
q_s(\rho)\equiv1-\frac{2}{\rho}=1-\frac{r_s}{r}\,,
\ee
we write the weak-coupling expansion in the horizon-preserving form
\be
f=q_s\biggl[1+\sum_{N\geq1}\alpha^N\widehat f_N(r)\biggr]^2,
\qquad
h=q_s\biggl[1+\sum_{N\geq1}\alpha^N\widehat h_N(r)\biggr]^{-2},
\qquad
\phi=\phi_0+\sum_{N\geq1}\alpha^N\widehat\phi_N(r).
\label{GBbackgroundExpansion}
\ee
Provided that the quantities in brackets remain regular and nonzero at
$\rho=2$, the explicit factors of $q_s$ preserve both the location and the
simple character of the horizon order by order. We take $\phi_0$ to be the
full asymptotic scalar value and absorb all order-by-order constant shifts
into it, so that $\widehat\phi_N(\infty)=0$.
At each perturbative order, horizon regularity removes horizon-singular
homogeneous modes, while asymptotic flatness, time normalization, and the
scalar convention above remove the inadmissible asymptotic modes. Once $M$ and
$\phi_0$ are specified, these conditions fix the remaining integration constants.
At first order, we obtain
\be
 \widehat f_1=\widehat h_1=0,
 \qquad
 \widehat\phi_1=
 \frac{2(3\rho^2+3\rho+4)\xi_1}
 {3\eta M^2\rho^3}.
\label{GBfirstOrderBackground}
\ee
From the asymptotic form
$\phi=\phi_0+Q_\phi/r+{\cal O}(r^{-2})$,
Eq.~(\ref{GBfirstOrderBackground}) gives
$Q_\phi=2\alpha\xi_1/(\eta M)+{\cal O}(\alpha^2)$.
Horizon regularity fixes $Q_\phi$ in terms of the mass and coupling
parameters, rather than leaving it as an independent integration constant.
This branch therefore carries secondary scalar hair.
Since $\widehat f_1=\widehat h_1=0$, the metric remains Schwarzschild through
${\cal O}(\alpha)$, and its leading backreaction arises at
${\cal O}(\alpha^2)$. The beyond-Horndeski interactions enter only at higher
orders, as shown by the power counting below. Hence, the second-order
coefficients coincide with those of the pure sGB theory:
\begin{align}
\widehat h_{2{\rm GB}}={}&
 \frac{\xi_1^2}{120\eta M^4\Mpl^2\rho^6}
 \left(147\rho^5+174\rho^4+228\rho^3-1624\rho^2
 -3488\rho-7360\right),
\label{GBh2Background}\\
\widehat f_{2{\rm GB}}={}&-
 \frac{\xi_1^2}{120\eta M^4\Mpl^2\rho^6}
 \left(147\rho^5+294\rho^4+548\rho^3+56\rho^2
 -416\rho-1600\right),
\label{GBf2Background}\\
\widehat\phi_{2{\rm GB}}={}&
\frac{\xi_1\xi^{(2)}_0}{450\eta^2M^4\rho^6}
\left[1095(\rho^5+\rho^4)+1460\rho^3+2190\rho^2
+1344\rho+800\right].
\label{GBphi2Background}
\end{align}
For a strictly linear GB coupling, $\xi^{(2)}_0=0$, and hence
$\widehat\phi_{2{\rm GB}}=0$ with the above asymptotic convention.

Two ingredients determine the perturbative order at which the
beyond-Horndeski interactions modify the background: the powers of the
scalar gradient in the interaction vertices and the explicit weak-coupling
scaling of $F_4$ and $F_5$. On the sGB branch,
\be
\phi'={\cal O}(\alpha),\qquad X={\cal O}(\alpha^2)\,.
\ee
Counting only the explicit scalar-gradient factors at this stage, we obtain
\be
U_4={\cal O}(F_4\alpha^4),\qquad
U_5={\cal O}(F_5\alpha^5),\qquad
U_{4,\phi'}={\cal O}(F_4\alpha^3),\qquad
U_{5,\phi'}={\cal O}(F_5\alpha^4)\,.
\label{GBbeyondHorndeskiPowerCounting}
\ee
The metric building blocks in Eq.~(\ref{GLPVMetricBuildingBlocks}) constructed
from $U_4$ and $U_5$ inherit the first two orders in
Eq.~(\ref{GBbeyondHorndeskiPowerCounting}), whereas $U_{4,\phi'}$ and
$U_{5,\phi'}$ enter the leading beyond-Horndeski contribution to the radial
scalar current.
Equation~(\ref{GBbeyondHorndeskiInteractions}) supplies the additional
weak-coupling factors
$F_4={\cal O}(\alpha{\cal A}_{\rm bH})$ and
$F_5={\cal O}(\alpha^2{\cal A}_{\rm bH})$, where
$\xi^{(1)}(\phi)={\cal O}(1)$ on the branch considered here. Combining these
factors with the component-dependent scaling of ${\cal A}_{\rm bH}$ yields
the onset orders summarized below.

We denote by $F_5[F_4]$ the linked quintic beyond-Horndeski function fixed
by Eq.~(\ref{GBcompatibleF5}). For $\phi_0\ne0$, the following table lists
the lowest power of $\alpha$ at which each component of
${\cal A}_{\rm bH}$ produces a direct quartic or linked-quintic source for
the metric and scalar background equations. Here
$(f_{\rm bg},h_{\rm bg})$ and $\phi_{\rm bg}$ denote the metric and scalar
background functions, respectively. Since the leading differential
operators are of order $\alpha^0$, each background correction generically
begins at the same order as its corresponding source:
\be
\begin{array}{|c|cc|cc|}
\hline
&\multicolumn{2}{c|}{\text{direct }F_4\text{ source}}
&\multicolumn{2}{c|}{\text{linked }F_5[F_4]\text{ source}}\\ \hline
{\cal A}_{\rm bH}\text{ component}
&(f_{\rm bg},h_{\rm bg})&\phi_{\rm bg}
&(f_{\rm bg},h_{\rm bg})&\phi_{\rm bg}\\ \hline
\mu_\phi\phi^{m_\phi}
&\alpha^5&\alpha^4&\alpha^7&\alpha^6\\ \hline
\mu_X X^n
&\alpha^{2n+5}&\alpha^{2n+4}
&\alpha^{2n+7}&\alpha^{2n+6}\\ \hline
\end{array}
\label{GBbeyondHorndeskiOnset}
\ee
The linked-quintic columns display only the contributions directly sourced
by $F_5[F_4]$. At these orders, the deformation-dependent coefficients can
also contain mixed recursive contributions involving lower-order pure-sGB
fields and quartic corrections generated earlier.

We now derive the leading scalar-background correction generated by the
compatible quartic--quintic deformation and thereby verify the direct-$F_4$
scalar onset orders in Eq.~(\ref{GBbeyondHorndeskiOnset}). At zeroth order in
$\alpha$, the metric functions satisfy $f=h=q_s$, so that $s=1$ and
$q_s+r q_s'=1$. The first-order pure-sGB scalar profile obeys
\be
\frac{{\rm d}\widehat\phi_1}{{\rm d}\rho}
=-\frac{2\xi_1P(\rho)}{\eta M^2\rho^4},
\qquad
P(\rho)\equiv\rho^2+2\rho+4\,.
\label{GBleadingScalarDerivative}
\ee
Evaluating Eqs.~(\ref{JrGLPV}) and~(\ref{PGLPV}) on the zeroth-order metric
and imposing Eq.~(\ref{GBcompatibleF5}) gives the following
beyond-Horndeski contributions to the radial scalar current and ${\cal P}$:
\begin{align}
J^r_{\rm bH}={}&
\frac{4q_s^2\phi'^3}{r^2}
\left(2F_4+XF_{4,X}\right)
-\frac{8\alpha\xi_{,\phi}}{\Mpl^2}
\frac{q_s'q_s^3\phi'^4}{r^2}
\left(5F_4+2XF_{4,X}\right),\\
{\cal P}_{\rm bH}={}&
-\frac{2q_s^2F_4\phi'^4}{r^2}
+\frac{8\alpha\xi_{,\phi}}{\Mpl^2}
\frac{q_s'q_s^3F_4\phi'^5}{r^2}.
\label{GBbeyondHorndeskiCurrent}
\end{align}
The first term in each expression is generated directly by the quartic
interaction, whereas the second is generated by its compatibility-induced
quintic partner.

For any background quantity $Y$, $Y_{\rm full}$ denotes its value in
Eq.~(\ref{GBactionMain}), whereas $Y_{\rm sGB}$ denotes its value in the
pure-sGB limit $F_4=F_5=0$. At fixed $M$ and $\phi_0$, we define
$\Delta Y\equiv Y_{\rm full}-Y_{\rm sGB}$.
This notation includes both direct and background-induced effects and should
not be confused with the determinants $\Delta_K$ and $\Delta_G$. In
particular,
$\phi_{\rm bg}=\phi_{\rm sGB}+\Delta\phi$, where $\Delta\phi$ is a static
background correction rather than the linear perturbation $\delta\phi$.
At the leading deformation-dependent order, the differential operator
acting on $\Delta\phi$ is supplied by the canonical term $G_2=\eta X$.
Indeed, Eq.~(\ref{Cicoefficients}) gives
$C_6^{(\eta X)}=C_9^{(\eta X)}=\eta X$, so the canonical contribution to
the radial current is $\eta h\phi'$. Its leading variation is therefore
$\eta q_s\Delta\phi'$.

Horizon regularity requires the current norm
$J_\mu J^\mu=(J^r)^2/h$ to remain finite. Since $h\to0$ at $r=r_s$, the
total radial current must vanish there on both the full and pure-sGB
backgrounds. For regular $\Delta\phi'$, the terms
$\eta q_s\Delta\phi'$ and $J^r_{\rm bH}$ also vanish as $r\to r_s$.
Consequently, the integration constant in the equation for
$\Delta\phi$ is zero. Setting $h=q_s$ and $s=1$ in the leading differential
operator and integrating Eq.~(\ref{EphiGLPV}) from $r_s$ to $r$ then yields
\be
r^2\eta q_s\Delta\phi'
+r^2J^r_{\rm bH}
=-\int_{r_s}^{r}\tilde r^{\,2}
\frac{\partial{\cal P}_{\rm bH}}{\partial\phi}\,
{\rm d}\tilde r .
\label{GBbeyondHorndeskiScalarQuadrature}
\ee
Equation~(\ref{GBbeyondHorndeskiScalarQuadrature}) contains the direct source
terms from both members of the compatible pair. It is used below only at the
leading quartic order; at higher orders, the recursively induced corrections
must also be included.

For generic nonzero $\phi_0$, both
$\partial{\cal P}_{\rm bH}/\partial\phi$ and the linked quintic current are
subleading at the first quartic order.  The leading scalar response is
therefore sourced only by the first term of $J^r_{\rm bH}$ in
Eq.~(\ref{GBbeyondHorndeskiCurrent}).
At the relevant leading order, either component of ${\cal A}_{\rm bH}$ can
be written in the unified form
\be
 F_4=\alpha C_pX^p\,,
\ee
where $(p,C_p)=(0,\mu_\phi\phi_0^{m_\phi})$ for the $\phi^{m_\phi}$ component and
$(p,C_p)=(n,\mu_X)$ for the $X^n$ component.
Since $X={\cal O}(\alpha^2)$ and $\phi'={\cal O}(\alpha)$, the contribution
$F_4=\alpha C_pX^p$ is of order $\alpha^{2p+1}$, and the associated
quartic current, proportional to $F_4\phi'^3$, is of order
$\alpha^{2p+4}$.  Equation~(\ref{GBbeyondHorndeskiScalarQuadrature}) then
implies $\Delta\phi={\cal O}(\alpha^{2p+4})$.

We denote the ${\cal O}(\alpha^{2p+4})$ coefficient by
$\Delta\widehat\phi_{2p+4}^{(p)}$, whose subscript records the power of
$\alpha$ and whose superscript labels the generating component.
Equations~(\ref{GBbackgroundExpansion}) and
(\ref{GBleadingScalarDerivative}) give
$\phi'=(\alpha/M)\,{\rm d}\widehat\phi_1/{\rm d}\rho
+{\cal O}(\alpha^2)$.
Substituting this expression into the leading quartic term of
$J^r_{\rm bH}$ and using $X=-q_s\phi'^2/2$ in
$F_4=\alpha C_pX^p$, for which
$2F_4+XF_{4,X}=(p+2)F_4$,
Eq.~(\ref{GBbeyondHorndeskiScalarQuadrature}) reduces to
\be
\frac{{\rm d}}{{\rm d}\rho}
\Delta\widehat\phi_{2p+4}^{(p)}=
\frac{2^{p+5}(p+2)(-1)^p C_p\xi_1^{2p+3}}
{\eta^{2p+4}M^{6p+10}}
\frac{(\rho-2)^{p+1}P(\rho)^{2p+3}}
{\rho^{9p+15}}.
\label{GBbeyondHorndeskiScalarDerivative}
\ee
Imposing
$\Delta\widehat\phi_{2p+4}^{(p)}(\infty)=0$ and integrating
Eq.~(\ref{GBbeyondHorndeskiScalarDerivative}), the leading correction takes
the form
\be
\Delta\phi=\alpha^{2p+4}
\Delta\widehat\phi_{2p+4}^{(p)}+\cdots\,,
\label{GBbeyondHorndeskiScalarSolutions}
\ee
where
\be
\Delta\widehat\phi_{2p+4}^{(p)}
=-\frac{2^{p+5}(p+2)(-1)^p C_p\xi_1^{2p+3}}
{\eta^{2p+4}M^{6p+10}}I_p(\rho),
\qquad
I_p(\rho)\equiv\int_\rho^\infty
\frac{(\varrho-2)^{p+1}P(\varrho)^{2p+3}}
{\varrho^{9p+15}}\,{\rm d}\varrho .
\label{GBbeyondHorndeskiScalarCoefficient}
\ee
For every nonnegative integer $p$, $I_p(\rho)$ is a finite sum of inverse
powers of $\rho$. It is regular at $\rho=2$ and vanishes as
$\rho\to\infty$. The factor $(-1)^p$ originates from $X^p$, since $X<0$
for $\rho>2$. For $\phi_0\ne0$,
Eqs.~(\ref{GBbeyondHorndeskiScalarSolutions}) and
(\ref{GBbeyondHorndeskiScalarCoefficient}) give the leading quartic
correction to the scalar background generated by either component of
${\cal A}_{\rm bH}$.

If $\phi_0=0$, then $\phi={\cal O}(\alpha)$, and the
$\mu_\phi\phi^{m_\phi}$ component gives
$F_4={\cal O}(\alpha^{m_\phi+1})$ and
$F_5={\cal O}(\alpha^{m_\phi+2})$. The quartic and linked-quintic current
terms in the scalar equation therefore begin at
$\alpha^{m_\phi+4}$ and $\alpha^{m_\phi+6}$, respectively, and the explicit
$\phi$-derivative sources begin at the same orders. The corresponding
contributions to $\Delta\phi$ have the same onset orders. The metric sources
contain one additional scalar gradient and hence begin at
$\alpha^{m_\phi+5}$ and $\alpha^{m_\phi+7}$, respectively.
The $\mu_X X^n$ component is unaffected by setting $\phi_0=0$ and retains the
four onset orders shown in Eq.~(\ref{GBbeyondHorndeskiOnset}).

For $\phi_0\ne0$, Eq.~(\ref{GBbeyondHorndeskiOnset}) gives the leading
metric corrections
\be
(\Delta f,\Delta h)=
{\cal O}(\alpha^5\mu_\phi)
+{\cal O}(\alpha^{2n+5}\mu_X).
\label{GBbeyondHorndeskiMetricOnset}
\ee
The linked-quintic corrections begin two orders later and therefore do not
enter the leading orders displayed in
Eq.~(\ref{GBbeyondHorndeskiMetricOnset}). For a single quartic generator
$F_4=\alpha C_pX^p$, the leading metric source scales as
$F_4\phi'^4={\cal O}(\alpha^{2p+5})$.
The corresponding metric
corrections therefore take the form
\be
\Delta f=\alpha^{2p+5}\Delta f_{2p+5}^{(p)}+\cdots,
\qquad
\Delta h=\alpha^{2p+5}\Delta h_{2p+5}^{(p)}+\cdots.
\label{GBbeyondHorndeskiMetricSolutions}
\ee
The functions $\Delta f_{2p+5}^{(p)}$ and
$\Delta h_{2p+5}^{(p)}$ are the first deformation-dependent metric
coefficients and admit exact quadrature representations. 
For notational simplicity, we define $N_p\equiv2p+5$ and henceforth write
$\Delta f_{N_p}$ and $\Delta h_{N_p}$ in place of
$\Delta f_{2p+5}^{(p)}$ and $\Delta h_{2p+5}^{(p)}$, respectively.
We denote the corresponding
changes in the hatted coefficients of
Eq.~(\ref{GBbackgroundExpansion}) by
$\Delta\widehat f_{N_p}$ and $\Delta\widehat h_{N_p}$. Since no
deformation-dependent metric correction occurs below order
$\alpha^{N_p}$, expanding Eq.~(\ref{GBbackgroundExpansion}) gives
$\Delta f_{N_p}=2q_s\Delta\widehat f_{N_p}$ and
$\Delta h_{N_p}=-2q_s\Delta\widehat h_{N_p}$.

To derive the leading metric coefficients, we introduce the source terms
entering their quadrature representation. For the Einstein--Hilbert term,
$G_4^{\rm EH}=\Mpl^2/2$, the contributions to the two background equations
are
\be
{\cal E}_{00}^{\rm EH}
=-\frac{\Mpl^2}{r^2}\bigl[r(h-1)\bigr]',
\qquad
{\cal E}_{11}^{\rm EH}
=\frac{\Mpl^2}{r^2}\left(rh\frac{f'}{f}+h-1\right).
\label{GBbeyondHorndeskiEinsteinEquations}
\ee
At order $\alpha^{N_p}$, the background coefficients determined at lower
orders enter as known sources. We collect the terms proportional to $C_p$
in Eqs.~(\ref{EGLPV00}) and~(\ref{EGLPV11}) and define
\be
0=\delta{\cal E}_{00}^{\rm EH}[\Delta h_{N_p}]
+{\cal S}_{00}^{(N_p)},
\qquad
0=\delta{\cal E}_{11}^{\rm EH}[\Delta f_{N_p},\Delta h_{N_p}]
+{\cal S}_{11}^{(N_p)}.
\label{GBbeyondHorndeskiMetricSources}
\ee
Here $\delta{\cal E}_{ab}^{\rm EH}$ denotes the linear response of the
Einstein--Hilbert equations to the deformation-induced metric coefficients.
It is obtained by substituting
$f=q_s+\varepsilon\Delta f_{N_p}$ and
$h=q_s+\varepsilon\Delta h_{N_p}$ into
${\cal E}_{ab}^{\rm EH}$ and retaining the terms proportional to the
bookkeeping parameter $\varepsilon$.
The sources
${\cal S}_{ab}^{(N_p)}$ contain all remaining terms at the same order,
including the direct beyond-Horndeski contributions and those induced by
lower-order fields, such as the scalar correction in
Eq.~(\ref{GBbeyondHorndeskiScalarSolutions}).

Using $q_s+rq_s'=1$ and the relations between the metric and hatted
coefficients, Eqs.~(\ref{GBbeyondHorndeskiEinsteinEquations}) and
(\ref{GBbeyondHorndeskiMetricSources}) reduce to
\be
(rq_s\Delta\widehat h_{N_p})'
=-\frac{r^2}{2\Mpl^2}{\cal S}_{00}^{(N_p)},
\qquad
\Delta\widehat f_{N_p}'=
\frac{\Delta\widehat h_{N_p}}{rq_s}
-\frac{r{\cal S}_{11}^{(N_p)}}{2q_s\Mpl^2}.
\label{GBbeyondHorndeskiMetricEquations}
\ee
Regularity of $\Delta\widehat h_{N_p}$ at $r=r_s$ removes the homogeneous
integration constant in the first equation, while
$\Delta\widehat f_{N_p}(\infty)=0$ fixes the asymptotic normalization of
the time coordinate. The resulting quadratures are
\be
\Delta\widehat h_{N_p}(r)
=-\frac{1}{2q_s(r)r\Mpl^2}
\int_{r_s}^{r}\tilde r^{\,2}
{\cal S}_{00}^{(N_p)}(\tilde r)\,{\rm d}\tilde r,\qquad
\Delta\widehat f_{N_p}(r)
=-\int_r^\infty\left[
\frac{\Delta\widehat h_{N_p}(\tilde r)}
{\tilde r q_s(\tilde r)}
-\frac{\tilde r\,{\cal S}_{11}^{(N_p)}(\tilde r)}
{2q_s(\tilde r)\Mpl^2}\right]{\rm d}\tilde r .
\label{GBbeyondHorndeskiMetricQuadratures}
\ee
For finite sources, the first quadrature determines the horizon value of
$\Delta\widehat h_{N_p}$. The regular horizon limits of the
order-$\alpha^{N_p}$ background equations imply
$({\cal S}_{00}^{(N_p)}+{\cal S}_{11}^{(N_p)})_s=0$, thereby canceling the
apparent $1/q_s$ pole in the second quadrature and ensuring that
$\Delta\widehat f_{N_p}$ remains regular. The source falloffs inherited from
$\phi'={\cal O}(r^{-2})$ also ensure convergence at infinity. Once the
lower-order fields are known, the full background
equations~(\ref{EGLPV00}) and~(\ref{EGLPV11}), together with the building
blocks in Eq.~(\ref{GLPVMetricBuildingBlocks}), determine
${\cal S}_{00}^{(N_p)}$ and ${\cal S}_{11}^{(N_p)}$.
Equation~(\ref{GBbeyondHorndeskiMetricQuadratures}) then yields the complete
order-$\alpha^{N_p}$ coefficients, including all recursively induced
contributions.

For $\phi_0\ne0$, Eqs.~(\ref{GBbeyondHorndeskiScalarSolutions}) and
(\ref{GBbeyondHorndeskiMetricSolutions}) show that a quartic generator
produces scalar and metric corrections beginning at
$\alpha^{2p+4}$ and $\alpha^{2p+5}$, respectively. In the ordering
$((f_{\rm bg},h_{\rm bg}),\phi_{\rm bg})$ used in
Eq.~(\ref{GBbeyondHorndeskiOnset}), $p=0$ gives
$(\alpha^5,\alpha^4)$ for the $\mu_\phi\phi^{m_\phi}$ component, whereas
$p=n$ gives $(\alpha^{2n+5},\alpha^{2n+4})$ for the $\mu_X X^n$ component.
These results reproduce the two direct-$F_4$ columns of
Eq.~(\ref{GBbeyondHorndeskiOnset}). Together with
Eqs.~(\ref{GBfirstOrderBackground})--(\ref{GBphi2Background}), they specify
a horizon-regular and asymptotically flat construction through the leading
direct-$F_4$ deformation orders. The linked-$F_5$ columns follow from the
power counting above.

\subsection{Exterior stability}
\label{GBexteriorstabilitysubsec}

We first establish sufficient local no-ghost and high-frequency
gradient-stability conditions for the weakly coupled pure-sGB solution and
then show that they persist under the compatible beyond-Horndeski deformation.
We work in the BH exterior $r>r_s$, where $f,h>0$, including the one-sided
horizon limit $r\to r_s^+$.  The background results of
Sec.~\ref{GBbackgroundsubsec} and the direct counting below show that the
deformation does not affect the principal coefficients through
${\cal O}(\alpha^2)$.  Substituting
Eqs.~(\ref{GBfirstOrderBackground})--(\ref{GBphi2Background}) into
Eq.~(\ref{Xdefinition}) gives
\be
X=-\frac{2\xi_1^2(\rho-2)P(\rho)^2}
 {\eta^2M^6\rho^9}\alpha^2+{\cal O}(\alpha^3),
\label{GBexteriorX}
\ee
On the same background, the odd-parity coefficients $\mF$, $\mG$, and
$\mH$ take the form
\begin{align}
\mF={}&\Mpl^2-\frac{16\xi_1^2}
{\eta M^4\rho^6}(2\rho^3+\rho^2+2\rho-36)\alpha^2
+{\cal O}(\alpha^3),
\label{GBFExteriorCoefficient}\\
\mG={}&\Mpl^2+\frac{16\xi_1^2}
{\eta M^4\rho^6}(\rho^2+2\rho+4)\alpha^2
+{\cal O}(\alpha^3),
\label{GBGExteriorCoefficient}\\
\mH={}&\Mpl^2+\frac{16\xi_1^2}
{\eta M^4\rho^6}(\rho^3-8)\alpha^2
+{\cal O}(\alpha^3)\,.
\label{GBoddExteriorCoefficients}
\end{align}
The general definition~(\ref{evenKDefinition}) also gives
\be
 \mK=\frac{2\xi_1^2P(\rho)^2}
 {\eta M^4\rho^6}\alpha^2+{\cal O}(\alpha^3).
\label{GBKExterior}
\ee
For $\eta>0$, the leading coefficient of $\mK$ is positive.  Its
${\cal O}(\alpha^2)$ scaling does not by itself signal strong coupling: the
kinetic determinant in Eq.~(\ref{evenKineticPFactorizations}) contains the
finite combination $\mK/\phi'^2=\eta r^2/2+{\cal O}(\alpha)$, which approaches
the canonical scalar normalization as $\alpha\to0$.
Substituting the background solution into
Eqs.~(\ref{evenRadialModeSpeeds}) and~(\ref{evenAngularSpeeds}) yields
\begin{align}
c_{r,{\rm grav}}^{2(+)}={}&\frac{\mG}{\mF}
=1+\frac{32\xi_1^2(\rho-2)
(\rho^2+3\rho+8)}{\eta M^4\Mpl^2\rho^6}\alpha^2+{\cal O}(\alpha^3),
\label{GBexteriorTensorRadialSpeed}\\
c_{r,{\rm sc}}^{2(+)}={}&1+{\cal O}(\alpha^2),
\label{GBexteriorScalarRadialSpeed}\\
c_{\Omega \pm,{\rm even}}^{2(+)}={}&
1 \pm \delta_\Omega(\rho)+{\cal O}(\alpha^2),\qquad\quad
\delta_\Omega(\rho)=
\frac{24|\xi_1|}{M^2\Mpl\rho^3}
\sqrt{\frac{2}{\eta}} |\alpha|\,.
\label{GBexteriorSpeeds}
\end{align}
The tensor radial speed in Eq.~(\ref{GBexteriorTensorRadialSpeed}) is common
to the odd- and even-parity sectors.  The second line is the scalar radial
speed; for exactly linear $\xi$, its correction starts at
${\cal O}(\alpha^4)$ \cite{Minamitsuji:2022mlv}.  The two branches in
Eq.~(\ref{GBexteriorSpeeds}) are the squared angular speeds of the coupled
gravitational--scalar even-parity modes.

For $\ell\geq2$, the odd-parity stability conditions require $\mF>0$,
$\mG>0$, and $\mH>0$.  The even-parity sector additionally requires
$\mK>0$ and positivity of $c_{r,{\rm sc}}^{2(+)}$ and both
$c_{\Omega\pm,{\rm even}}^{2(+)}$.
A sufficient set of pure-sGB exterior conditions is
\be
 \eta>0,\qquad
 \epsilon_{\rm GB}\equiv
 \frac{\xi_1^2}{\eta M^4\Mpl^2}\alpha^2 \ll1,\qquad
 \delta_\Omega(2)=
 \frac{3|\xi_1|}{M^2\Mpl}\sqrt{\frac{2}{\eta}} |\alpha|<1\,.
\label{GBexteriorConditions}
\ee
Provided that the omitted higher-order corrections remain sufficiently small
for all $\rho\geq2$, these inequalities ensure the required no-ghost and
gradient-stability conditions throughout the exterior.  For nonlinear
$\xi(\phi)$, the dimensionless combinations involving $\xi^{(j)}_0$ that
enter the remainders but are not included in $\epsilon_{\rm GB}$ must also
remain perturbatively small.  Since $\delta_\Omega(\rho)$ decreases
monotonically for $\rho\geq2$, its horizon value gives the strongest angular
condition, with
$\delta_\Omega(2)^2=18\epsilon_{\rm GB}$.  If $\delta_\Omega(2)$ is close to
unity, the magnitude of the ${\cal O}(\alpha^2)$ remainder in
$c_{\Omega-,{\rm even}}^{2(+)}$ at the horizon must be much smaller than
$1-\delta_\Omega(2)$; for exactly linear $\xi$, a sufficient condition is
$\epsilon_{\rm GB}\ll1-\delta_\Omega(2)$.

The even-parity monopole and dipole obey the corresponding local no-ghost and
radial gradient-stability conditions under the same weak-coupling assumptions.
On the leading pure-sGB exterior background, the coefficients defined in
Eqs.~(\ref{evenMonopoleCoefficients}) and
(\ref{evenDipoleCoefficients}) reduce to
\be
{\cal K}_0=\frac{\eta r^2}{2q_s}\left[1+{\cal O}(\alpha^2)\right],\qquad
{\cal G}_0=-\frac{\eta r^2q_s}{2}\left[1+{\cal O}(\alpha^2)\right],
\label{GBmonopoleExteriorCoefficients}
\ee
and
\be
{\cal K}_1=\frac{q_s\xi_1^2P(\rho)^2}{18\eta M^2\rho^2}\alpha^2
+{\cal O}(\alpha^3),\qquad
{\cal G}_1=-q_s^2{\cal K}_1\left[1+{\cal O}(\alpha^2)\right].
\label{GBdipoleExteriorCoefficients}
\ee
Since $\eta>0$, $\xi_1\ne0$, and $q_s>0$ in the exterior, the leading terms
imply ${\cal K}_0>0$, ${\cal G}_0<0$, ${\cal K}_1>0$, ${\cal G}_1<0$, and
$c_{r,0}^{2(+)}=c_{r,1}^{2(+)}=1+{\cal O}(\alpha^2)>0$
at each fixed $r>r_s$ for sufficiently small nonzero $\alpha$.
The limit ${\cal K}_1\to0$ as $\alpha\to0$ does not by itself signal
strong coupling. Indeed, Eqs.~(\ref{GBbackgroundExpansion}) and
(\ref{GBleadingScalarDerivative}) give $\phi'={\cal O}(\alpha)$ at fixed
$r$. Hence, $\phi'\to0$ in this limit, and the dipole gauge
$\delta\phi=0$ becomes inadmissible; a gauge retaining $\delta\phi$ has a
regular canonical-scalar limit.

We next show that the compatible beyond-Horndeski deformation preserves these
conditions for sufficiently small $|\alpha|$.  For fixed $\mu_\phi$ and
$\mu_X$, and for generic $\phi_0\ne0$,
Eqs.~(\ref{GBbeyondHorndeskiInteractions}) and
(\ref{GBexteriorX}) give
\be
 F_4={\cal O}(\alpha\mu_\phi)
     +{\cal O}(\alpha^{2n+1}\mu_X),\qquad
 F_5={\cal O}(\alpha^2\mu_\phi)
     +{\cal O}(\alpha^{2n+2}\mu_X).
\label{GBbeyondHorndeskiExteriorCounting}
\ee
At fixed background fields, substituting Eq.~(\ref{GBcompatibleF5}) into the
odd-parity coefficients gives the direct contributions
\be
 \delta_{\rm dir}\mH=-8X^2F_4
 \left(1-\frac{8\alpha\xi_{,\phi}h\phi'}{\Mpl^2r}\right),\qquad
 \delta_{\rm dir}\mG=-8X^2F_4
 \left(1-\frac{4\alpha\xi_{,\phi}h\phi'f'}{\Mpl^2f}\right),\qquad
 \delta_{\rm dir}\mF=0.
\label{GBbeyondHorndeskiOddCorrections}
\ee
Here $\delta_{\rm dir}$ denotes the explicit contribution of the compatible
pair evaluated on the undeformed sGB background; it excludes the indirect
contribution induced by the deformation of the background fields.  The unit
terms in parentheses
come from $F_4$.  Since $X={\cal O}(\alpha^2)$, they scale as
$X^2F_4={\cal O}(\alpha^5\mu_\phi)
+{\cal O}(\alpha^{2n+5}\mu_X)$.  The remaining terms come from the linked
$F_5$ interaction.  The relevant background factors are finite in the
one-sided horizon limit, so these terms carry an additional factor
$\alpha\phi'={\cal O}(\alpha^2)$ and begin at
${\cal O}(\alpha^7\mu_\phi)
+{\cal O}(\alpha^{2n+7}\mu_X)$.

We now include the background response.  We use the convention
$\Delta Y\equiv Y_{\rm full}-Y_{\rm sGB}$ introduced in
Sec.~\ref{GBbackgroundsubsec} also for the perturbation coefficients and
squared speeds below, with the two terms evaluated on their respective
backgrounds.  Thus, although
$\delta_{\rm dir}\mF=0$ in
Eq.~(\ref{GBbeyondHorndeskiOddCorrections}), $\Delta\mF$ is generally nonzero.
The full changes obey
\be
\frac{\Delta\mF}{\Mpl^2},\quad
\frac{\Delta\mG}{\Mpl^2},\quad
\frac{\Delta\mH}{\Mpl^2}
={\cal O}(\alpha^5\mu_\phi)
+{\cal O}(\alpha^{2n+5}\mu_X).
\label{GBbeyondHorndeskiTensorOnset}
\ee
Here $\Delta\mF$ is entirely background-induced, whereas $\Delta\mG$ and
$\Delta\mH$ contain both direct and background-induced contributions.  All
three enter at higher order than the ${\cal O}(\alpha^2)$ pure-sGB corrections in
Eqs.~(\ref{GBFExteriorCoefficient})--(\ref{GBoddExteriorCoefficients}) and
therefore preserve the signs of $\mF$, $\mG$, and $\mH$ at each fixed
$r>r_s$ for sufficiently small $|\alpha|$.

The even-parity scalar mode requires a relative comparison because
Eq.~(\ref{GBKExterior}) gives $\mK_{\rm sGB}={\cal O}(\alpha^2)$.  From
Eqs.~(\ref{evenReductionDefinitions}), (\ref{evenGLPVPDefinitions}), and
(\ref{evencoeffa}), one finds $\delta_{\rm dir}\mH_\vartheta=0$, whereas both
$\delta_{\rm dir}\mH$ and $\delta_{\rm dir}\mu_{\rm ev}/r$ are
${\cal O}(X^2F_4)$.  These relations imply
$\delta_{\rm dir}{\cal P}_1={\cal O}(X^2F_4)$.  The background onsets derived
in Sec.~\ref{GBbackgroundsubsec} do not lower this order, so
$\Delta{\cal P}_1={\cal O}(\alpha^5\mu_\phi)
+{\cal O}(\alpha^{2n+5}\mu_X)$.  Together with
Eq.~(\ref{GBbeyondHorndeskiTensorOnset}), this gives
$\Delta\mK=2\Delta{\cal P}_1-\Delta\mF
={\cal O}(\alpha^5\mu_\phi)+{\cal O}(\alpha^{2n+5}\mu_X)$.
Dividing by $\mK_{\rm sGB}$ lowers the power of $\alpha$ by two.  Expanding
the exact scalar radial speed in Eq.~(\ref{evenRadialModeSpeeds}) and the
angular eigenvalue equation~(\ref{evenAngularDispersion}) around the pure-sGB
solution gives speed changes of the same relative order.  The
${\cal O}(|\alpha|)$ pure-sGB angular splitting does not alter this counting.
We therefore obtain
\be
 \frac{\Delta\mK}{\mK_{\rm sGB}},\quad
 \Delta c_{r,{\rm sc}}^{2(+)},\quad
 \Delta c_{\Omega\pm,{\rm even}}^{2(+)}
 = {\cal O}(\alpha^3\mu_\phi)
 +{\cal O}(\alpha^{2n+3}\mu_X).
\label{GBbeyondHorndeskiSpeedOnset}
\ee
These corrections vanish as $\alpha\to0$ and are subleading to the
${\cal O}(|\alpha|)$ pure-sGB angular splitting in
Eq.~(\ref{GBexteriorSpeeds}).  For $\phi_0=0$, the
$\mu_\phi\phi^{m_\phi}$ contribution begins at still higher orders, as shown
in Sec.~\ref{GBbackgroundsubsec}.  For $\ell=0,1$, the relative changes in
${\cal K}_\ell$ and ${\cal G}_\ell$ induced by $F_4$ start at
${\cal O}(\alpha^3\mu_\phi)+{\cal O}(\alpha^{2n+3}\mu_X)$ or later, while
the linked-$F_5$ contributions begin two orders later.
Hence, for each fixed $r>r_s$, the compatible beyond-Horndeski corrections
preserve the signs required for monopole and dipole stability when
$|\alpha|$ is sufficiently small.

For $\ell=0,1$, we use ${\cal K}_{\ell,{\rm sGB}}$ and
${\cal G}_{\ell,{\rm sGB}}$ for the corresponding pure-sGB values of
${\cal K}_\ell$ and ${\cal G}_\ell$.  To formulate a single sufficient
condition throughout the exterior, we define
\be
 \epsilon_{\rm bH}^{(+)}\equiv\sup_{r\ge r_s}\max\left\{\max_{{\cal Q}=\mF,\mG,\mH}\frac{|\Delta{\cal Q}|}{\Mpl^2},~\left|\frac{\Delta\mK}{\mK_{\rm sGB}}\right|,~|\Delta c_{r,{\rm sc}}^{2(+)}|,~\max_{\sigma=\pm}|\Delta c_{\Omega\sigma,{\rm even}}^{2(+)}|,~\max_{\ell=0,1}\left\{\left|\frac{\Delta{\cal K}_\ell}{{\cal K}_{\ell,{\rm sGB}}}\right|,\left|\frac{\Delta{\cal G}_\ell}{{\cal G}_{\ell,{\rm sGB}}}\right|\right\}\right\}\,.
\label{GBbeyondHorndeskiEvenBound}
\ee
The last term is the maximum relative correction among the $\ell=0,1$
coefficients.  Their pure-sGB values at the retained order are given in
Eqs.~(\ref{GBmonopoleExteriorCoefficients}) and
(\ref{GBdipoleExteriorCoefficients}).  Ratios whose denominators vanish as
$\alpha\to0$ are evaluated at nonzero $\alpha$ and understood through their
finite small-$\alpha$ limits.  The low-multipole ratios are evaluated for
$r>r_s$ and extended to the horizon by finite one-sided limits.  At each fixed $r$,
the outer maximum selects the largest correction in
Eq.~(\ref{GBbeyondHorndeskiEvenBound}), and the supremum bounds it over the
entire exterior.  Equation~(\ref{GBexteriorConditions}) gives
$0\leq\delta_\Omega(2)<1$.  At leading order, the angular minus branch has
the smallest squared speed at the horizon,
$1-\delta_\Omega(2)$.  Provided that the background expansion of
Sec.~\ref{GBbackgroundsubsec} remains valid throughout the exterior, a
sufficient additional requirement is therefore
\be
 \epsilon_{\rm bH}^{(+)}
 \ll 1-\delta_\Omega(2).
\label{GBfullExteriorConditions}
\ee

For the regular polynomial beyond-Horndeski ansatz considered here, the
deformation-induced background corrections are regular at the horizon and
decay at spatial infinity.  Moreover, $X\to0$ at both boundaries, and
Eq.~(\ref{GBbeyondHorndeskiOddCorrections}) shows that the direct
beyond-Horndeski contributions to $\mG$ and $\mH$ vanish at least as $X^2$,
while $\delta_{\rm dir}\mF=0$.  Together with the pure-sGB stability
conditions, these properties ensure that the coefficient functions in the
small-$\alpha$ expansions of all quantities inside the maximum in
Eq.~(\ref{GBbeyondHorndeskiEvenBound}) are continuous for $r>r_s$, have
finite one-sided horizon limits, and remain bounded as $r\to\infty$.
The supremum is therefore finite.  Equations~
(\ref{GBbeyondHorndeskiTensorOnset}) and
(\ref{GBbeyondHorndeskiSpeedOnset}), together with the low-multipole counting,
then imply $\epsilon_{\rm bH}^{(+)}\to0$ at fixed $M$, $\phi_0$, $\mu_\phi$,
and $\mu_X$ as $\alpha\to0$.  Together with the pure-sGB remainder
requirements stated below Eq.~(\ref{GBexteriorConditions}),
Eqs.~(\ref{GBexteriorConditions}) and (\ref{GBfullExteriorConditions}) ensure
all local odd- and even-parity no-ghost and high-frequency radial and angular
gradient-stability conditions throughout the exterior for sufficiently small
$|\alpha|$.

\subsection{Interior EFT breakdown scale}
\label{GBinteriorcriticalsubsec}

The weak-coupling expansion used in Secs.~\ref{GBbackgroundsubsec} and
\ref{GBexteriorstabilitysubsec} can lose perturbative control deep inside the
horizon.  We first estimate this breakdown in pure sGB using the background curvature
and the odd-parity coefficient $\mH$, which controls the angular-gradient
part of the tensor principal symbol.  We then ask whether direct
beyond-Horndeski corrections to $\mH$ and to the radial principal coefficient
$\mG$ move the breakdown to a larger radius.
For this quantitative estimate, we restrict the coupling function
$\xi(\phi)$ to be exactly linear in $\phi$,
\be
 \xi(\phi)=\xi_0+\xi_1\phi\,,
\ee
so that $\xi_{,\phi}=\xi_1$ is constant and
$\xi^{(j)}(\phi)=0$ for $j\geq2$.  This restriction avoids an
uncontrolled Taylor expansion of $\xi(\phi)$ in the deep interior, where the
scalar displacement grows and is no longer parametrically small near the
breakdown scale derived below.
We introduce the dimensionless scalar
$\Phi=\sqrt{2\eta}\,\phi/\Mpl$, normalized such that its kinetic term
inside the overall $\Mpl^2/2$ factor takes the conventional form
$-(\nabla\Phi)^2/2$, and the effective linear GB coupling
\be
 \alpha_{\rm GB}\equiv
 \frac{\sqrt{2}\,\alpha\xi_1}{\sqrt{\eta}\,\Mpl},\qquad
 x\equiv\frac{r}{r_s},\qquad
 \widehat\alpha_{\rm GB}\equiv\frac{\alpha_{\rm GB}}{r_s^2}\,,
\label{GBcanonicalCoupling}
\ee
where $r_s=2M$ denotes the horizon radius.
The horizon-regular and asymptotically flat scalar profile was obtained in
Eq.~(\ref{GBfirstOrderBackground}).  More explicitly, substituting
$\phi=\phi_0+\alpha\widehat\phi_1+\cdots$ and $\rho=r/M=2x$
into Eq.~(\ref{GBleadingScalarDerivative}), and then using
Eq.~(\ref{GBcanonicalCoupling}), gives
\begin{equation}
 \frac{{\rm d}\Phi}{{\rm d}x}=-4\widehat\alpha_{\rm GB}
 \frac{x^2+x+1}{x^4}+{\cal O}(\widehat\alpha_{\rm GB}^3).
\label{GBinteriorScalarDerivative}
\end{equation}
Equation~(\ref{GBbeyondHorndeskiTensorOnset}) shows that the compatible
beyond-Horndeski pair does not contribute to $\mH$ through the order
displayed below.  For the linear sGB coupling,
$G_4=\Mpl^2/2$, $G_{4,X}=G_{5,\phi}=0$, and
$G_{5,X}=-4\alpha\xi_1/X$.
Substituting these relations together with $X=-h\phi'^2/2$ and
$h=1-1/x+{\cal O}(\widehat\alpha_{\rm GB}^2)$ 
into Eq.~(\ref{mH}) yields
\begin{equation}
 \frac{\mH}{\Mpl^2}=1-16\widehat\alpha_{\rm GB}^2
 \frac{1-x^3}{x^6}+{\cal O}(\widehat\alpha_{\rm GB}^4).
\label{GBinteriorH}
\end{equation}
Equations~(\ref{GBinteriorScalarDerivative}) and~(\ref{GBinteriorH}) are
expansions in $\widehat\alpha_{\rm GB}$ that retain the complete $x$
dependence at each displayed order, rather than near-horizon expansions in $x-1$.  Their continuation to $0<x<1$, where $r$ is timelike at the displayed order, is valid only while the small-$\widehat\alpha_{\rm GB}$ hierarchy remains 
perturbatively controlled.

To estimate where the small-coupling expansion of $\mH$ loses
perturbative control, we define $r_c\equiv x_c r_s$, where $x_c>0$ is
determined by
$\mH(r_c)=\mH(x_c r_s)=0$ after retaining Eq.~(\ref{GBinteriorH}) through
${\cal O}(\widehat\alpha_{\rm GB}^2)$.  Solving this condition gives
\be
\begin{aligned}
 x_c^3&=-8\widehat\alpha_{\rm GB}^2
 +4|\widehat\alpha_{\rm GB}|
 \sqrt{1+4\widehat\alpha_{\rm GB}^2}
 =4|\widehat\alpha_{\rm GB}|
 +{\cal O}(\widehat\alpha_{\rm GB}^2),\\
 r_c&=(4|\alpha_{\rm GB}|r_s)^{1/3}
 \left[1+{\cal O}(|\widehat\alpha_{\rm GB}|)\right].
\end{aligned}
\label{GBcriticalRadius}
\ee
An independent estimate of the same parametric breakdown scale follows
from the curvature expansion.  For the perturbative linear-sGB solution
of Ref.~\cite{Sotiriou:2014pfa}, with
$\alpha_{\rm GB}^2=2\alpha^2\xi_1^2/(\eta\Mpl^2)$ as defined in
Eq.~(\ref{GBcanonicalCoupling}), the dominant terms in its formal
small-$r$ continuation are
\be
 R_{\rm GB}^2=\frac{48M^2}{r^6}
 +\alpha_{\rm GB}^2\left[
 \frac{26880M^4}{r^{12}}
 +{\cal O}\!\left(\frac{M^3}{r^{11}}\right)\right]
 +{\cal O}(\alpha_{\rm GB}^4).
\label{GBcurvatureExpansion}
\ee
After matching the canonical scalar normalization, $\alpha_{\rm GB}$ is
$\sqrt{2}$ times the coupling denoted by $\alpha$ in
Ref.~\cite{Sotiriou:2014pfa}.  Thus, when expressed in terms of
$\alpha_{\rm GB}^2$, the coefficient of the $r^{-12}$ term is one half of
that quoted there.  The ratio of the leading curvature correction in
Eq.~(\ref{GBcurvatureExpansion}) to the Schwarzschild term is
$560\alpha_{\rm GB}^2M^2/r^6$.  This ratio becomes unity at
\be
 r_{\rm pert}=560^{1/6}(|\alpha_{\rm GB}|M)^{1/3}
 =1.44\,r_c
 \left[1+{\cal O}(|\widehat\alpha_{\rm GB}|)\right].
\label{GBperturbativeRadius}
\ee
Moving inward from the horizon, the background expansion therefore loses
perturbative control at $r\sim r_{\rm pert}$, before the truncated
expression for $\mH$ vanishes at $r=r_c$.  Since the two radii have the
same parametric scaling, Eqs.~(\ref{GBcriticalRadius}) and
(\ref{GBperturbativeRadius}) consistently locate the breakdown at
$r={\cal O}(r_c)$.  In particular, the formal zero of $\mH$ lies outside
the perturbatively controlled region and does not by itself establish a
physical instability.

We next examine whether the compatible beyond-Horndeski pair can cause the
perturbative expansion to break down at a radius larger than
$r_{\rm pert}$.  Since
$r_{\rm pert}/r_s={\cal O}(|\widehat\alpha_{\rm GB}|^{1/3})\ll1$, the
region near this scale lies in the deep BH interior, $x\ll1$.  Integrating
Eq.~(\ref{GBinteriorScalarDerivative}) and retaining the dominant
$x^{-3}$ term gives
\begin{equation}
 \phi_{\rm lead}(r)=\phi_0+
 \frac{8\alpha\xi_1M}{3\eta r^3},\qquad
 |X_{\rm lead}(r)|=
 \frac{64\alpha^2\xi_1^2M^3}{\eta^2r^9}.
\label{GBinteriorPowerCounting}
\end{equation}
The second relation follows from
$\phi'_{\rm lead}=-8\alpha\xi_1M/(\eta r^4)$,
$h\simeq-2M/r$, and $X=-h\phi'^2/2$.
Substituting these profiles and $f\simeq h\simeq-2M/r$ into the first two
relations in Eq.~(\ref{GBbeyondHorndeskiOddCorrections}) gives
\begin{equation}
 \frac{|\delta_{\rm dir}\mH|}{\Mpl^2}\simeq
 \frac{8|X^2F_4|}{\Mpl^2}
 \left|1-\frac{r_c^6}{r^6}\right|,\qquad
 \frac{|\delta_{\rm dir}\mG|}{\Mpl^2}\simeq
 \frac{8|X^2F_4|}{\Mpl^2}
 \left(1+\frac{r_c^6}{2r^6}\right),
\label{GBbeyondHorndeskiInteriorCorrections}
\end{equation}
where $X$ and $F_4$ on the right-hand sides are evaluated on the leading
profiles, and we have used
$r_c^6=128\alpha^2\xi_1^2M^2/(\eta\Mpl^2)$ at leading order.
The terms independent of $r_c^6/r^6$ arise from the quartic interaction,
whereas those proportional to this ratio arise from its
compatibility-induced quintic partner.  In the deep-interior region down
to the estimated pure-sGB breakdown scale,
\begin{equation}
 r_{\rm pert}\leq r\ll r_s,\qquad
 \frac{r_c^6}{r^6}\leq
 \frac{r_c^6}{r_{\rm pert}^6}=\frac{4}{35}
 \left[1+{\cal O}(|\widehat\alpha_{\rm GB}|)\right].
\label{GBquinticExteriorOfBreakdown}
\end{equation}
Thus, at leading order, the quintic interaction changes the direct quartic
contribution by at most a fraction $4/35$ in $\mH$ and $2/35$ in $\mG$
down to $r=r_{\rm pert}$.  It therefore cannot introduce an independent,
parametrically larger breakdown radius, and no separate quintic matching
scale is required.

We define a single direct beyond-Horndeski scale $r_{\rm bH}$ as the largest
positive radius at which the common quartic contribution
$8|X^2F_4|/\Mpl^2$ in
Eq.~(\ref{GBbeyondHorndeskiInteriorCorrections}) becomes unity.  For
$F_4=\alpha(\mu_\phi\phi^{m_\phi}+\mu_X X^n)$, this condition gives
\begin{equation}
 1=\frac{8|\alpha|}{\Mpl^2}
 \left[\frac{64\alpha^2\xi_1^2M^3}
 {\eta^2r_{\rm bH}^9}\right]^2
 \left|\mu_\phi
 \left(\phi_0+
 \frac{8\alpha\xi_1M}{3\eta r_{\rm bH}^3}\right)^{m_\phi}
 +\mu_X
 \left[\frac{64\alpha^2\xi_1^2M^3}
 {\eta^2r_{\rm bH}^9}\right]^n\right|\,,
\label{GBbeyondHorndeskiRadius}
\end{equation}
which accounts for interference between the two generators and covers both
the hair-dominated and $|\phi_0|$-dominated regimes.
For $\mu_\phi=0$, it reduces to
\begin{equation}
 r_{\rm bH}^{\,9(n+2)}=
 \frac{8|\alpha\mu_X|}{\Mpl^2}
 \left[\frac{64\alpha^2\xi_1^2M^3}{\eta^2}\right]^{n+2}.
\label{GBbeyondHorndeskiXRadius}
\end{equation}
An admissible root of Eq.~(\ref{GBbeyondHorndeskiRadius}) must satisfy
$r_{\rm bH}\ll r_s$, and it changes the matching scale only when
$r_{\rm bH}>r_{\rm pert}$.  For such a root, the direct
principal-coefficient estimate is
\begin{equation}
 r_{\rm EFT}^{({\rm dir})}\equiv
 \max\left(r_{\rm pert},r_{\rm bH}\right),
\label{GBtotalBreakdownRadius}
\end{equation}
where $r_{\rm bH}$ is omitted if no admissible positive root exists.  A root
with $r_{\rm bH}\leq r_{\rm pert}$ leaves
$r_{\rm EFT}^{({\rm dir})}=r_{\rm pert}$, whereas a formal root with
$r_{\rm bH}={\cal O}(r_s)$ lies outside the domain of
Eq.~(\ref{GBinteriorPowerCounting}) and must be reassessed using the full
background and principal coefficients.

Equation~(\ref{GBtotalBreakdownRadius}) combines the pure-sGB breakdown
estimate with the direct beyond-Horndeski corrections to $\mH$ and $\mG$
evaluated on the leading sGB profiles in
Eq.~(\ref{GBinteriorPowerCounting}).  A complete estimate must also test the
small-$\alpha$ expansions of the deformation-corrected background and the
remaining even-parity perturbation equations.

As a concrete illustration, consider a pure $X$-dependent deformation
$F_4=\alpha\mu_X X^n$.  At fixed $\mu_X$,
Eqs.~(\ref{GBperturbativeRadius}) and
(\ref{GBbeyondHorndeskiXRadius}) give
$r_{\rm pert}\propto|\alpha|^{1/3}$ and
$r_{\rm bH}\propto|\alpha|^{(2n+5)/[9(n+2)]}$, respectively.  Hence,
$r_{\rm bH}/r_{\rm pert}\propto
|\alpha|^{-(n+1)/[9(n+2)]}$, while $r_{\rm bH}/r_s\to0$ as
$|\alpha|\to0$.  A parametrically controlled hierarchy
$r_{\rm pert}<r_{\rm bH}\ll r_s$ can therefore arise, in which the
beyond-Horndeski interaction sets the first interior breakdown scale while
remaining perturbatively small throughout the exterior.

More generally, the overall breakdown scale of the perturbative treatment is
set by the largest of $r_{\rm EFT}^{({\rm dir})}$ and any additional breakdown
radii identified by these checks. Below this scale, at
least one expansion used in the background or local stability analysis is
unreliable.  If the action~(\ref{GBactionMain}) is treated as a complete
classical theory, the relevant equations must be solved without truncating in
$\alpha$.  If it is instead treated as an EFT truncation, omitted
higher-dimensional operators may no longer be suppressed, requiring their
inclusion or matching to an ultraviolet completion.  This loss of
perturbative control does not, by itself, imply a physical instability or
alter the exterior-stability result of
Sec.~\ref{GBexteriorstabilitysubsec}.

\section{Conclusions}
\label{consec}

We have developed a unified framework for the background and linear
perturbations of static, spherically symmetric BHs with a radial scalar
profile in general quartic--quintic GLPV theories, including Horndeski
as a subclass.  For backgrounds satisfying the field 
equations, we derived the complete odd- and even-parity quadratic 
actions and obtained their local principal characteristics 
on regular, nondegenerate branches. 
The formulation applies whether $t$ or $r$ 
is timelike and yields the no-ghost
conditions and the radial and angular squared propagation speeds in the
high-frequency limit.  These are necessary local criteria and do not by
themselves establish full finite-frequency mode stability.

In Sec.~\ref{backgroundgeneralsec}, we derived the three metric equations and
the scalar-field equation for the background metric~(\ref{metric}) with
$\phi=\phi(r)$, allowing general $(\phi,X)$ dependence in all $G_i$, $F_4$,
and $F_5$ while keeping $F_4$ and $F_5$ independent. 
The derivation assumes only $f/h>0$ and therefore applies to both
$f,h>0$ and $f,h<0$.  On branches with $\phi'\ne0$, the radial diffeomorphism
identity~(\ref{backgroundBianchiIdentity}) relates the scalar-field equation
to the metric equations, providing a nontrivial consistency check of the full
set of GLPV background equations.

In Sec.~\ref{oddsec}, we presented the complete quadratic action for
odd-parity perturbations.
For $\ell\geq2$, the sector contains a single propagating tensor degree of
freedom, and the local no-ghost and high-frequency gradient-stability
conditions reduce to $\mF>0$, $\mG>0$, and $\mH>0$ in both the $f,h>0$ and
$f,h<0$ regions.  When $r$ is timelike, the squared propagation speed along 
the spacelike $t$ direction is the reciprocal of its radial counterpart in the $f,h>0$
region. The odd-parity vector harmonic vanishes identically for $\ell=0$.  The
$\ell=1$ dipole sector contains no local propagating degree of freedom; on a
regular branch, its stationary solution describes infinitesimal slow rotation,
and no additional high-frequency stability condition arises.

In Sec.~\ref{evensec}, we obtained the complete even-parity quadratic action
together with the five unreduced perturbation equations.  For $\ell\geq2$,
eliminating three nondynamical variables on a regular, nondegenerate branch
yields a two-field system with two local propagating degrees of freedom,
corresponding to coupled tensor and scalar modes.  In a region where $t$ is
timelike, the exact kinetic factorization shows that $\mK>0$ is the only
additional no-ghost condition once the odd-parity conditions are imposed.
The exact factorization of the radial characteristic polynomial further shows
that the odd- and even-parity tensor polarizations have the same squared radial
propagation speed, including for nonzero $F_4$ and $F_5$.  The regular
$\ell=0$ and $\ell=1$ even-parity sectors each contain one local propagating
degree of freedom, with separate no-ghost and radial gradient-stability
conditions.

When $r$ is timelike, the parent-action analysis establishes the local
equivalence of the unreduced and reduced formulations. The radial
characteristic polynomial takes the reciprocal form of its $t$-timelike
counterpart, while the tensor characteristics in the two parity sectors
continue to coincide.  For angular propagation, 
a nonzero antisymmetric mixing $\Upsilon_\infty$
produces $c_{\Omega+}^2={\cal O}(L)$ and
$c_{\Omega-}^2={\cal O}(L^{-1})$; the high-frequency branch obeys
$\Omega_+^2\propto k_\Omega^4$ and has unbounded phase and group velocities.
Away from accidental prefactor zeros, two finite, nonzero, scale-independent
angular cones require the Horndeski-related compatibility
condition~(\ref{GLPVcompatibility}).  This condition aligns the quartic and
quintic covariant degeneracy directions and permits a single local disformal
transformation removing both beyond-Horndeski interactions, although its
regular invertibility and the remaining stability conditions must be checked
separately.

In Sec.~\ref{horizonInstabilitySec}, we established local obstructions for
$X_s\equiv X(r_s)\ne0$ in one-sided exterior neighborhoods of simple outer
horizons in three settings, identifying singular,
degenerate, and fine-tuned loci separately. First, every nontrivial branch of
the exact quadratic-GLPV solution of Ref.~\cite{Bakopoulos:2022bho} is either
locally unstable or has a degenerate radial tensor cone near the horizon.
Second, a negative-square factorization yields a generic no-go result for
shift- and reflection-symmetric quadratic GLPV theories.  Third, we considered
general functions $G_i(\phi,X)$ $(i=2,\ldots,5)$, $F_4(\phi,X)$, and
$F_5(\phi,X)$ satisfying Eq.~(\ref{GLPVcompatibility}).  After excluding
branches with singular or degenerate tensor characteristics, the remaining
regular branch admits a locally invertible disformal map to Horndeski.  
On this branch, the scalar radial speed generically diverges unless
$\bar\kappa_r=0$.  If $\bar\kappa_r=0$ and
$\bar\kappa_{\rm H}\bar\zeta_{\rm H}\ne0$, the negative-square
relation~(\ref{fullGLPVStrictNegativeProduct}) nevertheless implies that at
least one no-ghost or angular gradient-stability condition is violated.
Since the coupling functions are otherwise arbitrary, this
provides a broad extension of the Horndeski near-horizon result of
Ref.~\cite{Minamitsuji:2022vbi} to the compatible quartic--quintic GLPV class.
The exceptional loci identified in Secs.~\ref{generalGLPVNoGoSubsec} and
\ref{fullGLPVNoGoSubsec} require separate analyses, and no conclusion about
their stability follows from the present results.

In Sec.~\ref{sGBsec}, we studied the complementary weakly coupled sGB branch
with $X_s=0$.  For analytic power-law deformations in $F_4$, the compatibility
condition uniquely fixes the linked $F_5$, whose contributions begin two
orders later than the corresponding quartic ones.  We constructed a
horizon-regular background expansion and obtained quadratures for the leading
scalar and metric corrections.  If this expansion and the omitted remainders
remain controlled, Eqs.~(\ref{GBexteriorConditions}) and
(\ref{GBfullExteriorConditions}) ensure all local no-ghost and high-frequency
radial and angular gradient-stability conditions in both parity sectors,
including the even $\ell=0,1$ sectors, throughout the exterior for
sufficiently small $|\alpha|$.

For an exactly linear GB coupling, the curvature expansion loses perturbative
control at $r\sim r_{\rm pert}$, before the formal zero of $\mH$ at $r=r_c$.
The linked quintic interaction introduces no independent parametrically
larger matching scale.  Combining the pure-sGB breakdown scale with the direct
beyond-Horndeski corrections to the principal coefficients gives the direct
EFT matching radius
$r_{\rm EFT}^{({\rm dir})}=\max(r_{\rm pert},r_{\rm bH})$ when an admissible
$r_{\rm bH}$ exists. The actual matching scale must also include any larger
breakdown radius arising from the deformation-corrected background or the
even-parity perturbation expansion.  This breakdown indicates a loss of
perturbative control, not a physical instability, and leaves the exterior
result unchanged.

The no-ghost and high-frequency gradient-stability criteria obtained here
provide a systematic first test for candidate BH solutions in GLPV theories.
A key direction for future work is to determine whether stable,
asymptotically flat BHs with static scalar hair exist beyond the sGB class
considered here.  The complete quadratic actions and perturbation equations
also provide a basis for computing the quasinormal-mode spectra of branches
that pass these local tests, including weakly coupled sGB BHs with compatible
beyond-Horndeski deformations.  Such calculations require solving the full
finite-frequency perturbation equations with the appropriate horizon and
asymptotic boundary conditions, beyond the local eikonal analysis performed
here.

\section*{Acknowledgements}

ST thanks the members of Chulalongkorn University for their warm 
hospitality during his stay.
ST acknowledges support from JSPS KAKENHI Grant Nos.~26K07090 and
26H00847 and from the Waseda University Grant for Special Research Projects
(No.~2026C-486).

\appendix

\section{Coefficients in the second-order action of even-parity perturbations}
\label{AppEvenCoefficients}

The coefficients in Eq.~(\ref{evenStotal}) are listed below.  Every function
and its derivatives are evaluated on the background at $(\phi(r),X(r))$. 
Here $X=-h\phi'^2/2$, and $s$ is the
positive quantity defined in Eq.~(\ref{positiveSDefinition}); no separate
assumption on the signs of $f$ and $h$ is made.
In addition to $\vartheta_1$ defined in Eq.~(\ref{evenReductionDefinitions}), we
introduce the following GLPV combination
\be
\vartheta_2 \equiv
\frac{h^2\phi'^2}{2rs}
\left[F_4(f'r+2f)+6F_5f'h\phi' \right]\,.
\label{eventheta2}
\ee
The coefficients multiplying $H_0$ are
\begin{align}
a_1={}& \frac{f}{s}\Biggl[
h^2r\left(G_{5,\phi X}-2G_{4,XX}\right)\phi'^3
+\frac12h^3G_{5,XX}\phi'^4
-h\phi'^2\left\{\frac32hG_{5,X}
+r^2\left(G_{4,\phi X}-\frac12G_{3,X}\right)-\frac12G_{5,X}\right\}
\notag\\
&\hspace{1.5cm}
+2hr\left(G_{4,X}-G_{5,\phi}\right)\phi'
+r^2G_{4,\phi}\Biggr]+\frac{fh^2\phi'^3}{s}
\left(2F_{4,X}hr\phi'^2-8F_4r+3F_{5,X}h^2\phi'^3
-15F_5h\phi'\right), 
\notag\\
a_2={}&a_1'
-\frac{a_1(f'h+h'f)}{2fh}
-\left(\frac{\phi''}{\phi'}-\frac{f'}{2f}\right)a_1
+\frac{r}{\phi'}\left(\frac{f'}{f}-\frac{h'}{h}\right)a_4
+L\vartheta_1,\notag\\
a_3={}&-\frac12\phi'a_1-ra_4,\qquad
a_4=\frac{f}{2s}\mH,\notag\\
a_5={}&a_2'-a_1''-L\vartheta_1',\qquad
a_6=-\frac{s}{2\phi'}
\left(\mH'+\frac{\mH}{r}-\frac{\mF}{r}\right)
-\frac{(h'\phi'+2h\phi'')\vartheta_1}{2h\phi'},\notag\\
a_7={}&a_3',\qquad
a_8=-\frac{a_4}{2h}-\frac12\phi'\vartheta_1,\qquad
a_9=a_4'+\left(\frac1r-\frac{f'}{2f}\right)a_4.
\label{evencoeffa}
\end{align}
On a background solution, $a_4'$ is not an independent coefficient: it obeys
Eq.~(\ref{evenTensorRadialRelations}).  Since $a_4$ contains the full GLPV
quantity $\mH$ of Eq.~(\ref{mH}), the $F_4$ and $F_5$ terms must not be added
again when this on-shell relation is used.
The coefficients involving $H_1$ are
\begin{align}
b_1=\frac{a_4}{2f},\qquad
b_2=-\frac{2a_1}{f},\qquad
b_3=-\frac{2}{f}(a_2-a_1')+\frac{2L\vartheta_1}{f},\qquad
b_4=-\frac{2a_3}{f},\qquad
b_5=-2b_1.
\label{evencoeffb}
\end{align}
The $c_i$ coefficients are
\begin{align}
c_1={}&-\frac{a_1}{fh}+\frac{2r\vartheta_1}{f},
\notag\\
c_2={}&\frac{f}{s}\Biggl\{
f'\Biggl[
\frac{r^2}{2f}\Biggl\{\frac{h}{2}\left(-3G_{3,X}+8G_{4,\phi X}\right)\phi'^2
-\frac{h^2}{2}\left(-G_{3,XX}+2G_{4,\phi XX}\right)\phi'^4-G_{4,\phi}\Biggr\}\notag\\
&-\frac{h\phi'r}{f}\Biggl\{\frac{h^2}{2}\left(2G_{4,XXX}-G_{5,\phi XX}\right)\phi'^4
-\frac{h}{2}\left(12G_{4,XX}-7G_{5,\phi X}\right)\phi'^2+3G_{4,X}-3G_{5,\phi}\Biggr\}\notag\\
&+\frac{h\phi'^2}{4f}\left\{G_{5,XXX}h^3\phi'^4-G_{5,XX}h(10h-1)\phi'^2
+3G_{5,X}(5h-1)\right\}\Biggr]\notag\\
&+r^2\phi'\left\{\frac12G_{2,X}-G_{3,\phi}
-\frac{h}{2}\left(G_{2,XX}-G_{3,\phi X}\right)\phi'^2\right\}\notag\\
&+2r\Biggl\{\frac{h}{2}\left(-3G_{3,X}+8G_{4,\phi X}\right)\phi'^2
-\frac{h^2}{2}\left(-G_{3,XX}+2G_{4,\phi XX}\right)\phi'^4-G_{4,\phi}\Biggr\}\notag\\
&-\frac{h^3}{2}\left(2G_{4,XXX}-G_{5,\phi XX}\right)\phi'^5
+\frac{h}{2}\left\{2(6h-1)G_{4,XX}+(1-7h)G_{5,\phi X}\right\}\phi'^3\notag\\
&-(3h-1)\left(G_{4,X}-G_{5,\phi}\right)\phi'\Biggr\}
+\frac{20h^2\phi'^3}{s}\Biggl[
\frac{3}{40}F_{5,XX}f'h^3\phi'^5+\frac{1}{20}h^2F_{4,XX}(f'r+f)\phi'^4\notag\\
&-\frac{21}{20}F_{5,X}f'h^2\phi'^3-\frac{11}{20}F_{4,X}h(f'r+f)\phi'^2
+\frac{21}{8}F_5f'h\phi'+F_4(f'r+f)\Biggr],
\notag\\
c_3={}&-\frac12s\,r^2
\frac{\partial{\cal E}_{11}}{\partial\phi},
\notag\\
c_4={}&\frac14s\Biggl[
\frac{h\phi'f'}{f}\left\{2G_{4,X}-2G_{5,\phi}
-h\left(2G_{4,XX}-G_{5,\phi X}\right)\phi'^2
-\frac{h\phi'}{r}\left(3G_{5,X}-hG_{5,XX}\phi'^2\right)\right\} \notag\\
&+4G_{4,\phi}+2h\left( G_{3,X}-2G_{4,\phi X}\right)\phi'^2
+\frac{4h}{r}\left(G_{4,X}-G_{5,\phi}\right)\phi'
-\frac{2h^2}{r}\left(2G_{4,XX}-G_{5,\phi X}\right)\phi'^3\Biggr]
\notag\\
&+\frac{h\phi'^3}{rs}\Biggl[
\frac32F_{5,X}f'h^2\phi'^3+F_{4,X}h\left(\frac12f'r+f\right)\phi'^2
-\frac{21}{2}F_5f'h\phi'-5F_4\left(\frac12f'r+f\right)\Biggr],\notag\\
c_5={}&-h\phi'c_4-\frac{h}{2r}s\,\mG-\frac{f'}{2f}a_4-\vartheta_2 \phi'^2,
\notag\\
c_6={}&\frac{f'\phi'}{8f}a_1+\frac{f'r}{2f}a_4-\frac14\phi'c_2+\frac12h\phi'rc_4
+\frac14hs\,\mG+\frac12r\vartheta_2 \phi'^2.
\label{evencoeffc}
\end{align}
The $d_i$ coefficients are
\begin{align}
d_1={}&\frac{a_4}{2f},\qquad
d_2=2hc_4+2\vartheta_2 \phi',\notag\\
d_3={}&-\frac{1}{r^2}\left(\frac{2\phi''}{\phi'}+\frac{h'}{h}\right)a_1
+\frac{2f}{(f'r-2f)\phi'}\left(
\frac{2\phi''}{h\phi'r}+\frac{f'^2}{f^2}
-\frac{f'h'}{fh}-\frac{2f'}{fr}
+\frac{2h'}{hr}+\frac{h'}{h^2r}\right)a_4\notag\\
&+\frac{f'r-2f}{fr}\frac{\partial a_4}{\partial\phi}
+\frac{s}{\phi'r^2}\mF-\frac{fs}{(f'r-2f)\phi'}
\left(\frac{f'}{fr}+\frac{2\phi''}{\phi'r}
+\frac{h'}{hr}-\frac{2}{r^2}\right)\mG
-\frac{3F_5fh\phi'^3(h'\phi'+2h\phi'')}
{r^2s},\notag\\
d_4={}&\frac{hs}{2r^2}\mG.
\label{evencoeffd}
\end{align}
Finally, the $e_i$ coefficients are
\begin{align}
e_1={}&\frac{1}{\phi'fh}\left[
\left(\frac{f'}{f}+\frac{h'}{2h}\right)a_1
-2a_1'+a_2-2rha_6\right]+\frac{f\left[(h'r+2h-L)\vartheta_1+2hr\vartheta_1'\right]
-2f'hr\vartheta_1}{\phi' f^2 h},\notag\\
e_2={}&-\frac{1}{2\phi'}\left(
\frac{f'}{f}a_1+2c_2+4hrc_4\right)
-2r\vartheta_2,\notag\\
e_3={}&\frac{1}{2}s\,r^2
\frac{\partial{\cal E}_{\phi}}{\partial\phi},
\notag\\
e_4={}&\frac{c_4'}{\phi'}
-\frac{f'a_4'}{2fh\phi'^2}
-\frac{s}{2r\phi'^2}\mG'
+\frac{a_1}{h\phi'r^2}
\left(\frac{\phi''}{\phi'}+\frac{h'}{2h}\right)+\frac{a_4}{4h\phi'^2}\Biggl[
\frac{(f'r-6f)f'}{f^2r}
+\frac{h'(f'r+4f)}{rfh}
-\frac{4f(h'\phi'+2h\phi'')}
{\phi'h^2r(f'r-2f)}\Biggr]\notag\\
&+\frac{h'}{2h\phi'}c_4
-\frac{f'r-2f}{2fhr\phi'}\frac{\partial a_4}{\partial\phi}
+\frac{(f'hr-f)\mF}
{2r^2s\,h^2\phi'^2}+\frac{s}{2r\phi'^2h}
\left[
\frac{f(h'\phi'+2h\phi'')}
{h\phi'(f'r-2f)}
+\frac{2f-f'hr}{2fr}
\right]\mG\notag\\
&-\frac{h (rf' + 2f)\vartheta_1+r \phi' \vartheta_2
(f' h r - 2 f h - 2f)}
{2\phi'^2 rh^3(f'r-2f)} (h' \phi' + 2 h \phi'')\,.
\label{evencoeffe}
\end{align}

We note that our normalization of ${\cal E}_{\phi}$ defined in
Eq.~(\ref{EphiGLPV}) is smaller by a factor of two than
that of Ref.~\cite{Kase:2023mho}.  Accordingly, the coefficient multiplying
$\partial{\cal E}_{\phi}/\partial\phi$ in $e_3$ is twice as large, namely
$1/2$ instead of $1/4$ in Ref.~\cite{Kase:2023mho}.

\bibliographystyle{mybibstyle}
\bibliography{bib}

\end{document}